\documentclass[twocolumn,prb,superscriptaddress,showpacs,footinbib]{revtex4}

\usepackage[utf8]{inputenc}
\usepackage[T1]{fontenc}
\usepackage{times}
\usepackage[american]{babel}
\usepackage{graphicx}
\usepackage{xcolor}
\usepackage{booktabs}
\usepackage{amsmath, amssymb}
\usepackage{braket}    
\usepackage{xspace}
\usepackage[unicode]{hyperref}
\usepackage{appendix}
\renewcommand{\vec}{\boldsymbol}
\renewcommand{\vec}{\boldsymbol}
\newcommand{\db}[2][]{\text{d}^{#1}#2}

\newcommand{\rhos}{\ensuremath{\gamma_\text{s}}}

\newcommand{\avr}[1]{\braket{#1}}
\newcommand{\Avr}[1]{\Braket{#1}}
\newcommand{\abs}[1]{|#1|}
\newcommand{\al}{\ensuremath{\alpha}}
\newcommand{\be}{\ensuremath{\beta}}

\renewcommand{\Re}{\operatorname{Re}}
\renewcommand{\Im}{\operatorname{Im}}

\begin{document}

%%
%% FRONTMATTER
%%
\title{Correlation effects and collective excitations in bosonic bilayers: role of quantum statistics, superfluidity and dimerization transition}
\date{\today}

%%%% REVTEX ONLY
\author{A.~Filinov}
\email{filinov@theo-physik.uni-kiel.de}
\affiliation{Institut für Theoretische Physik und Astrophysik,
Christian-Albrechts-Universitat, Leibnizstr. 15, D-24098 Kiel, Germany}
\affiliation{Joint Institute for High Temperatures RAS, Izhorskaya Str.~13, 125412 Moscow, Russia}
%%%% REVTEX ONLY

\begin{abstract}

A two-component two-dimensional (2D) dipolar bosonic system in the bilayer geometry is considered. By performing quantum Monte Carlo simulations in a wide range of layer spacings we analyze in detail the pair correlation functions, the static response function, the kinetic and interaction energies. By reducing the layer spacing we observe a transition from weakly to strongly bound dimer states. The transition is accompanied by the onset of short-range correlations, suppression of the superfluid response, and rotonization of the excitation spectrum. A dispersion law and a dynamic structure factor for the {\em in-phase} (symmetric) and {\em out-of-phase} (antisymmetric) collective modes, during the dimerization, is studied in detail with the stochastic reconstruction method and the method of moments. The antisymmetric mode spectrum is most strongly influenced by suppression of the inlayer superfluidity (specified by the superfluid fraction $\gamma_s=\rho_s/\rho$).  In a pure superfluid/normal phase only an acoustic/optical(gapped) mode is recovered. In a partially superfluid phase, both are present simultaneously, and the dispersion splits into two branches corresponding to a normal and a superfluid component. The spectral weight of the acoustic mode scales linearly with $\gamma_s$. This weight transfers to the optical branch when $\gamma_s$ is reduced due to formation of  dimer states.  In summary, we demonstrate how the interlayer dimerization in dipolar bilayers can be uniquely identified by static and dynamic properties.

\end{abstract}


%%%% REVTEX ONLY
\maketitle
%%%% REVTEX ONLY
%%
%% TEXTBODY
%%

\section{Introduction}\label{intro}

A stack of vertically coupled layers with dipolar Bose-Einstein condensates is a remarkable physical system with interesting static and dynamic properties. The intralayer and interlayer correlations play here a dominant role and determine the character of  the static response and the screening properties. The interlayer attractive dipolar interaction, in the bilayer (or multilayer) case, results in a variety of interesting physical phenomena, which include a peculiar behavior of the scattering cross-section and bound states,~\cite{kl1,zinner,arm1,bar1} interlayer pairing and superfluidity.~\cite{pik1,pik2,pot1} Even more intriguing is the behavior of the collective modes as they can be directly detected with experimental probes that couple to the particle density operators. A variety of experimental techniques existing in  solid state physics can help to perform such analyses: inelastic electron-scattering spectroscopy,~\cite{Liu,Kram,Eber} frequency-domain far-infrared or microwave spectroscopy\cite{Allen} and inelastic light-scattering spectroscopy.~\cite{Abst,Pin,Erik}

From the experimental side the use of layered and quasi-2D coupled BECs has a number of key advantages. It has been proposed recently,~\cite{Goral,DeMille,Barn,Mich,Wang} that the use of thin layers is an effective way to control the three-body losses and to significantly reduce the parameter space of the dynamical instability, the main problem encountered in experiments with dipolar gases. 

Recent experimental achievements with layered ultra-cold polar molecules~\cite{mirand} motivates our current studies of quasi-2D dipolar bosonic bilayers in strongly interacting regime. The bilayer and multilayered geometries have a great potential for realization of new exotic phases. One prominent example is the formation of vertically aligned chains composed of particles from different layers. Some preliminary theoretical predictions on the distribution and the length of chains based on the thermodynamic considerations has been presented recently.~\cite{Arm12} The next intriguing question is how quantum statistics will influence properties of the chains in a degenerate regime, critical temperature of BEC and superfluid transition. The dynamical properties are of high interest as well. The spectrum of collective density excitations will be strongly modified by the inter-chain dynamics and intra-chain modes, being in strong dependence with variation of interlayer distance and a strength of interlayer coupling.~\cite{Arm12} 

Similar to the predictions for a single 2D layer,~\cite{ast1,buch,fil2011} dipolar bilayer and multilayers are expected to undergo a crystallization transition at high densities.~\cite{ramos} All these possibilities, existing in multilayers, form an interesting topic for future analyses.

In our current studies we consider a bilayer geometry as the simplest case, where all mentioned effects can be analyzed in detail, without additional complications due to multilayer effects. It has been predicted theoretically that a dipolar bilayer undergoes a number of phase transition with variation of interlayer spacing. One prominent example is the formation of a two-particle complex -- a dimer.~\cite{wang,tre} It will be demonstrated below that a transition from weakly to strongly bound dimers significantly modifies all thermodynamic characteristics. The energetics (and the binding energy) of a single dimer state can be analyzed to a large extent analytically,~\cite{kl1,zinner} however, similar studies at finite densities and temperatures are more complicated.~\cite{Arm12} In particular, variation of the interlayer coupling leads to a quantum phase transition from single to two-component (pair) superfluidity, as was predicted by quantum Monte Carlo simulations.~\cite{kuklov,macia} The ground state properties and the spectrum of collective excitations has been analyzed recently using the hypernetted-chain Euler-Lagrange and the correlated basis function methods.~\cite{hufn} Of special interest are the bilayers with high population imbalance, where more complicated many-body states have been analyzed.~\cite{Klawunn,Pupillo,Matv}

In our present studies we consider a translationally invariant system where all dipoles are polarized perpendicular to the plane of 2D confinement and, as the result, within each layer they experience only repulsive interaction. The dimerization is possible due to the interlayer coupling, and for considered interlayer spacings bound states are limited by two-particle states. We perform path integral Monte Carlo simulations (PIMC) in the grand canonical ensemble~\cite{prokof} to get access to thermodynamic properties and the superfluid response. To characterize collective density excitations we reconstruct the dynamic structure factor via the stochastic optimization method.~\cite{fil2012} The results are  compared with the sum-rule approach.~\cite{lipp,Ark2010} This comparison is aimed to clarify whether the method of moments is suitable to provide adequate description of cold Bose systems in a wide range of parameters. In particular, we demonstrate that fulfillment of the third  power moment of the spectral function, which includes interlayer static correlations, is crucial to correctly account the dimerization transition in the bilayer geometry.

The paper is organized as follows. In Sec.~\ref{model} we introduce the model of a 2D dipolar bilayer. In Sec.~\ref{dens} the density response function for a multicomponent system and its relation with the dynamic structure factor is specified. In Sec.~\ref{mom1} we present the method of moments and apply it to determine the dispersion relation of collective modes. In Sec.~\ref{statprop} we analyze the static thermodynamic characteristics and their dependence on the interlayer spacing and the dipole coupling strength. The structure of the pair distribution functions and energy characteristics allows to identify and follow in detail the dimerization transition.  The latter has a strong effect on the inlayer superfluidity. In Sec.~\ref{exc_sec} we present our results for the dynamic structure factor $S_\pm(q,\omega)$. For the mass-symmetric bilayer, the excitation spectrum can be analyzed in terms of its eigenmodes -- the in-phase and out-of-phase density oscillations. In this representation the density response function is a diagonal matrix and interpretation of the observed spectral features significantly simplifies. We discuss a connection of the excitation branches present in $S_\pm(q,\omega)$ with a superfluid and normal components of a Bose gas. The results are compared with the predictions from the method of moments (Sec.~\ref{mom1}). We finally draw our conclusions in Sec.~\ref{concl}.

\section{Model of two-component bilayer}\label{model}

The model bilayer system consists of quasi-2D planes with bosonic particles of the same mass and dipole moment. We consider a polarized system, when all dipolar moments are oriented perpendicular to the planes. The dipole-dipole pair interaction is purely repulsive within the same layer, while it can be both repulsive and attractive for bosons from different layers depending on their spatial separation. The contact pair interaction, being important for simulations in 3D geometry, can be neglected in the quasi-2D geometry being completely screened by the intralayer repulsion. 

The inlayer particle density is controlled by the chemical potential $\mu$, being a free parameter in  our simulations. We analyze a symmetric bilayer and take $\mu_1=\mu_2$. The effects due to the density imbalance are not considered.

The Hamiltonian can be written in a general form as for a two-component system
\begin{align}
  &\hat H=\sum_{\al=1}^2 \hat{H}_{\al}+\sum\limits_{i=1}^{N_1} \sum_{j=1}^{N_2}\hat{V}_{12},\\  &\hat{H}_{\al}=\sum\limits_{i=1}^{N_\al} \left[- \frac{\hbar^2}{2 m_\al} \nabla_i^2 +\hat{V}_{\al \al}\right],\label{Ham2}
 \end{align}
where $m_{\alpha}$ ($\al=1,2$) are the particle masses of two kinds, and $V_{\al \al}$ and $V_{12}$ are the intra- and interlayer interaction potentials, correspondingly. For the polarized system we use
\begin{align}
V_{\al \al}=\frac{p_{\al}^2}{r^3}   \quad (\al=1,2),\quad
 V_{12}=p_1 p_2\, \frac{r^2- 2 z^2}{(r^2+z^2)^{5/2}},
\end{align}
where $r$ is the in-plane (projected) two-particle relative distance, and $z$ is the interlayer separation. For a symmetric bilayer we take, $p_1=p_2=p$ and $m_1=m_2=m$.

We use the same length- and energy-units as in Ref.~\cite{fil2012}: $a=1/\sqrt{\rho_\al}$ ($\rho_\al$ is the inlayer density) and $E_0=\hbar^2/m_\al a_\al^2$. In the case of different particle species, one of the components is used as a reference system.

The (reduced) parameters varied in the simulations are: i) temperature, $T=k_BT/E_0$, ii) the interlayer spacing, $d=z/a$, iii) the effective length of dipole interaction, $a_d=m_\al p_\al^2/\hbar^2$, or the equivalent parameter, $D=a_d/a$. 

In the beginning of each simulation with the specified parameters $\{d,D,T\}$, the chemical potential, $\mu=\mu/E_0$, is adjusted such that the average inlayer density matches the relation $\rho_\al(T,d,D,\mu)\approx 1$. This ensures that the used length scale coincides with the average interparticle distance of the reference component, i.e. $ a_\al \approx 1$. Alternatively, the input parameters $\{T,d,D,\mu\}$ have to be rescaled to satisfy $\rho_\al=1$ (or $a_\al=1)$.

The simulation have been performed for three dipole coupling strengths, $D=\{0.1,1,5.5\}$, referenced in the following as weak, moderate and strong coupling regime. In the single layer they are distinguished by absence/presence of a roton feature in the excitation spectrum.~\cite{fil2012} As will be shown below, addition of a second layer effectively enhances the coupling strength to $D^\star \sim 8D$ and, can be used to control the depth of the roton minimum by variation of the interlayer spacing. Therefore, a bilayer geometry is quite favorable for experimental realizations of strongly correlated Bose gases. Here, a similar depth of the roton minimum, as in a single layer system, can be reached by a factor $\sim 64$ lower inlayer density. Possible experimental realizations include the atoms with large magnetic moment~\cite{dysp,erbium} ($^{164}$Dy, $^{168}$Er) and heteronuclear molecules~\cite{krb1,krb2,rbcs,lics,lik} (KRb, RbCs, LiCs, LiK).

The temperature is kept fixed at $T=1$. In this regime a single layer is $100\%$ superfluid.~\cite{fil2011} The average density and the particle number are controlled via the chemical potential $\mu$ and the simulation box size $L$. Both are chosen to accumulate  $N_{\al}=70\ldots 100$ particles per layer, i.e. $\rho_\al=N_\al/L^2\sim 1$ for $L=9$, to reduce finite size effects. The number of Monte Carlo samples used for the thermodynamic averages is $10^8\ldots 10^9$.

\section{Density response function}\label{dens}

The knowledge of the density response function and its poles provides information on the collective density excitations. Within the linear response theory one considers a weak external field which produces a perturbation via coupling to the density operator
 \begin{align}
  \hat{\rho}_\al(\vec{r})=\sum\limits_{i=1}^{N_\al} \delta (\vec{r}-\vec{r}_{j\al}),
 \end{align}
expressed in the momentum representation as
 \begin{align}
  &\hat{\rho}_{\vec q \al}^{+}=\sum\limits_{i=1}^{N_\al} e^{i \vec q \vec r_{i\al}}=\hat{\rho}_{-\vec q \al},\\
  &\hat{V}_{\al}^{\text{ext}}(\vec{q},t)=V(t)\hat{\rho}_{\vec q\al}^{+}.
 \end{align}
The index $\al$ denotes the particle type (layer) in a multi-component (multilayer) system.

The density-density response function is defined via the Green's function of two density operators
\begin{align}
\chi_{\al \be}(\vec{q},\omega+i\nu)=&-\frac{i}{\Omega\hbar} \int\limits_{0}^{\infty} \db t \, e^{i (\omega+i\nu) t}\Avr{[\rho_{\vec q \al}(t),\rho_{-\vec{q} \be}(0)]}.\label{densresp}
\end{align}
It forms a matrix for a multi-component system, which links the density response of a subsystem $\al$ with an external perturbation field $V^{\text{ext}}_{\al\be}(q,\omega)$ applied either to the same ($\be=\al$) or a different ($\be\neq \al$) subsystem. When $\al=\be$ ($\al\neq \be$) the applied density perturbation and the measured density response corresponds to the same (different) subsystem. 

With an external perturbation field, $V^{\text{ext}}_{\al\al}$ applied to a subsystem $\al$, the density excitations are induced also in other subsystems, as they are mutually dynamically coupled via the interaction fields $V_{\al\be}$. These perturbations act as additional (induced) external fields. The combined effect of all fields on a subsystem $\al$ can be written (within the linear response) in the form 
 \begin{align}
 \avr{\rho_\al^{\text{ind}}(\vec q,\omega)}=\sum\limits_{\be} \tilde{\chi}_{\al\be}(\vec q,\omega) V^{\text{ext}}_{\al\be}(\vec q,\omega),
 \label{chi_ab}
 \end{align}  
 which includes a contribution from both diagonal and off-diagonal elements of the density response matrix~(\ref{densresp}). As a result, in the density fluctuation spectrum of the subsystem $\al$ we see new resonances -- collective modes corresponding to the coupled density oscillations in different subsystems.\cite{w3sumref} In a strongly interacting bilayer (determined by the strength of $V_{12}$) this effect becomes important and leads to hybridization effects in the excitation spectra.
 
In an experiment, one measures the dynamic structure factor related with the density response function via the fluctuation-dissipation theorem (FDT)
\begin{align}
 \Im \chi_{\al\be}(\vec q,\omega)=-\pi \rho_{\al\be} (1-e^{-\beta \omega}) S_{\al\be}(\vec q,\omega).\label{fdt}
\end{align}

The main idea of the present approach is to obtain $S_{\al\be}(\vec q,\omega)$ not from the real time dynamics~(\ref{densresp}), but from the evolution of a quantum system in the imaginary-time. In this case, one needs to apply a special reconstruction procedure and solve the inverse problem specified by the relation between the spectral density $S_{\al\be}(\vec q,\omega)$ and the imaginary-time density-density correlation function defined for $\tau \in [0,\be]$
\begin{align}
G_{\al\be}(\vec q,\tau)=\frac{1}{\avr{N}}\avr{\hat{\rho}_{\vec q \al}(\tau) \hat{\rho}_{-\vec q \be}(0)}=\int\limits_{-\infty}^{\infty} d\omega \, e^{-\tau \omega} S(\vec q,\omega) \label{g2}.
\end{align}
The used approach is shortly reviewed in Appendix~\ref{SO}, and discussed more in detail  in Ref.~\cite{Mish,fil2012}

\section{Frequency power moments}\label{mom1}

In this section we introduce several sum rules valid for $\chi(q,\omega)$, and apply them to obtain a dispersion relation of collective modes, $\omega(q)$, in the two-resonance approximation. 

The well-known Feynman's ansatz  assumes that a single delta-peak resonance exhausts (provides the main contribution to) an excitation spectrum. Indeed, the involved parameters, an energy and a spectral weight, can be chosen to satisfy the $\mu_{0s}$ and $\mu_{1s}$ sum rules~[see Eq.~\ref{omegak2}] by the choice: $\omega^f(q)=\mu_{1s}(q)/\mu_{0s}(q)$ and $S^f(q)=\mu_{0s}(q)$. Finite temperature corrections can be included~\cite{fil2012} and become important in the phonon and roton part of the spectrum when $\hbar \omega^f(q) \lesssim k_B T$.

If additional power moments are available, they provide a significant improvement with respect to the Feynman approximation. The use of the inverse power moment gives a lower upper bound for the dispersion relation, $\omega^\chi(q)=\mu_{0s}/\mu_{-1s}$, and a better estimate of the roton energy.~\cite{string,fil2011,fil2012}

The third power moment $\mu_{3s}$ has been evaluated for a variety of classical and quantum systems, including liquid $^4$He,~\cite{puff,zamb2000} one-component classical plasmas in 2D and 3D geometry,~\cite{iwamo,Arkhip} binary mixtures of ions and electron systems in bilayers.~\cite{DeLu} Of special interest is the generalization to two- and multi-component systems, when particle species can differ in their mass and pair interaction potential.~\cite{w3sumref}

In the present work we generalize Feynman's ansatz by including a second mode, which represents either a second high-frequency branch or a multiexcitation continuum. It can be approximately represented by a second delta peak, carrying a finite spectral weight and contributing to all sum rules $\mu_k$. The developed approach (method of moments) is applied to the mass- and density-symmetric bilayer. We predict changes in the dispersion relation of collective modes as one varies the interlayer spacing $d$, temperature $T$, and the dipolar coupling parameter $D$. Obviously, not all effect can be captured within a simplified theory of moments based on the delta peak resonances. Nevertheless, even such a simplified treatment allows to capture main physics, and helps to understand hybridization of the modes due to the interlayer coupling.

In Sec.~\ref{exc_sec} this method will be compared with the microscopic data for $S(q,\omega)$ from the path-integral Monte-Carlo (PIMC) simulations for a wide range of physical parameters. In our recent paper~\cite{fil2012} $S(q,\omega)$ was reconstructed from the imaginary time density-density correlation function. This procedure does not involve a priori assumption on a shape of spectral density, and can provide accurate results comparable to experimental data, as was demonstrated recently  for liquid $^4$He.~\cite{reatto} 

Below we introduce the method of moments (MM) and derive our main results for one and two-component systems. In Sec.~\ref{exc_sec} results will be checked against a full dynamic structure factor.

\subsection{One component system}\label{one-comp}
The frequency power moments are introduced as integral properties of a spectral density. They are defined either via the imaginary part of the density-density response function, $\tilde \chi=-\Im [\chi]/\pi \rho$, as
 \begin{align}
  \mu_k(\vec q)=\Avr{\omega^k}(\vec q)=\int\limits_{-\infty}^{\infty}  \db \omega \, \omega^k \, \tilde\chi(\vec q,\omega)
  \label{omegak}
 \end{align}
or via the dynamic structure factor
 \begin{align}
  \mu_{ks}(\vec q)=\int\limits_{-\infty}^{\infty}  \db \omega \, \omega^k \, S(\vec q,\omega),
  \label{omegak2}
 \end{align}
mutually related with $\tilde \chi$ by the FDT~(\ref{fdt}).
 
The following results are obtained directly from Eq.~(\ref{densresp})
 \begin{align}
  &\mu_1(\vec q)=\Avr{\omega}(\vec q)=\Avr{[\dot{\hat\rho}_{\vec q},\hat\rho_{-\vec q}]},\\
  &\mu_3(\vec q)=\Avr{\omega^3}(\vec q)=\Avr{[\ddot{\hat\rho}_{\vec q},\dot{\hat\rho}_{-\vec q}]}.
 \end{align}
Using the Heisenberg equations they can be recast in the form of commutation relations with the Hamiltonian operator
 \begin{align}
  &\mu_1(\vec q)=\Avr{\left[[\hat\rho_{\vec q},\hat{H}],\hat\rho_{-\vec q}\right]},\\
  &\mu_3(\vec q)=\Avr{\left[[[\hat\rho_{\vec q},\hat H],\hat{H}],[\hat{H},\hat\rho_{-\vec q}]\right]}.
 \end{align}
 For a one-component system after substitution of~(\ref{Ham2}) [with $V_{12}=0$] we end up with the result
 \begin{align}
  &\mu_1(\vec q)=\rho\tilde\epsilon_{\vec q},\quad \rho=N/V,\label{w1}\\
  &\mu_3(\vec{q})=\tilde\epsilon_q^2 \left[\rho \tilde\epsilon_q/4+ (6/\mathcal{D})\, T(\vec{q})/V + \Avr{C(\vec{q})}\right],\label{w3}
 \end{align}
where $\tilde\epsilon_q=\hbar^2q^2/m$ and $\mathcal{D}$ is the system dimensionality. Several notes are necessary. The second term in~(\ref{w3}) has a meaning of a spatial component of  kinetic energy and depends on the momentum projection on a specific direction $\vec{e}_{q}=\vec{q}/q$
\begin{align}
  T(\vec{q})=\frac{\mathcal{D}}{2m} \Avr{\sum_{i=1}^N (\vec{e}_q\vec{p}_i)(\vec{e}_q\vec{p}_i)}.
 \end{align}
For a spatially isotropic system, after the angular averaging, this term reduces to the average kinetic energy, $T(q)=\Avr{E_{\text{kin}}}$.
  
The third term in Eq.~(\ref{w3}) is determined by the spatial distribution of particles and their pair interaction
 \begin{align}
 C(\vec{q})=&\frac{1}{V q^2} \sum_{i=1}^N \sum_{n\neq i}^N \left(e^{i \vec{q}(\vec{r}_i-\vec{r}_n)}-1\right)\nonumber \\ 
 &(\vec{e}_q\vec{\nabla}_i)(\vec{e}_q\vec{\nabla}_n) V_{in}(\vec{r}_i,\vec{r}_n).\label{cq}
 \end{align}
For polarized dipoles on a 2D plane (with repulsive isotropic interaction
$V_{11}=p^2/r^3$), this term after the angular averaging reduces to 
\begin{align}
 \Avr{C(q)}=\rho\int_0^{\infty} \db r\, r\, \rho g(r)\,  \frac{3 \pi p^2}{q^2 r^5} \large[ 3 - 3 J_0(qr)+ 5 J_2(qr) \large],\label{cq0}
 \end{align}
 where $\rho g(r)$ is the radial pair distribution function defined in the grand canonical ensemble as
 \begin{align}
 \rho g(r)=\frac{1}{\Avr{N}} \Avr{\sum_{i=1}^{N} \sum_{j=1}^{N}{}^{'} \delta(r-\left|\vec{r}_i-\vec{r}_j\right|)}\label{rho_nn}.
 \end{align} 
The long wavelength limit of~(\ref{cq0}) has been analyzed by Golden and Kalman~\cite{gold_dip} and can be expressed via the average interaction energy per volume
 \begin{align}
  \Avr{C(q\rightarrow 0)}=\frac{33 \pi \rho}{8} \int_0^{\infty} \db r\, r\, \rho g(r)\, \frac{p^2}{r^3}=\frac{33}{8} \frac{\Avr{E_{\text{int}}}}{V}.
 \end{align}
By taking the ratio of the power moments~(\ref{w1}),(\ref{w3}) we get an acoustic dispersion
 \begin{align}
   \omega(q\rightarrow 0)\leq \lim\limits_{q\rightarrow 0} \sqrt{\frac{\mu_3(q)}{\mu_1(q)}} \equiv cq,
   \label{cq1}
 \end{align}
with the sound speed defined as
 \begin{align}
  c=\sqrt{\frac{\hbar^2}{m}\left[ 3\Avr{\epsilon_{\text{kin}}}+ \frac{33}{8}\Avr{\epsilon_{\text{int}}}\right]} \label{cq2}.
 \end{align}
Here $\epsilon_{\text{kin}}=T(q)/\rho V$ and $\epsilon_{\text{int}}=\Avr{E_{\text{int}}}/\rho V$ are the kinetic and interaction energies per particle. 
Note, that in Eq.~(\ref{cq1}) we are only allowed to write an inequality, and, hence, the estimated isothermal sound speed provides an upper bound for the true sound speed of acoustic phonons.

The long wavelength limit of $\mu_3(q)$ gets a contribution from the kinetic energy and the correlation part. On the contrary, the large-$q$ behavior is determined by the first term in~(\ref{w3}), with the scaling $\propto q^6$. Hence the free-particle excitations, with the energy $\tilde \epsilon_q$, dominate the third moment at large momenta.

 \subsection{Two-mode solution}\label{twomode}
In this section we generalize the canonical Feynman ansatz.
For the density-density response function we write a two-mode ansatz in the form with two delta functions
\begin{align}
 \tilde\chi(\vec q,\omega)=-\frac{\Im [\chi(\vec q,\omega)]}{\pi\rho}=\sum_{i=L,H} S^i [\delta(\omega-\omega^i)-\delta(\omega+\omega^i)].
 \label{chiF}
\end{align}
With this definition the density prefactor drops out in the sum rules~(\ref{w1}) and (\ref{w3}), while the third moment now contains the kinetic and interaction energy per particle. The high-energy mode $\omega^H$ represents either an additional quasi-particle excitation branch or a combined effect including a multiexcitation continuum. The $q$-dependence is omitted in the used notations: $\omega^i= \omega^i(\vec q)$ and $S^i= S^i(\vec q)$. 

The substitution of (\ref{chiF}) in~(\ref{omegak}) [for $k=-1,0,1,3$] defines a closed system of equations with respect to the free-parameters of the two mode ansatz
\begin{align}
 &\mu_{0s}=S^L\coth \frac{\beta\omega^L}{2}+S^H\coth \frac{\beta\omega^H}{2},\nonumber\\
 &\mu_{1}=S^L \omega^L+S^H \omega^H,\nonumber\\
 &\mu_{-1}=S^L/\omega^L+S^H/\omega^H,\nonumber\\
 &\mu_3=S^L (\omega^L)^3+S^H (\omega^H)^3.\label{f-sums-equation}
\end{align}
Note, that the zero moment $\mu_{0s}$ is evaluated from~(\ref{omegak2}) and via the FDT includes finite temperature effects. Otherwise, due to the antisymmetry property, $\tilde\chi(q,-\omega)=-\tilde\chi(q,\omega)$, all even-power moments are exactly zero. The odd-moments of (\ref{omegak}) and (\ref{omegak2}) obey an exact relation
\begin{align}
 \frac{1}{2}\, \mu_{(2k+1)}=\mu_{(2k+1)s}.
\end{align}

The system of equations~(\ref{f-sums-equation}) can be solved numerically. First, we introduce a definition of several upper bounds for the dispersion relation $\omega^L(q)$
\begin{align}
 &\omega^{\chi}(\vec q)=\frac{\mu_{0s}}{\mu_{-1s}}=\frac{S(\vec q)}{\abs{\Re [\chi(\vec q,0)]/2\rho}},\label{w_chi0}\\
 &\omega^{f}(\vec q)=\frac{\mu_{1s}}{\mu_{0s}}=\frac{\hbar^2 q^2}{2m \, S(\vec q)},\\
 &\omega^{\mu_3}(\vec q)=\sqrt{\frac{\mu_{3s}}{\mu_{1s}}}=\sqrt{\frac{\mu_{3s}}{\hbar^2 q^2/2m}},
 \label{3upbound}
\end{align}
which satisfy the inequality~\cite{lipp}
\begin{align}
  \omega^{\chi}(\vec q) \leq \omega^{f}(\vec q) \leq \omega^{\mu_3}(\vec q).
 \label{border}
\end{align}
For zero temperature this result is exact. For finite temperatures the order of the upper bounds can change.

As a next step, we express $\{\omega^H,S^L,S^H\}$ via the dispersion relation $\omega^L(q)$ considered as a free parameter. The  first three equations in~(\ref{f-sums-equation}) are reduced to the quadratic form with respect to $\omega^H$
\begin{align}
 A(\omega^L)\cdot [\omega^H]^2+ B(\omega^L,\omega^H)\cdot \omega^H +C(\omega^L)=0,
 \label{wH-eq}
\end{align}
with the prefactors defined as
\begin{align}
 &A(\omega^L)=1-\frac{\omega^L}{\omega^{\chi}}\coth \frac{\beta \omega^L}{2},\label{AA}\\
 &B(\omega^L,\omega^H)=\left(\frac{[\omega^L]^2}{\omega^{\chi}}-\omega^f\right)\coth \frac{\beta \omega^H}{2},\\
 &C(\omega^L)=\omega^L\left(\omega^f \coth \frac{\beta \omega^L}{2} -\omega^L\right)\label{BB}.
\end{align}
In a first approximation we can substitute, $\coth \beta \omega^H/2\approx 1$. In a quantum case temperature is typically low, $k T \lesssim \hbar \omega^H$, and $B(\omega^L,\omega^H)$ shows only a weak dependence on $\omega^H$. Next, Eq.~(\ref{wH-eq}) is solved numerically
\begin{align}
 (\omega^H)^{(n)}=\max \left[\frac{-B^{(n)} \pm \sqrt{B^{(n)2}-4 A\, C}}{2A}\right],\label{omegaH} 
\end{align}
by successive iterations and using the standard Newton's method. The iterations start from the zero-order: $B^{(0)}=\left((\omega^L)^2/\omega^{\chi}-\omega^f\right)$, when  temperature effects are neglected. For $\beta \omega^H \gg 1$ the energy $(\omega^H)^{(n)}$ converges in few iterations. A corresponding solution is obtained for each wavenumber $q$.

Next, the spectral weights of two modes can be expressed in terms of the frequencies $\omega^L$ and $\omega^H \equiv(\omega^H)^{(n)}$ as
\begin{align}
 &S^H=S(q) \frac{\omega^f \coth \frac{\beta \omega^L}{2} -\omega^L}{\omega^H \coth \frac{\beta \omega^L}{2} -\omega^L \coth \frac{\beta \omega^H}{2}},\\
 &S^L=\frac{S(q)- S^H\coth \frac{\beta \omega^H}{2}}{\coth \frac{\beta \omega^L}{2}}.
 \label{sH-sL}
\end{align}
We perform substitution in the third moment $\mu_3$ in Eq.~(\ref{f-sums-equation}),
and evaluate the deviation from the reference value given by~(\ref{w3}). 

Finally, we scan over different input frequencies $\omega^L$, and repeat all the steps in~(\ref{wH-eq})-(\ref{sH-sL}) to find an optimal value $\omega^{L\star}$ which provides best agreement with~(\ref{w3}).

In two special cases ($A=0$ or/and $C=0$), the solution of~(\ref{wH-eq}) becomes degenerate:  $\omega^H=\omega^L$. In this case, we use a one-mode ansatz and set $S^H=0$. For the dispersion relation $\omega^L(q)$ we get two possible  solutions
\begin{align}
 &A=0: \quad \omega^L=\omega^{\chi} \tanh \frac{\beta \omega^L}{2},\\
 &C=0: \quad \omega^L=\omega^{f} \coth \frac{\beta \omega^L}{2}.
\end{align}
Both results coincide with the upper bounds for $\omega(q)$ derived in Ref.~\cite{fil2012}

\subsection{Two-component system}\label{twocomp}

The method of moments, discussed above, can be directly transferred to a two-component system.
Similar to Sec.~\ref{one-comp}, first we need to evaluate the power moments of the density response matrix, using corresponding commutation relations between the density operator and the two-component Hamiltonian~(\ref{Ham2}). With the spectral density defined as
 \begin{align}
\chi_{\al\be}(\vec q,\omega)=-\Im [\chi_{\al\be}(\vec q,\omega)]/\pi,
 \end{align}
the commutation relations now read
\begin{align}
  &\Avr{\omega}_{\al\be}(\vec q)=\Avr{\left[[\hat\rho_{\vec q \al},\hat{H}],\hat\rho_{-\vec q \be}\right]},\\
  &\Avr{\omega^3}_{\al\be}(\vec q)=\Avr{\left[[[\hat\rho_{\vec q \al},\hat H],\hat{H}],[\hat{H},\hat\rho_{-\vec q \be}]\right]}.
\end{align}
Their evaluation results in the following sum-rules
\begin{align}
&\mu_{1,\al\be}(\vec q)=\rho_\al \tilde\epsilon_{q\al} \delta_{\al\be},\label{wab}\\
&\mu_{3,\al\al}(\vec{q})= \mu_{3\al}(\vec{q})-\tilde\epsilon_{q \al}^2 \Avr{C_{12}^{(1)}(\vec{q})},\label{w3aa}\\
&\mu_{3,\be\be}(\vec{q})= \mu_{3\be}(\vec{q})-\tilde\epsilon_{q \be}^2 \Avr{C_{12}^{(1)}(\vec{q})},\\
&\mu_{3,12}(\vec{q})= \tilde\epsilon_{q 1}\, \tilde\epsilon_{q 2}\Avr{C_{12}^{(2)}(\vec{q})}.\label{w3ab}
\end{align}
Here $\mu_{3\al(\be)}$ is the third moment~(\ref{w3}) of a one-component Hamiltonian $\hat H_{\al(\be)}$. The free-particle contribution is specified by $\tilde\epsilon_{q \al(\be)}=\hbar^2 q^2/m_{\al(\be)}$.
The correlation terms are expressed as ($\al\neq \be$)
 \begin{align}
 C_{\al\be}^{(1)}(\vec{q})= &\frac{1}{Vq^2} \sum\limits_{j=1}^{N_\al}\sum\limits_{i=1}^{N_\be}(\vec{e}_q\vec{\nabla}_{j\be})(\vec{e}_q\vec{\nabla}_{i\al}) V_{\al\be}(\vec{r}_{i\al},\vec{r}_{j\be}),\\
 C_{\al\be}^{(2)}(\vec{q})=& \frac{1}{V q^2} \sum\limits_{j=1}^{N_\al}\sum\limits_{i=1}^{N_\be} e^{-i\vec{q}(\vec{r}_{i\al}-\vec{r}_{j\be})} \nonumber \\
 & (\vec{e}_q\vec{\nabla}_{j\be})(\vec{e}_q\vec{\nabla}_{i\al}) V_{\al\be}(\vec{r}_{i\al},\vec{r}_{j\be}),
 \end{align}
and have an explicit dependence on the inter-component interaction potential $V_{12}$.
 
For polarized dipoles in a bilayer, the correlation terms simplify to one-dimensional integrals
\begin{align}
&\Avr{C_{12}^{(1)}(q)}=\bar{\rho}\int\limits_0^{\infty} \db r \, r \, g_{12}(r)\, [2 f_1(r) -f_2(r)],\label{c1}\\
&\Avr{C_{12}^{(2)}(q)}= \bar{\rho}\int\limits_0^{\infty} \db r \, r \, g_{12}(r)\times\,  \nonumber\\
   & \hspace{1.5cm} [2 J_0(qr) f_1(r)-(J_0(qr)-J_2(qr)) f_2(r)],\label{c2}\\
 &f_1(r)= \frac{3\pi p^2}{q^2\tilde{r}^5} \left(1- \frac{5 d^2}{\tilde{r}^2}\right),\;
 f_2=\frac{15\pi p^2 r^2}{q^2\tilde{r}^7} \left(1- \frac{7 d^2}{\tilde{r}^2}\right),   
\end{align}
where $\bar{\rho}=\frac{1}{2}(\rho_\al+\rho_\be)$ is the average density, $r$ is the in-plane interparticle distance and $\tilde{r}=\sqrt{r^2+d^2}$ includes the inter-layer spacing $d$. In the long wavelength limit ($q=0$), the correlation functions in~(\ref{c1}) and~(\ref{c2}) coincide.

The interlayer ($\al\neq \be$) radial pair distribution function introduced above 
measures the pair correlations between the layers
\begin{align}
 \rho_{\al\be}\, g_{\al\be}(r)=&\frac{1}{\Avr{N}}\Avr{\sum\limits_{i=1}^N \sum\limits_{j=1}^N \delta(r-\left|\vec{r}_{i\al}-\vec{r}_{i\be}\right|)}\nonumber\\
             =&\frac{2}{\Avr{N}}\Avr{\sum\limits_{i=1}^{N_\al} \sum\limits_{j=1}^{N_\be} \delta(r-\left|\vec{r}_{i\al}-\vec{r}_{i\be}\right|)}.\label{gab}
\end{align}       
It has the meaning of the conditional probability and is normalized to the average particle number in both layers, $\Avr{N}=\Avr{N_\al+N_\be}$. The offdiagonal density element $\rho_{\al\be}$ is defined by the limit, $\lim\limits_{r\rightarrow \infty} g_{\al\be}(r)=1$. This distribution contains important information on many-body effects and quantum statistics. The brackets denote the grand canonical ensemble average. The particle numbers $N_\al$ and $N_\be$ fluctuate around their mean values specified by the chemical potential.

The intra ($\al=\be$)  and interlayer ($\al\neq \be$)  static structure factors are defined via the corresponding Fourier transform
\begin{align}
S_{\al\be}(\vec{q})=&\delta_{\al\be}+ \int \db \vec{r} e^{i\vec{q}\vec{r}} \rho_{\al\be} [g_{\al\be}(\vec{r})-1],
 \label{Snm}     
\end{align}  
and can be simplified in a 2D spatially isotropic homogeneous system 
\begin{align}
S_{\al\be}(q)=\int\limits_0^{\infty} \db r \, \rho_{\al\be}\, [g_{\al\be}(r)-1] \, 2\pi r  \,J_0(qr).\label{Snm1}     
\end{align}

In the long wavelength limit the compressibility sum rule holds
\begin{align}
 S_{\al\be}(0)=\frac{2(\Avr{N_\al N_\be}-\Avr{N_\al} \Avr{N_\be})}{\Avr{N_\al+N_\be}}=\rho_{\al\be}\, k_B T\, \kappa_{\al\be}.
 \label{Snm2}
\end{align}
Simulations in the grand canonical ensemble allow to explicitly estimate $S_{\al\be}(0)$ and $\kappa_{\al\be}$ via the particle number fluctuations. The relation~(\ref{Snm2}) can be used as a test of Eq.~(\ref{Snm1}), which involves an extrapolated behavior of $\rho_{\al\be} g_{\al\be}(r)$ at distances beyond the simulation cell, see Eq.~(\ref{rhoint}).

The static limit of the density response function can be evaluated via integration of the density-density correlation function in the imaginary time
\begin{align}
&G_{\al\be}(\vec q,\tau)=\frac{2}{\avr{N_\al+N_\be}}\Avr{\hat{\rho}_{\vec q \al}(\tau) \hat{\rho}_{-\vec q \be}(0)},\label{g1}\\
&\frac{\Re [\chi_{\al\be}(\vec q,\omega=0)]}{2 \rho_{\al\be}}=-\int_0^{\beta} G_{\al\be}(\vec q,\tau) \, d \tau.
\label{gg2}
\end{align}
According to the definition above
\begin{align}
 G_{\al\be}(\vec q,\tau=0)=S_{\al\be}(\vec q).\label{gsk}
\end{align}
As a result, in the long wavelength limit we obtain
\begin{align}
\frac{S_{\al\be}(0)}{k_B T} =\frac{-\Re [\chi_{\al\be}(0,\omega=0)]}{2 \rho_{\al\be}}.
 \label{chinm}
\end{align}

In Sec.~\ref{statprop} we compare $S_{\al\be}(q)$ obtained independently via Eq.~(\ref{Snm}) and the direct estimator, Eqs.~(\ref{g1}),(\ref{gsk}), used at a set of wavenumbers, $\vec{q}_n=2 \pi n\vec{e}_q/L $. Some small deviations between these two estimators are mainly observed at low $q$ and originate from the interpolation formula used for $\rho_{\al\be}\, g_{\al\be}(r)$ at $r\geq L/2$
\begin{align}
 \rho_{\al\be} \, g_{\al\be}(r)=\rho_{\al\be} [1+a\,e^{-br}\,\sin(cr-d)].\label{rhoint}
\end{align}
The later includes several fit parameters $\{a,b,c,d\}$ adjusted to match the pair distribution function within the simulation cell close to the cell boundary.

\section{Thermodynamic properties of dipolar bilayers}\label{statprop}

In this section we discuss thermodynamic properties of dipolar bilayers obtained for  several coupling strength $D$. The free parameter is the interlayer spacing $d$. The observed changes in the static and thermodynamic properties will be used later in Sec.~\ref{exc_sec} for the discussion of collective excitation spectra. 

\subsection{Moderate coupling: $D=1$\; ($1.4 \leq U_0 \leq 3.3$)}\label{D1coup}

A single layer of bosonic dipoles at the coupling parameter $D=1$ shows a weak rotonization of the dispersion of collective longitudinal density modes.~\cite{fil2012} The intralayer correlations play an important role and their accurate treatment requires to go beyond the mean-field. At the same time, the Berezinskii-Kosterlitz-Thouless (BKT) temperature for the normal fluid-superfluid transition reaches its maximum value,~\cite{fil2011} $T_c^{\text{BKT}}=1.4$. At the temperature considered here ($T=1$), both layers are fully superfluid, once the layer spacing is sufficiently large.

\begin{figure}
 \begin{center} 
 \hspace{-0.35cm}\includegraphics[width=0.5\textwidth]{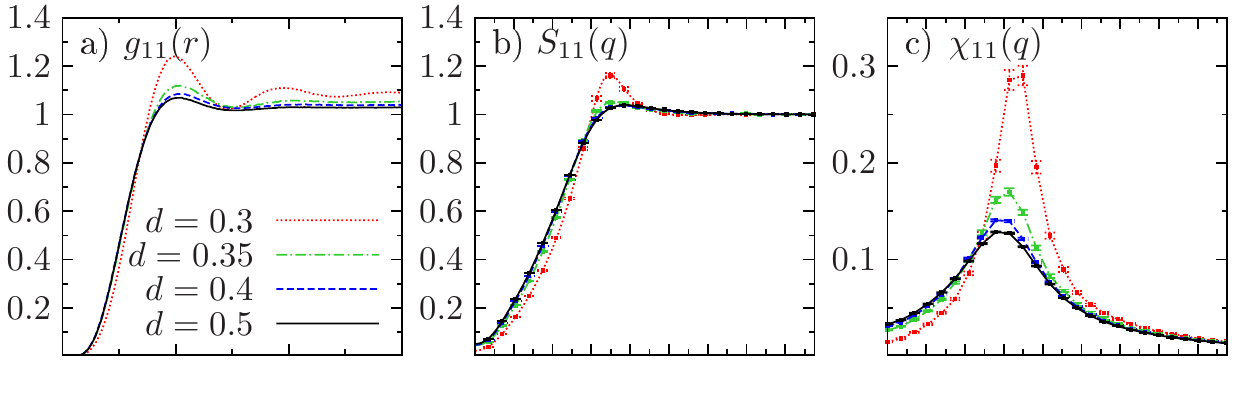}\\
 %{../../../quasi2d/New/d55ed40/data/2d/m1new/d1/gr11sk11chi11nox}\\ 
 \vspace{-0.25cm}
 \hspace{-0.35cm}\includegraphics[width=0.5\textwidth]{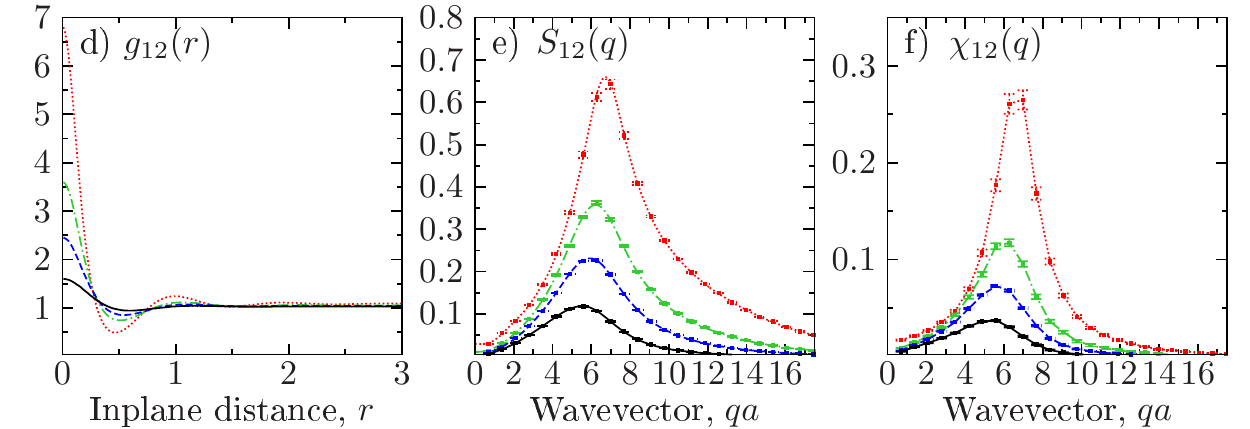}
 %{../../../quasi2d/New/d55ed40/data/2d/m1new/d1/gr12sk12chi12} 
 \end{center}
\vspace{-0.6cm} 
 \caption{a),d) Intra- and interlayer PDFs $g_{\al\be}(r)$. b),e) Static structure factor $S_{\al\be}(q)$. c),f) Density response function $\chi_{\al\be}(q)\equiv \abs{\Re \chi_{\al\be}(q,\omega=0)}/2\rho_{\al\be}$, Eq.~(\ref{gg2}). The interlayer spacing: $0.3\leq d\leq 0.5$. Simulation parameters: $D=1$, $\mu=20$, $V(L^2)=81$ and $T=1$. In panels b),e) the dotted(dashed) lines, on the top of the data for $S_{\al\be}(q)$ (symbols with errorbars), represent the Fourier transform~(\ref{Snm1}) of the interpolated radial intra(inter)-layer correlation functions $g_{\al\be}(r)$, Eqs.~(\ref{rho_nn}),(\ref{gab}) and~(\ref{rhoint}).}
 \label{fig:grsk-D1m1}
 \end{figure}

{\em Static properties.}\label{statD1}
Fig.~\ref{fig:grsk-D1m1} illustrates the induced changes in the static characteristics of a dipolar bilayer, once the interlayer spacing is varied. The intralayer PDF $g_{11}(r)$ reveals strong short-range correlations identified by the correlation hole at the origin. On the contrary, there is no sign of long-range spatial correlations: $g_{11}(r)$ becomes flat already after the first correlation shell ($r\gtrsim 2$). This situation changes when both layers are brought to a close vicinity ($d \sim 0.3$). Here, the first time, $g_{11}(r)$ starts to exhibit an oscillatory behavior, however, strongly damped. 

% Fig1
 
The strength of the interlayer correlations can be read out from the behavior of $g_{12}(r)$. When the layer spacing is continuously reduced from $d=0.5$ to $d=0.3$, we observe development of a peak at $r=0$, see Fig.~\ref{fig:grsk-D1m1}d. Dipoles from different layers demonstrate a tendency to a pairwise vertical alignment. The head-to-tail alignment within each layer
is excluded in our model by zero thickness of the layers. Such a possibility in physical systems depends on the ratio between the scattering length $a_s$ (represented as the radius of a hard core potential) and the oscillator length of vertical confinement, $l_z=\sqrt{\hbar/m_{\al}\omega_z}$. Our current model corresponds to a quasi-2D geometry when $a_s\gtrsim l_z$.

The sharp peak $g_{12}(0)$ observed at $d\lesssim 0.34$ is interpreted as a formation of strongly localized dimer states. Its halfwidth characterizes the inplane dimer size and is about a factor four smaller then the average interparticle distance $a\approx 1$ (in our units). Due to a spatial localization of dimers (note a well pronounced dip in $g_{12}$ around $r\sim 0.5$), they can be approximately treated as composite particles with double mass. The inlayer correlations in this regime can be characterized by a new effective dipole coupling $D^{\star} > D$, enhanced due to the dimer-dimer interaction. The excitation spectrum is expected to reveal a more pronounced roton feature. A clear signal of a roton is the oscillatory behavior of $g_{11}(r)$, as observed in Fig.~\ref{fig:grsk-D1m1}a for $d=0.3$.

The bound state formation is accompanied by the increase of the inlayer density: note, a systematic increase in the asymptotic value of $g_{11}(r)$ at large $r$. The static characteristics are also modified. In particular, the second and the third correlation shells in $g_{11}(r)$ are formed at $d\lesssim 0.34$. This is a clear trend that strongly localized dimers form a more ordered structure with short-range correlations but the system still remains in a homogeneous gas phase.

\begin{figure}
 \begin{center} 
\hspace{-.0cm}\includegraphics[width=0.47\textwidth]{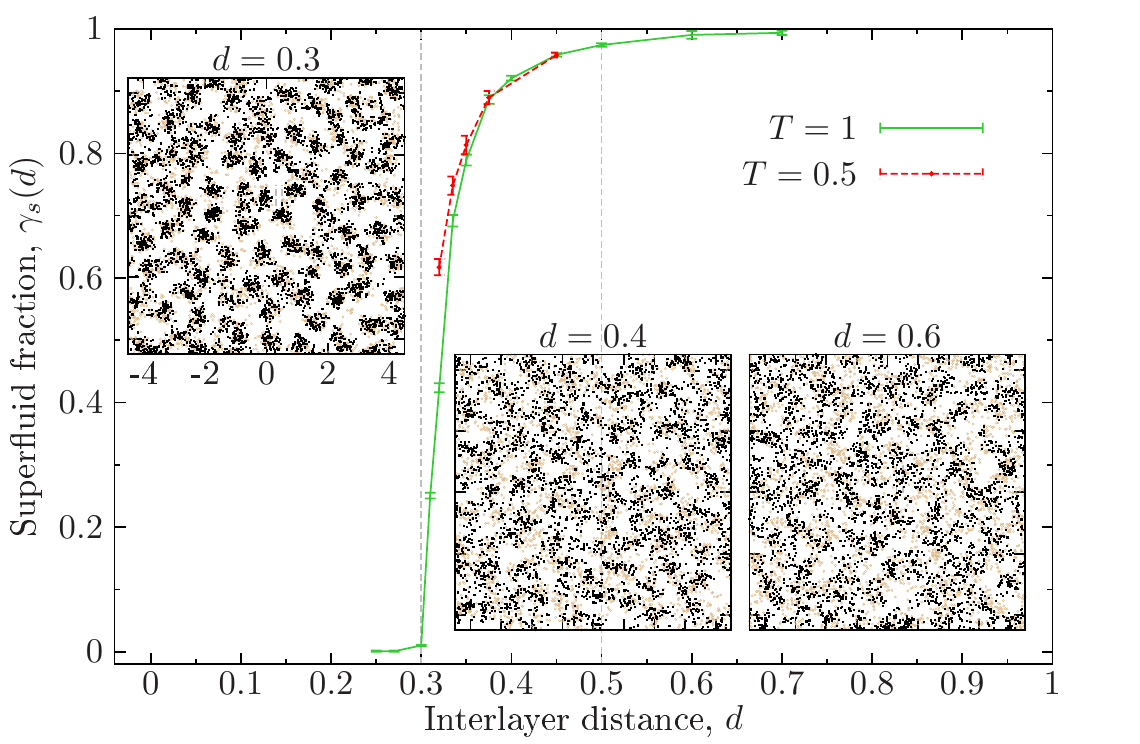}
%{../../../quasi2d/New/d55ed40/data/2d/m1new/d1/Ebind/confsuperl0}
\end{center}
 \vspace{-0.7cm} 
\caption{The $d$-dependence of the inlayer superfluid fraction $\gamma_s(d,T)$ at temperatures $T=1(0.5)$. Coupling $D=1$. The values $\gamma_{s\al}(T)$, measured in each layer $\al$, coincide within the statistical errors. The simulated system contains $\Avr{N}\sim 160$ particles, with $\Avr{N_{\al}}\sim 80$ particles per layer. The finite-size corrections can change absolute values, but not the observed trend in $\gamma_s(d)$. The insets show instantaneous density snapshots from PIMC at $d=0.3,0.4$ and $0.6$ (each quantum particle is presented as a cloud of $\sim 100$ beads). Two colors (black/brown) are used to distinguish top and bottom layer.}
\label{fig:D1m1-superf}
 \end{figure}

Additional information is provided by the static structure factor $S_{\al\be}(q)$. A broad peak around the wavenumber $qa\sim 2 \pi$ (corresponding to the inverse mean interparticle distance) is present both in $S_{11}(q)$ and $S_{12}(q)$. A significant broadening of $S_{12}(q)$ at $d \leq 0.34$ shows that there is a strong correlation between density perturbations in both  layers, and these correlations survive in a broad range of excitation momenta $q$. In its turn, possibility for a momentum transfer between the layers means, that the kinetic energy of excited quasiparticles is comparable with the interlayer interaction energy, $q^2/2m \lesssim D/d^3$. For large $q$ such a possibility exists only in the dimer phase. Once $d$ is decreased, the range of momenta $q$ where $S_{12}(q) \neq 0$ increases. This trend is illustrated in Fig.~\ref{fig:grsk-D1m1}e. For simple estimate, we can choose $d=0.5$ and $d=0.3$. The characteristic interlayer interaction is given by $D/d^3 \sim 8$ and $37$, correspondingly. These values should be compared with the energy of collective in-phase density excitations at $qa\gtrsim 10$ in Fig.~\ref{fig:two-m1D1}. Obviously, for $d=0.5$ and  $qa\gtrsim 10$ the interlayer coupling is weak, $D/d^3\ll \omega^L_{+}(q)\sim 25$, and, therefore, $S_{12}(q)$  decreases fast at large wavenumbers. In contrast, for $qa \lesssim 8$ the condition  $\hbar \omega(q)  \lesssim D/d^3$ is satisfied, and both $S_{12}$ and $\chi_{12}$ take a non-zero value. For $d=0.3$, the above condition is satisfied in a broad range of momenta. The observed decay of $S_{12}(q)$ for small $q$ is related with the general momentum-scaling of the quasiparticle density in 2D.

{\em Superfluid response.} 
To further illustrate the effect of the interlayer coupling on the dimerization process, we present in Fig.~\ref{fig:D1m1-superf} snapshots of the particle density taken at $d=0.3, 0.4$ and $0.6$. They are supplemented by the $d$-dependence of the inlayer superfluid fraction $\gamma_s(d)=\rho_{s\al}(d)/\rho_{\al}(d)$. The density snapshots support the dimerization scenario at small $d$ and demonstrate a significant spatial localization of particles in both layers. The net effect is a reduction of the quantum spatial coherence and a monotonic decrease of the superfluid fraction $\gamma_s$, accompanied by the increase of the peak height $g_{12}(0)$. At $d \leq 0.34$, the effect of spatial ordering can be clearly observed in the density snapshots (still only on the microscopic scale opposite to a long-range correlations in a crystal), while the superfluid response fast drops to zero. As will be discussed below, in this regime strongly localized dimers are formed and the inlayer superfluidity completely vanishes. 

Qualitatively, our observations can be explained as follows. Below some layer spacing $d$ the dimer size becomes smaller than the inlayer average interparticle spacing, $\sigma_d < a$, and dimers can be treated as new composite particles. The system is characterized by a new effective coupling parameter which is larger then in a single layer, $D^{\star}=m^{\star}p^{\star 2}/\hbar^2 a  \sim 8 D$. The factor 2 comes from the mass, $m^\star=2m$, and the factor 4 from the dimer-dimer interaction (involving four particles). In this regime, the phase diagram will be similar to the one of 2D bosonic dipoles,~\cite{fil2012} with the crystallization transition at $D^{\star}\gtrsim 17(2)$. For the inlayer coupling  $D=1$ and $D^{\star} \sim 8$, we are below this critical value. Hence, the oscillations of $g_{11}(r)$, as observed in Fig.~\ref{fig:grsk-D1m1}a, should not be ascribed to the onset of the crystallization transition, but rather to the formation of strongly correlated gas phase, which becomes again superfluid at lower temperatures.

To check this possibility, we repeated our simulations at twice lower temperature, $T=0.5$. The comparison of $T=1$ and $T=0.5$ has not revealed significant differences. The slope of the PDFs remains nearly the same. We conclude that the dimers will remain in the gas phase down to the ground state ($T=0$) at least for the layer spacing $d \gtrsim 0.25$. For smaller $d$, we can not exclude formation of more complicated bound states, like trimers. This scenario becomes energetically favorable in the bilayers with a strong density imbalance.~\cite{Klawunn,Pupillo} %The effect should be similar to one observed in the electron-hole systems in solid state physics~[].

Now we turn to a discussion of the superfluid phase present at $T \leq T_c$. For composite dipoles $T_c$ can be estimated from the single layer data:~\cite{fil2011} $T_c/[\hbar^2/m a^2]\sim 1.2$ for $D=8$. For composite dipoles ($m^{\star}=2m$) the temperature (in our units) is by a factor two smaller: $T_c/[\hbar^2/m^\star a^2]\sim 0.6$. This explains the zero superfluidity in our simulations at $T=1$ and $d\lesssim 0.3$, see Fig.~\ref{fig:D1m1-superf}. Additional simulations at $T=0.5$ have shown that a finite superfluid response in the dimer phase is restored.

\begin{figure}
 \begin{center} 
\hspace{1.0cm}\includegraphics[width=0.49\textwidth]{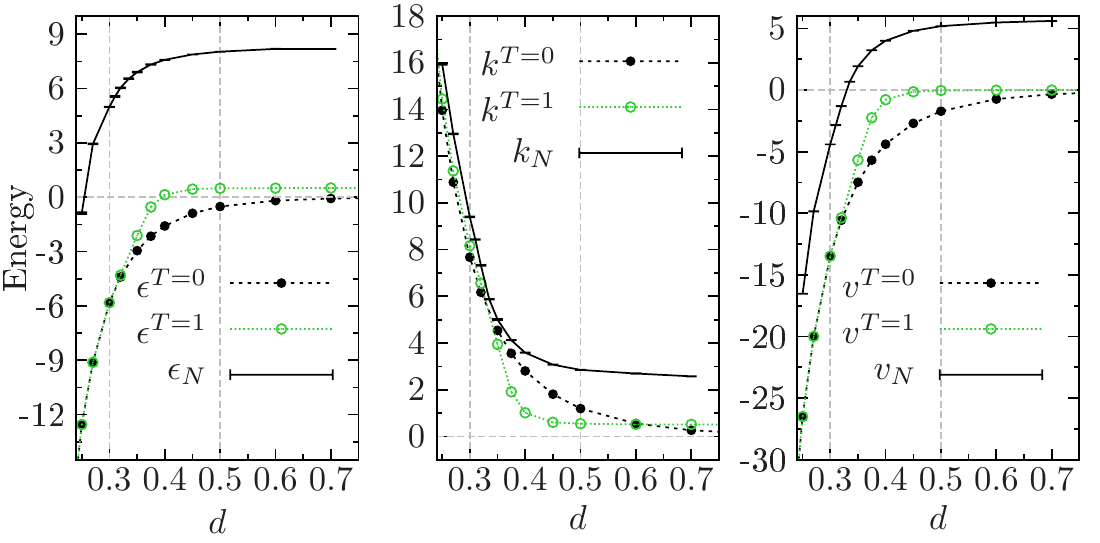}
%{../../../quasi2d/New/d55ed40/data/2d/m1new/d1/Ebind/energym1d1b1}
\end{center}
 \vspace{-0.7cm} 
\caption{The $d$-dependence of total $\epsilon_N$, kinetic $k_N$ and potential energy $v_N$ (per  particle) from the many-body simulations at $T=1$. For comparison the solution of a single dimer problem is presented by $\epsilon^T, v^T$ and $k^T$ (the dimer energies are divided by $N=2$). The upper index $T$ denotes the temperature argument in the matrix squaring technique.~\cite{matrix}}
\label{fig:D1m1-energy}
 \end{figure}

{\em Energy characteristics.} The next feature which identifies the dimerization is a specific $d$-dependent slope of the energy characteristics $\{\epsilon_N,k_N,v_N\}$ shown in Fig.~\ref{fig:D1m1-energy}. Main changes are observed in the range $0.3\leq d \leq 0.5$. A fast increase of the kinetic energy $k_N$ is observed, and is attributed to the energy of zero-point fluctuations in spatially localized bound states. Simultaneously, the potential energy  $v_N$ drops to negative values for $d \lesssim 0.33$, indicating that the interlayer attraction dominates over the intralayer repulsion.

\begin{figure}
 \begin{center} 
\hspace{.0cm}\includegraphics[width=0.5\textwidth]{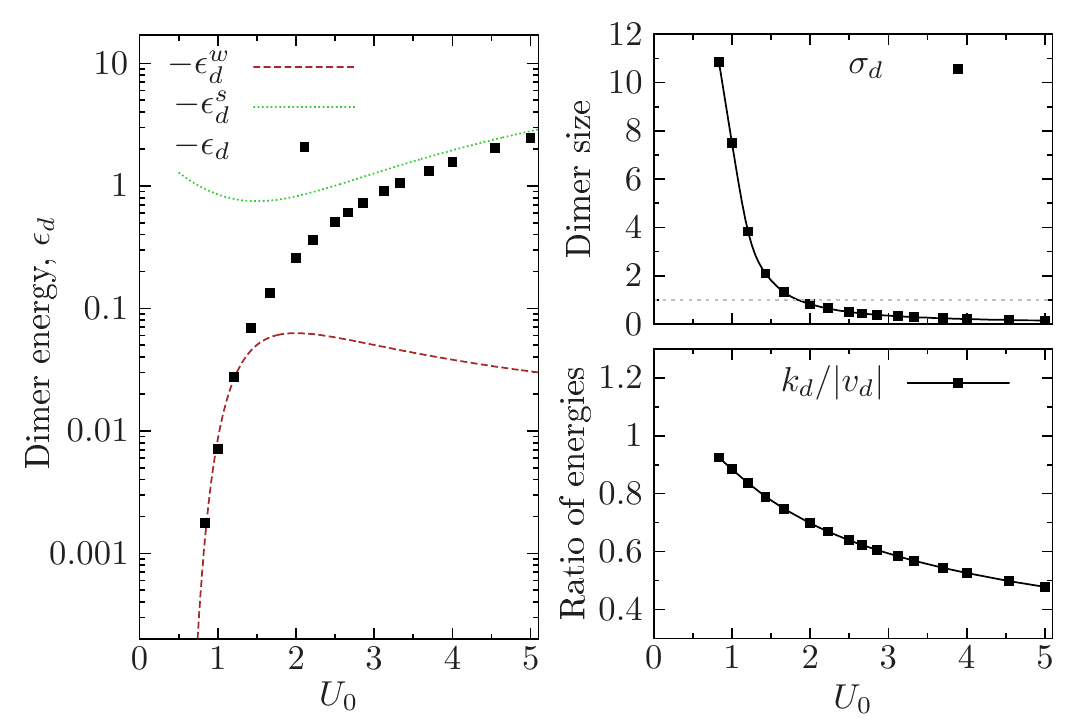}
%{../../../quasi2d/New/d55ed40/data/2d/m1new/d1/Ebind/energu0}
\end{center}
 \vspace{-0.8cm} 
\caption{{\em Left}: Dimer energy $\epsilon_d=\abs{\epsilon}/\epsilon_0$ as a function of the coupling strength $U_0= D/d$ and in the weak ($w$) and strong ($s$) coupling approximation, see Eqs.~(\ref{u01}) and~(\ref{u02}). {\em Right}: The mean dimer size, $\sigma_d=\sqrt{\avr{r^2}/a^2}$, and the ratio of kinetic to potential energy, $k_d/\abs{v_d}$, versus $U_0$.}
\label{fig:D1m1-energu0}
 \end{figure}

To explain the observed $d$-dependence, we compare the many-body results with a single dimer solution. The dimer problem in the bilayer geometry  has been already addressed before both numerically and analytically.~\cite{kl1,zinner} The present approach is based on the matrix squaring technique~\cite{matrix} and numerical evaluation of a two-body density matrix (DM) and its $\beta$-derivative. The obtained dimer energy can be compared with the analytical results of Ref.~\cite{kl1}

For {\em weak} coupling, the binding energy as a function of the interlayer coupling constant, $U_0=m p^2/\hbar^2 d$, takes the form
\begin{align}
 \epsilon_d^w/\epsilon_0 \approx \exp \left(-\frac{8}{U_0^2}\left[1-U_0 +\frac{U_0^2}{4}\left(\frac{5}{2}+\ln \frac{e^{\gamma}}{2} \right) \right] \right),
 \label{u01}
\end{align}
where $\epsilon_0=\hbar^2/m d^2$ and $\gamma\approx 0.577$ is the Euler constant. This result remains accurate~\cite{kl1} up to $U_0\lesssim 1.2$. 

In the opposite limit of {\em strong} coupling ($U_0\gg 1$) the dimer energy was determined by the variational calculations~\cite{var_ed}
\begin{align}
\epsilon_d^s/\epsilon_0 \approx -2U_0 +4 \sqrt{3 U_0/2}-15/4.
 \label{u02}
\end{align}
Both asymptotics~(\ref{u01}),(\ref{u02}) are presented in Fig.~\ref{fig:D1m1-energu0}(left panel) and compared with our numerical data. As expected both limits are nicely reproduced. The variational ansatz coincides with the numerics for $U_0\gtrsim 5$, while Eq.~(\ref{u01}) remains accurate up to $U_0 \sim 1.2$.

The energy characteristics of a single dimer are presented in Fig.~\ref{fig:D1m1-energy} with the curves $\epsilon^T, v^T$ and $k^T$. They are rescaled to the energy units of the many-body simulations as: $U_0=D/d$ and $\epsilon_d/E_0=(\epsilon_d/\epsilon_0) /d^2$.  The upper index $T$ indicates the temperature argument of the pair DM evaluated with the matrix squaring technique. The case $T=0$ denotes a low temperature limit when we reach convergence to ground state properties at finite temperatures.

As shows Fig.~\ref{fig:D1m1-energy} the main trend observed in $k_N, v_N$ and $\epsilon_N$ is also reproduced by a single dimer. This testifies that the pairwise interlayer correlations play here a dominant role. A difference and a shift in absolute values are due to many-body contributions. At $d > 0.6$, as we approach the limit of independent layers, the many-body results saturate at their single layer values. These are, obviously, zero in the single dimer case, apart from the kinetic energy which equals $k^T \sim k_BT$.

More information on the dimer states formed at $d \lesssim 0.5$ is presented in Fig.~\ref{fig:D1m1-energu0} (two right panels). Here the $U_0$-dependence of the mean dimer size, $\sigma_d/a=\sqrt{\avr{r^2}/a^2}$, and the ratio of the internal kinetic and the potential energy are shown. 
For $U_0 \lesssim 2$ we observe a fast divergence of $\sigma_d$ being a clear indication of the crossover from strongly to weakly bound dimers. For $U_0=1$ the dimer size equals $\sigma_d/a \approx 7.5$ and significantly exceeds the average interparticle spacing in  the many-body system. This state is characterized by nearly equal values of the kinetic and the potential energies. At $U_0=2$ (corresponds to the layer spacing $d=0.5$ at $D=1$) the binding energy equals $\epsilon_d/E_0 \approx -1.0355$. In our finite temperature simulations at $T=1$ such a state is thermodynamically unstable. In contrast, at $U_0=2.5,3.12$ and $5.0$ [$d=0.4,0.32$ and $0.2$] the dimer size reduces to $\sigma_d \approx 0.5,0.32$ and $0.14$. The binding energy takes the values $\epsilon_d/E_0 \approx -3.1732,-8.8042$ and $-61.4326$. In this regime, it becomes a dominant energy scale in the many-body simulations. We conclude, that, at least for $d\lesssim 0.32$, treatment of interlayer dimers as composite particles, characterized by $m^\star$ and $D^\star$, is well grounded. Alternatively, another criterion can be employed. A many-body system can not be treated as an ensemble of dimer states once $\sigma_d/a \gtrsim 1$. This holds for $U_0 \lesssim 2$ or the layer spacing $d \gtrsim 0.5$: see Fig.~\ref{fig:D1m1-energu0} where $\sigma_d=1$ is shown by a horizontal dotted line.

\begin{figure}
 \begin{center} 
\hspace{-1.0cm}\includegraphics[width=0.49\textwidth]{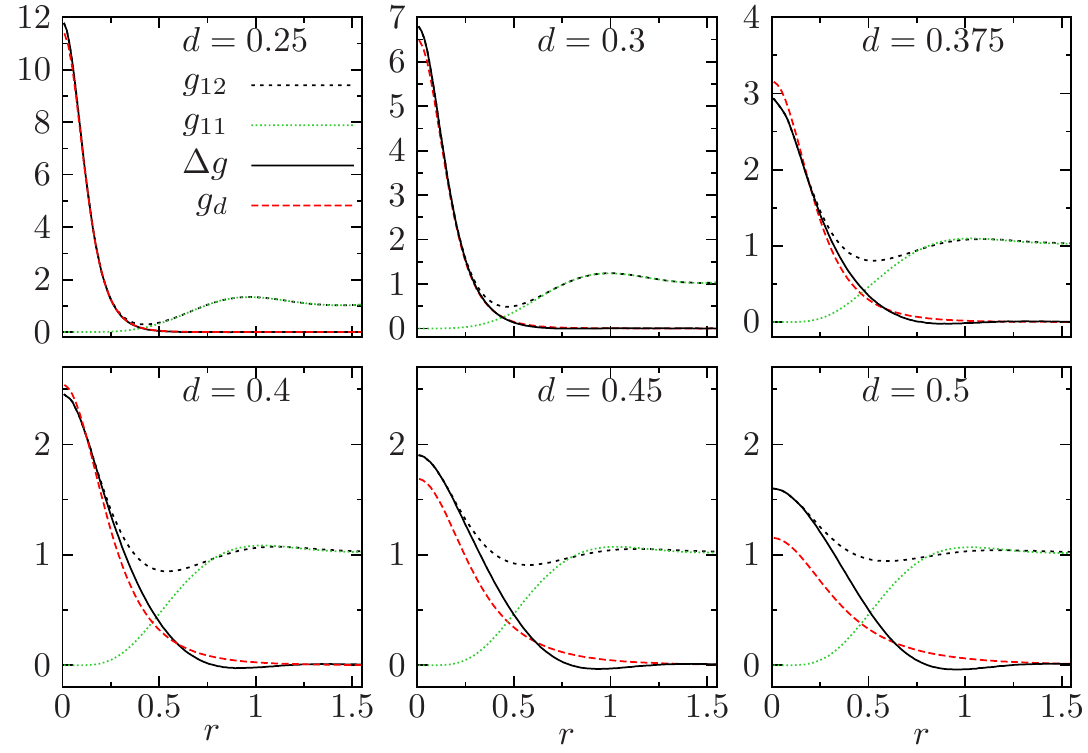}
%{../../../quasi2d/New/d55ed40/data/2d/m1new/d1/Ebind/rho12}
\end{center}
 \vspace{-0.7cm} 
\caption{The inter- and intralayer PDFs. Coupling $D=1$. The many-body results $g_{12}(r), g_{11}(r)$ and $\Delta g(r)=g_{12}(r)-g_{11}(r)$ are compared with the single dimer solution $g_d(r)$. The range of layer spacing $0.25 \leq d \leq 0.5$ corresponds to $2\leq U_0 \leq 4$ in Fig.~\ref{fig:D1m1-energu0}.}
\label{fig:D1m1-rho12}
\end{figure}

The influence of many-body effects on the dimer states can be nicely illustrated by the PDF $g_{12}(r)$. Its behavior near $r=0$ is determined by the two-body density matrix of two dipoles from different layers. The angular average defines the distribution function of a single dimer, $g_d(r)=\Avr{\rho_2(r,r,\beta)}$, with $r=\abs{\vec r_i -\vec r_j}$. For the dimers, which are thermodynamically stable, both distribution functions should coincide near the origin, i.e. $g_{12}(r)\approx g_d(r)|_{r \lesssim a}$.

To extract the dimer PDF from the many-body simulations we consider the difference, $\Delta g(r)=g_{12}(r)-g_{11}(r)$. We assume that the probability distribution of a particle in the first layer relative to all particles in the second layer, excluding one in a bound state, should be given by the intralayer PDF $g_{\al\al}(r)$. Once two-body interlayer correlations dominate over all other correlations, this picture is reasonable. The $d$-dependence of the binding energy suggests that this holds, at least, for $d \lesssim 0.32$.

\begin{figure}
 \begin{center} 
\includegraphics[width=0.52\textwidth]{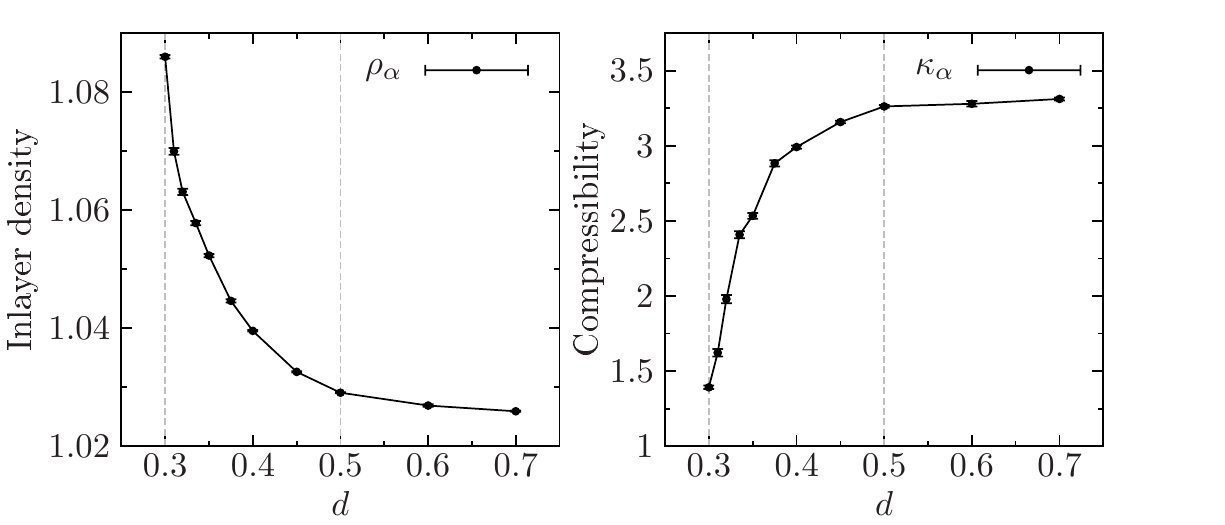}
%{../../../quasi2d/New/d55ed40/data/2d/m1new/d1/Ebind/densNd1}
\end{center}
 \vspace{-0.7cm} 
\caption{The $d$-dependence of the inlayer density $\rho_\al$ and the compressibility $\kappa_\al [10^2]$ for $D=1$.}
\label{fig:D1m10-sound}
 \end{figure}

Bound state properties can be modified at finite densities due to the intra and interlayer correlations. The internal properties of dipolar pairs will be reproduced by a single dimer solution, when $\abs{\epsilon_d} \gg D/a^3, k_B T$. For $D=1$ and $T=1$ this results in the estimate $\abs{\epsilon_b} \gg 1$. 

Fig.~\ref{fig:D1m1-rho12} presents the comparison between the dimer state in a many-body environment, $\Delta g(r)$, and the single dimer distribution $g_d(r)$. As expected, a nice agreement is observed below $d\sim 0.4$, due to the increase of the dimer binding energy and 
the spatial localization. In contrast, at $d=0.45$ and $0.5$ a new trend is present. At finite densities the average dimer size is reduced  compared to a free dimer case. The many-body environment acts in favor of the interlayer dimerization, as the repulsive intralayer interaction plays a stabilizing role for dipolar pairs.

To further characterize the dimerization, in Fig.~\ref{fig:D1m10-sound} we analyze the $d$-dependence of the inlayer density $\rho_{\al}$ and the isothermal compressibility $\kappa_\al$. In the long wavelength limit both are related to the static structure factor as
\begin{align}
 &S_{\al\al}(q=0)=1+\rho_{\al} \int\limits_V [g_{\al\al}(\vec{r})-1] \db \vec{r} \nonumber \\
 &=\frac{\avr{N_\al^2} -\avr{N_\al}^2}{\avr{N_\al}}=\rho_{\al} k_B T \, \kappa_{\al}.
 %\frac{k_B T}{m_\al c_\al^2},
 \label{kappa_aa}
\end{align}
The effect of the second layer comes into play below $d=0.5$. The inlayer density is steadily increasing with the reduction of $d$. For $d \lesssim 0.4$ particles from both layers are pairwise coupled and form composite bosons. For a fixed chemical potential $\mu$, each layer accommodates more particles, as it becomes energetically favorable due to an enhanced binding energy. This process is accompanied by reduction of the compressibility $\kappa_\al$. The system forms a strongly correlated  gas of dimers. This new phase is less compressible, whereas the particle number fluctuations in each layer~(\ref{kappa_aa}) are suppressed due to formation of bound states.

 \subsection{Strong coupling: $D=5.5$ \; ($5.5 \leq U_0 \leq 9.2$)}
 
We repeat our analysis for a strongly correlated bilayer. Some of the discussed features are similar to the $D=1$ case. The increased dipole coupling ($D=5.5$) sets a new energy scale, and as a result the dimerization transition shifts to a larger layer spacing. The coupling parameter in a single dimer problem increases to $5.5 \leq U_0(D/d) \leq 9.2$ for $0.6 \leq d \leq 1.0$, see Fig.~\ref{fig:D1m1-energu0}.

% ----------------------------------------------------
\begin{figure}
 \begin{center}
\hspace{0.0cm}\includegraphics[width=0.49\textwidth]{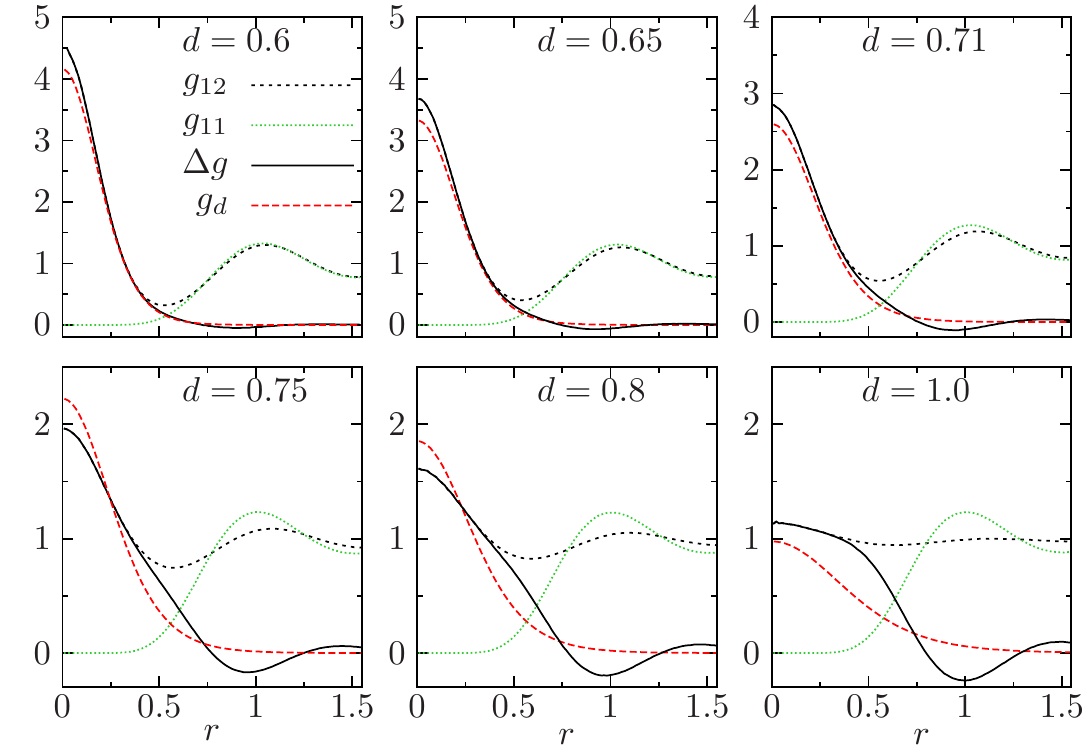}
%{../../../quasi2d/New/d55ed40/data/2d/m1new/d55/Ebind/rho12}
\end{center}
 \vspace{-0.7cm} 
\caption{The inter- and intralayer PDFs at several spacing for $D=5.5$. The many-body results $g_{12}(r), g_{11}(r)$ and $\Delta g(r)=g_{12}(r)-g_{11}(r)$ are compared with the single dimer solution $g_d(r)$. The range $0.6 \leq d \leq 1$ corresponds to $5.5 \leq U_0 \leq 9.17$ in Fig.~\ref{fig:D1m1-energu0}.}
\label{fig:D55m1-rho12}
\end{figure}

{\em Dimerization and superfluid response.}
The dimerization transition is illustrated in Fig.~\ref{fig:D55m1-rho12} and starts  around $d \approx 0.8$. At $d=0.71$ we already find is a nice agreement between the finite density result, $\Delta g$, and the single dimer, $g_d$. There is a qualitative agreement in the range, $0.75\leq d \leq 0.8$. The shape of $\Delta g$ is disturbed due to the overlap with neighboring particles. Here, we observe a destabilizing effect of the many-body environment. The peak height at the origin is reduced: $g_d(0)> \Delta g(0)$.  In contrast, for $d\leq 0.71$ we find an opposite trend: the inlayer interaction slightly enhances the spatial localization of  dimers. 

\begin{figure}
 \begin{center} 
\includegraphics[width=0.51\textwidth]{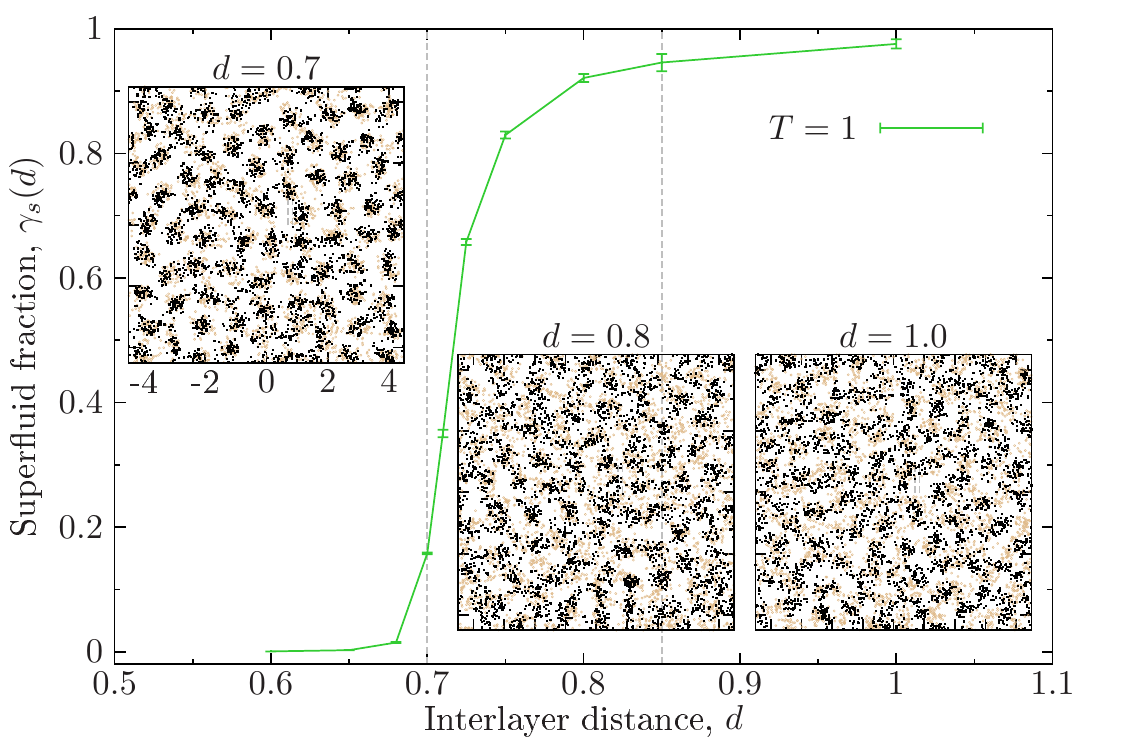}
%{../../../quasi2d/New/d55ed40/data/2d/m1new/d55/Ebind/confsuperl0}
\end{center}
 \vspace{-0.7cm} 
\caption{The $d$-dependence of the superfluid fraction $\gamma_s(d,T)$ at temperature $T=1$ and $D=5.5$.
The system size is $\Avr{N}\sim 160$ with $\Avr{N_{\al}}\sim 80$ particles per layer.  The insets show instantaneous density snapshots at $d=0.7,0.8$ and $1.0$. Two colors (black/brown) distinguish top and bottom layers.}
\label{fig:D55m1-superf}
 \end{figure}

The onset of dimerization at $d\sim 0.8$ correlates with a fast drop of the superfluid density, see Fig.~\ref{fig:D55m1-superf}. The superfluid fraction $\gamma_s$ already drops to zero at $d \approx 0.68$, when the average dimer size is reduced below half of the average interparticle distance and equals $\sigma_d/a \approx 0.37$. 
 
%----------------------------------------------------- 

\begin{figure}
 \begin{center} 
 \hspace{-0.35cm}\includegraphics[width=0.5\textwidth]{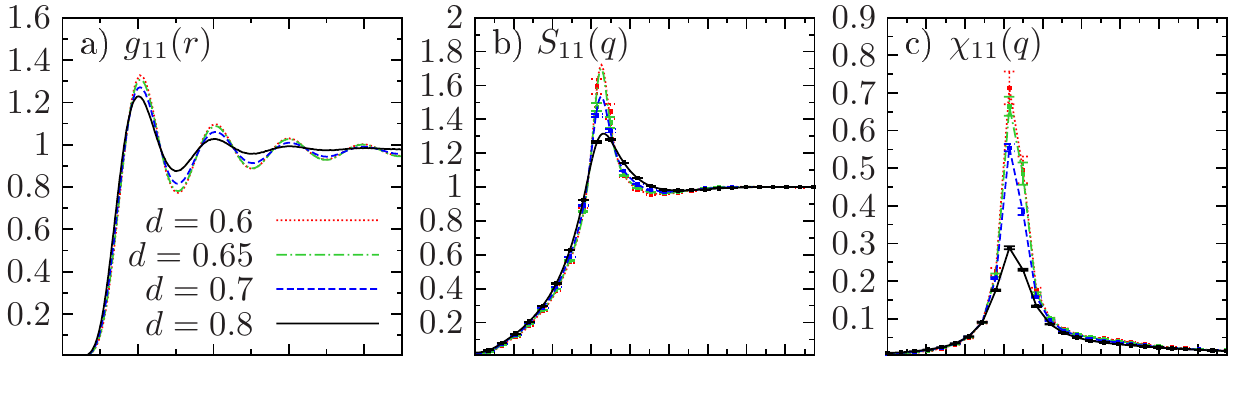}\\
 %{../../../quasi2d/New/d55ed40/data/2d/m1new/d55/gr11sk11chi11nox}\\ 
 \vspace{-0.25cm}
 \hspace{-0.35cm}\includegraphics[width=0.5\textwidth]{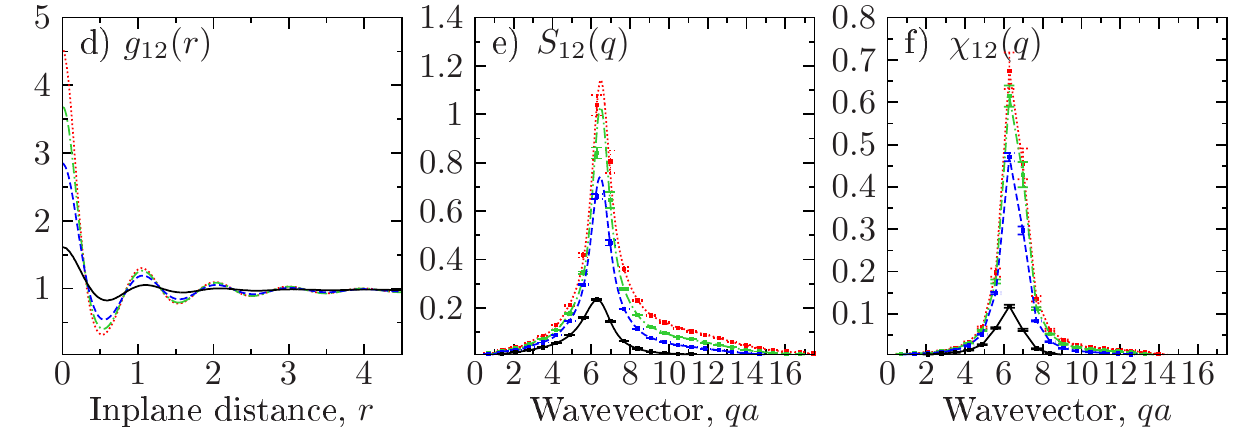}
 %{../../../quasi2d/New/d55ed40/data/2d/m1new/d55/gr12sk12chi12} 
 \end{center}
\vspace{-0.6cm} 
 \caption{a),d) Intra- and interlayer PDFs $g_{\al\be}(r)$. b),e) Static structure factor $S_{\al\be}(q)$. c),f) Density response function $\chi_{\al\be}(q)\equiv \abs{\Re \chi_{\al\be}(q,\omega=0)}/2\rho_{\al\be}$. The interlayer spacing: $0.6\leq d\leq 0.8$. Simulation parameters: $D=5.5$, $\mu=70$, $V(L^2)=81$ and $T=1$.}
 \label{fig:grsk-D55m1}
 \end{figure}

{\em Static properties.}
The instantaneous density snapshots in both layers are presented in Fig.~\ref{fig:D55m1-superf}. They nicely illustrate that the vertical alignment of particles from different layers dominates, especially at $d=0.7$. In this regime, the dimers can be treated as composite particles. A new effective coupling parameter, $D^{\star} \sim 8 D =44$, exceeds the critical value, $D^{\star}=17(1)$, required for the crystallization transition.~\cite{buch} The simulated temperature ($T=1$), however, is too high (by factor two) to observe a defect-free Wigner lattice. Still some pieces of a crystalline structure are present. For $d\leq 0.7$ the intralayer PDF $g_{\al\al}$ shows several well pronounced correlation shells, see Fig.~\ref{fig:grsk-D55m1}a. Both the static structure $S_{\al\al}$ and the density response function $\chi_{\al\al}$ are peaked around the wavevector, $q\approx 2 \pi/a $, corresponding to the inverse interparticle distance.

\begin{figure}
 \begin{center} 
\hspace{-.0cm}\includegraphics[width=0.49\textwidth]{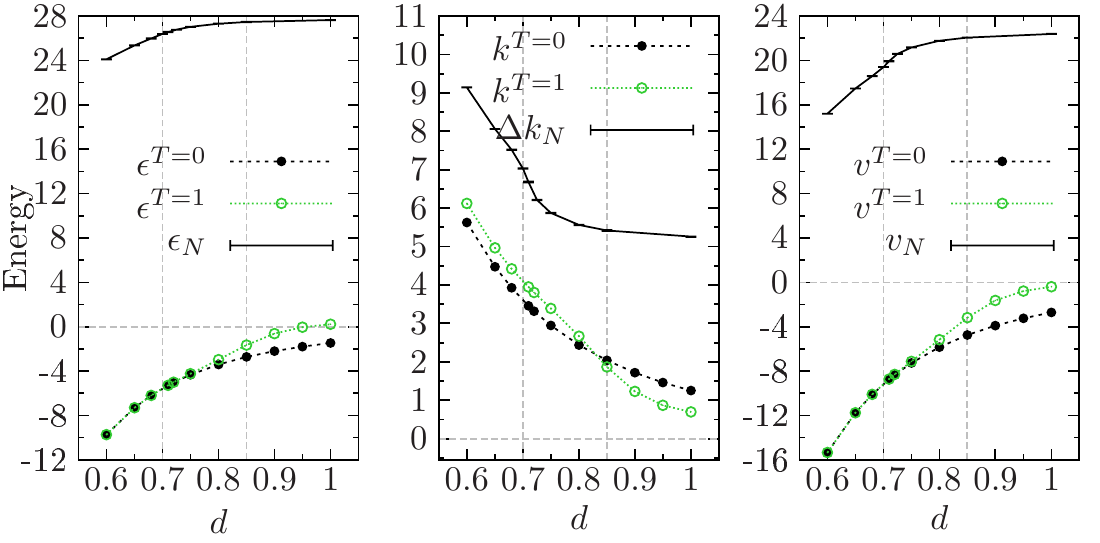}
%{../../../quasi2d/New/d55ed40/data/2d/m1new/d55/Ebind/energym1d55b1}
\end{center}
 \vspace{-0.7cm} 
\caption{The $d$-dependence of total, kinetic and potential energies, $\epsilon_N, k_N, v_N$ (per particle), from many-body simulations at $T=1$ for $D=5.5$.
For comparison the solution of a single dimer problem is presented by $\epsilon^T, v^T$ and $k^T$.}
\label{fig:D55m1-energy}
 \end{figure}

{\em Thermodynamic properties.}
The $d$-dependence of the total, potential and kinetic energies (per particle) is shown in Fig.~\ref{fig:D55m1-energy}.  All quantities show a noticeable change in the slope around $d\sim 0.75$. For larger $d$ the energies saturate at their single layer values. This demonstrates that both layers become nearly independent in a homogeneous superfluid  phase. For $d\lesssim 0.75$ the system enters in the molecular (dimer) phase. The $d$-dependence is dominated by the single dimer solution shown by $\{\epsilon^T,k^T,v^T\}$. The energy shift with respect to $\{\epsilon_N,k_N,v_N\}$ is due to the many-body contributions.

\begin{figure}
 \begin{center} 
\hspace{-.0cm}\includegraphics[width=0.52\textwidth]{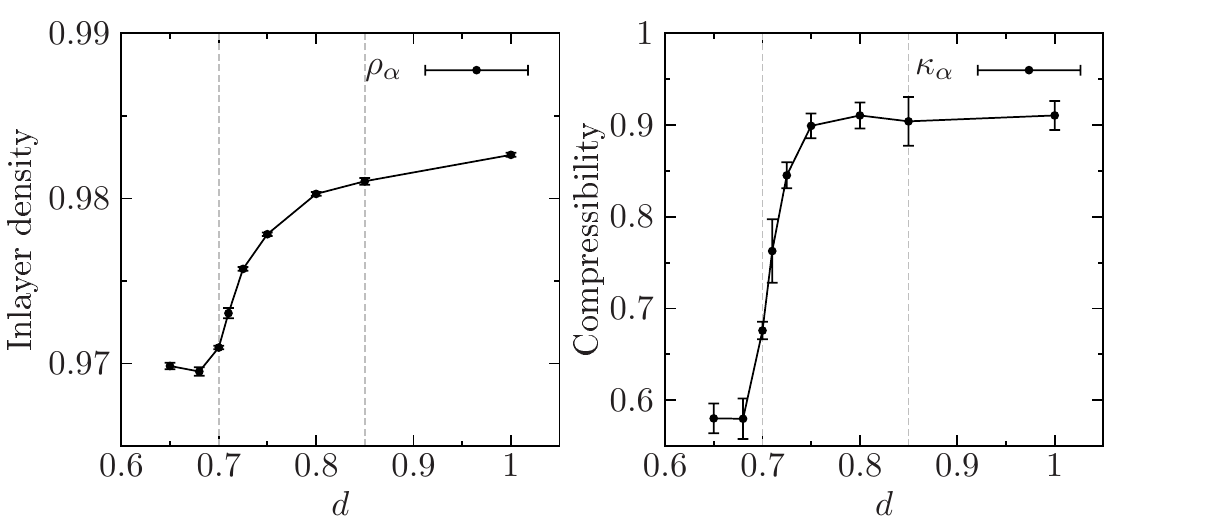}
%{../../../quasi2d/New/d55ed40/data/2d/m1new/d55/Ebind/densNd55}
\end{center}
 \vspace{-0.7cm} 
\caption{The $d$-dependence of the inlayer density $\rho_\al$ and the compressibility $\kappa_\al [ 10^2]$ for $D=5.5$.}
\label{fig:D55m10-sound}
 \end{figure}

The results for the density and compressibility are presented in Fig.~\ref{fig:D55m10-sound}. The range $0.68 \lesssim  d \lesssim 0.75$ corresponds to a transient region characterized by a partially superfluid phase. The density increases by one percent as the superfluid fraction $\gamma_s$ approaches $1$. For $0.65 \leq d < 0.68$ ($d > 0.75$) the inlayer density saturates at the equilibrium value in the normal (superfluid) phase. 

The inlayer compressibility is not influenced by the interlayer correlations in the superfluid regime with $\gamma_s > 0.9$. It only starts to decrease with the formation of the interlayer dimers at $d \lesssim 0.75$ and follows the $d$-dependence of $\gamma_s(d)$ in Fig.~\ref{fig:D55m1-superf}. With the formation of strongly bound states below $d\sim 0.6$ the inlayer compressibility fast reduces to zero due to a strong enhancement of the energy penalty for the independent particle number fluctuations in both layers.

\subsection{Weak coupling: $D=0.1$\; ($0.16 \leq U_0 \leq 1$)}

As a third case, we consider a weakly interacting system. Possible physical realizations include ensembles of Cr atoms.~\cite{atomic1,atomic2,pfau} The analyzed range of interlayer spacings, $0.1 \leq d\leq 0.6$, corresponds to the coupling parameter $0.16 \leq U_0 \leq 1$ characterized by a significantly reduced binding energy $\epsilon_d$, see Fig.~\ref{fig:D1m1-energu0}(left panel). The inlayer correlation energy compares or exceeds $\epsilon_d$. In this regime a single dimer state is strongly perturbed due to a many-body environment. For illustration we start the discussion from the dimer distribution function. 

\begin{figure}
 \begin{center} 
\hspace{0.0cm}\includegraphics[width=0.49\textwidth]{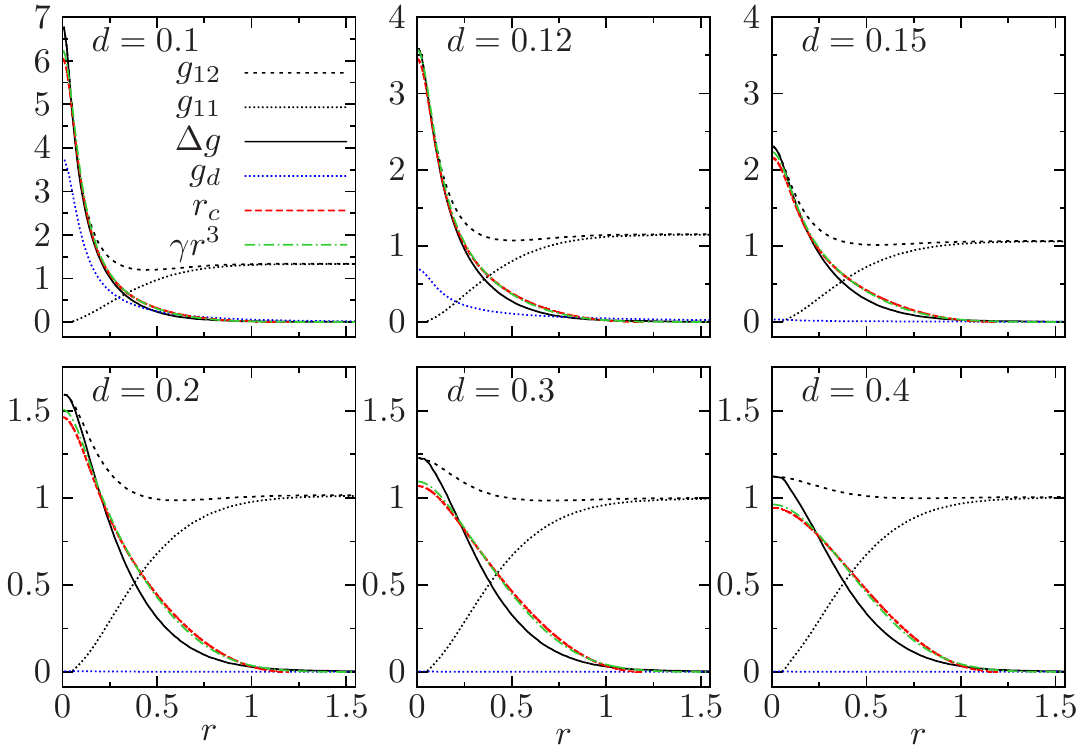}
%{../../../quasi2d/New/d55ed40/data/2d/m1new/d01/Ebind/rho12}
\end{center}
 \vspace{-0.7cm} 
\caption{Inter- and intralayer pair distribution functions for $D=0.1$. The many-body results, $g_{12}(r), g_{11}(r)$ and $\Delta g(r)=g_{12}(r)-g_{11}(r)$, are compared with a single dimer solution $g_d(r)$. Two additional curves take into account finite density effects via an  effective external potential: i) a hard wall of a radius $r_c$ (the curve ``$r_c$''); ii) a soft potential $V_b(r)=\gamma r^3 $ which mimics the intralayer dipole repulsion. The considered layer spacing, $0.1 \leq d \leq 0.4$, correspond to $0.25 \leq U_0 \leq 1$ in Fig.~\ref{fig:D1m1-energu0}.}
\label{fig:D01m1-rho12}
\end{figure}

{\em  Dimerization transition.}
In Fig.~\ref{fig:D01m1-rho12} the dimer distribution function at a finite density, $\Delta g$, is compared with the single dimer case, $g_d$. The interlayer spacing $d \geq 0.1$ results in $U_0 \leq 1$, and corresponds to a vanishingly small dimer binding energy $\epsilon_d(U_0)$, see Fig.~\ref{fig:D1m1-energu0}, when it can be well approximated by Eq.~(\ref{u01}). However, the results in Fig.~\ref{fig:D01m1-rho12} show that the dimer state predicted by $g_d(r)$  is significantly underestimated compared to $\Delta g(r)$. For $d=0.1$ the peak heights differ by factor two, $\Delta g(0)\approx 2 g_d(0)$. With the increase of $d$ the discrepancy only increases. As shows $g_d(r)$ in Fig.~\ref{fig:D01m1-rho12}, the single dimer becomes spatially delocalized for $d\geq 0.15$. In contrast, a pronounced dimerization peak is observed in $\Delta g$. In the range from  $d=0.1$ ($U_0=1$) to $d=0.15$ ($U_0=0.667$) the binding energy of a single dimer $\epsilon_d$ is reduced by almost three orders of magnitude. Hence, to reproduce the dimerization feature observed in $\Delta g$ one should go beyond the single dimer model and include finite density effects.

To generalize our dimer model we include the effect of other particles by an effective external field, and then solve the corresponding Bloch equation for the pair DM. Two cases are considered. First, we set a hard wall potential of radius $r_c$, which specifies the boundary condition: $\rho(r,r;\beta)=0$ for $r\geq r_c$. A particular choice of the $r_c$-value takes into account the density effect: a correlation hole around a dimer excludes the possibility to find other particles within a sphere of radius $r_c$. Hence, a pairwise repulsion between different dimers localize them to the spatial volumes (areas) $v\sim r_c^2$ (in 2D). A free parameter $r_c$ is chosen to agree with the peak height $\Delta g(r=0)$ at $d=0.1$. The value $r_c=1.2$ provides a reasonable choice, and is in agreement with the position of the first peak in the intralayer PDF $g_{\al\al}(r)$ for all $d\geq 0.1$. In the second case, we set a soft boundary potential: $V_b(r)|_{r>r_0}=\gamma (r^3-r_0^3)$ and $V_b(r)|_{r\leq r_0}=0$. The free parameters $\{r_0,\gamma\}$ are chosen to fit the shape of $\Delta g(r)$ for $d=0.1$. In the calculations with other $d$-values the parameters $\{r_0=0.5,\gamma=8\}$ are kept fixed.
 
In Fig.~\ref{fig:D01m1-rho12} both models are shown with the curves ``$r_c$'' and ``$\gamma r^3$''. We observe a good agreement with the many-body result  $\Delta g$, at least for $d \lesssim 0.2$, and a significant improvement over the free dimer model. 

\begin{figure}
 \begin{center} 
\includegraphics[width=0.51\textwidth]{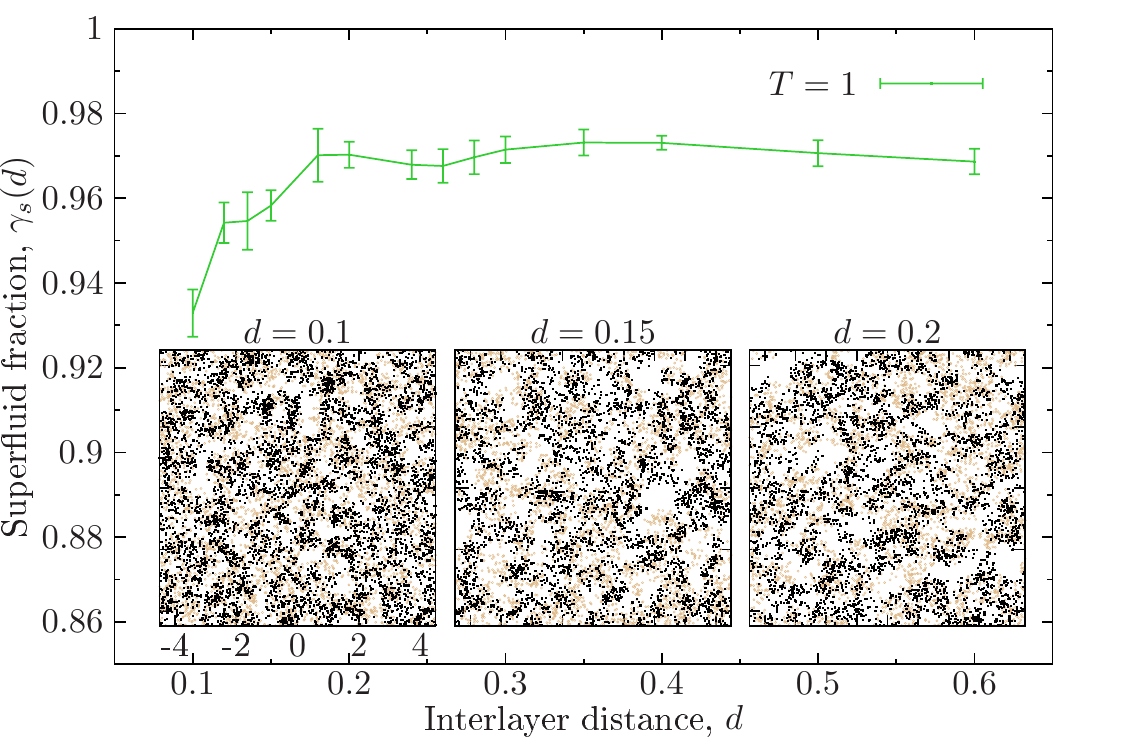}
%{../../../quasi2d/New/d55ed40/data/2d/m1new/d01/Ebind/confsuperl0}
\end{center}
 \vspace{-0.7cm} 
\caption{The $d$-dependence of the superfluid fraction $\gamma_s(d,T)$ for $T=1$ and $D=0.1$. The insets show the instantaneous density snapshots at $d=0.1,0.15$ and $0.2$. Two colors (black/brown) distinguish top and bottom layers.}
\label{fig:D01m1-superf}
 \end{figure}

{\em Superfluid response.}
The onset of dimerization observed for $d\lesssim 0.2$ only slightly reduces the superfluid density, see Fig.~\ref{fig:D01m1-superf}. This result is in a striking contrast to the case $D=1(5.5)$ in Figs.~\ref{fig:D1m1-superf},~\ref{fig:D55m1-superf}. The density snapshots show that particle ``clouds'' strongly overlap (see the insets in Fig.~\ref{fig:D01m1-superf}). Hence, the effect of spatial localization, as observed in Figs.~\ref{fig:D1m1-superf},~\ref{fig:D55m1-superf}, is not relevant and the intralayer spatial coherence is preserved. 

\begin{figure}
 \begin{center} 
 \hspace{-0.35cm}\includegraphics[width=0.5\textwidth]{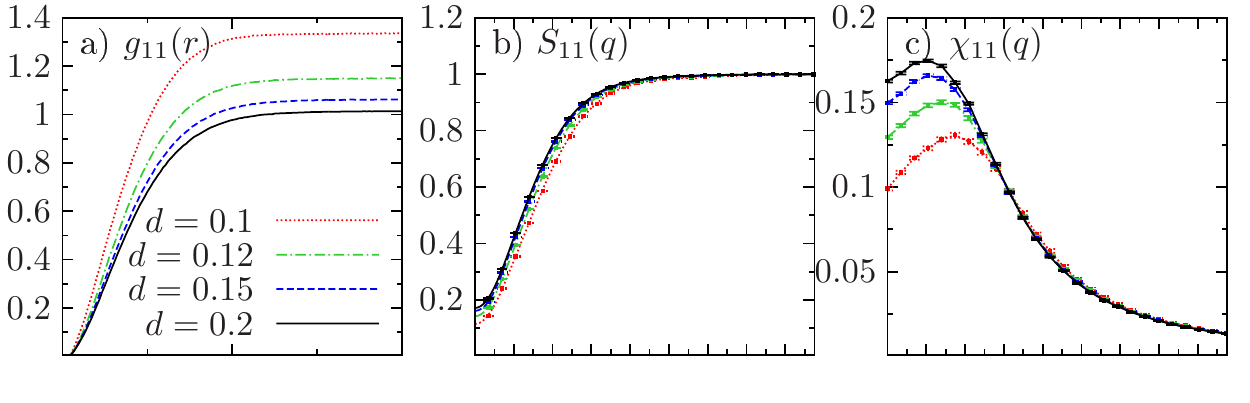}\\
 %{../../../quasi2d/New/d55ed40/data/2d/m1new/d01/gr11sk11chi11nox}\\ 
 \vspace{-0.25cm}
 \hspace{-0.35cm}\includegraphics[width=0.5\textwidth]{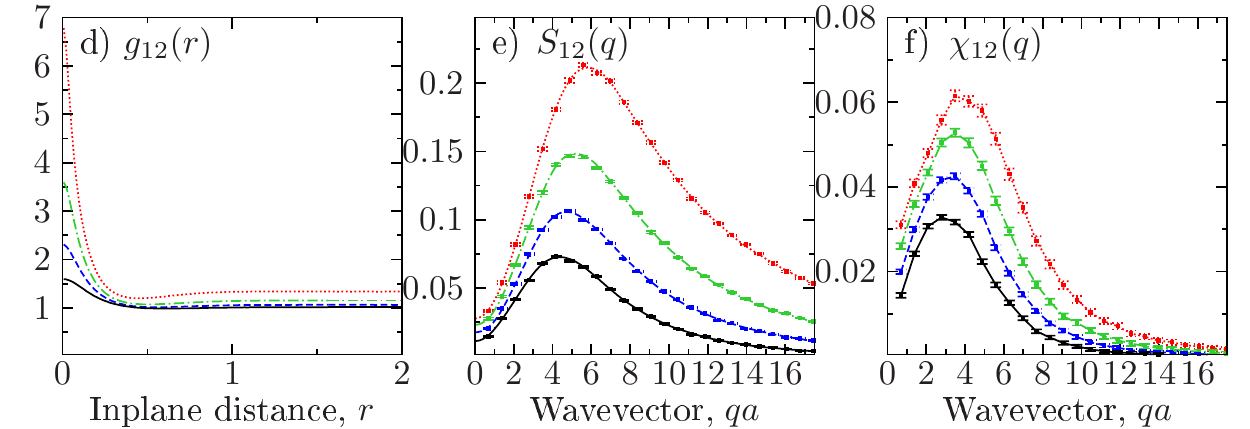}
 %{../../../quasi2d/New/d55ed40/data/2d/m1new/d01/gr12sk12chi12} 
 \end{center}
\vspace{-0.6cm} 
 \caption{a),d) Intra- and interlayer PDFs $g_{\al\be}(r)$. b),e) Static structure factor $S_{\al\be}(q)$. c),f) Density response function $\chi_{\al\be}(q)\equiv \abs{\Re \chi_{\al\be}(q,\omega=0)}/2\rho_{\al\be}$, Eq.~(\ref{gg2}). The interlayer spacing: $0.1\leq d\leq 0.2$. Simulation parameters: $D=0.1$, $\mu=4.8$, $V(L^2)=81$ and $T=1$.}
 \label{fig:grsk-D01m1}
 \end{figure}

{\em Static properties.} 
The variation of the layer spacing $d$, being of a minor importance for the superfluid response, results in noticeable changes in the static properties, see Fig.~\ref{fig:grsk-D01m1}. When $d$ is reduced, the intralayer density continuously increases (compare the asymptotic values of $g_{11}$). The interlayer response functions $S_{12}(q)$ and $\chi_{12}(q)$ develop a broad peak. The absolute value of the response function increases at small $d$. The origin of this effect is different from $D=1(5.5)$. In the former case the peak position shows only a weak $d$-independence and is close to the wavenumber $q\approx 2 \pi/a$. The formation of several correlation shells in $g_{\al\al}(r)$ validates that its origin is the  intralayer correlations and the spatial ordering (see Fig.~\ref{fig:grsk-D55m1}). This becomes possible due to formation of composite bosons with double mass and a larger dipole coupling $D^\star$. For $D=0.1$ the composite particle picture can not be directly applied. A single dimer state is not stable (at least for $d>0.1$) and bound states can exist due to the density effect, as discussed above. The constituents of dimers can exchange with neighboring particles in the same layer, as  follows from the density plots in Fig.~\ref{fig:D01m1-superf}. In the assumption that the composite particle picture is valid, a new effective coupling, $D^{\star} \sim 8 D=0.8$, is not large enough to induce a (quasi)long range spatial ordering similar to $D=1(5.5)$. A structure of $S_{12}$ and $\chi_{12}$, in Fig.~\ref{fig:grsk-D01m1}, demonstrates a strong dependence on the layer spacing. In contrast, the intralayer characteristics remain structureless, see $g_{11}$ and $S_{11}$ in Fig.~\ref{fig:grsk-D01m1}. The $d$-dependence of the effective intralayer coupling can be readout from a slope and a peak position of $\chi_{11}(q)$ by comparison with a single layer data.~\cite{fil2011} For $d=0.1$ and $D=0.1$ in the bilayer the slope of $\chi_{11}$ is similar to one in a single layer for $D=0.5$. This result is close to our estimate $D^\star\sim 0.8$ based on the dimer picture. 

% ---------------------------------------------------------------- 

\begin{figure}
 \begin{center} 
\hspace{-.0cm}\includegraphics[width=0.49\textwidth]{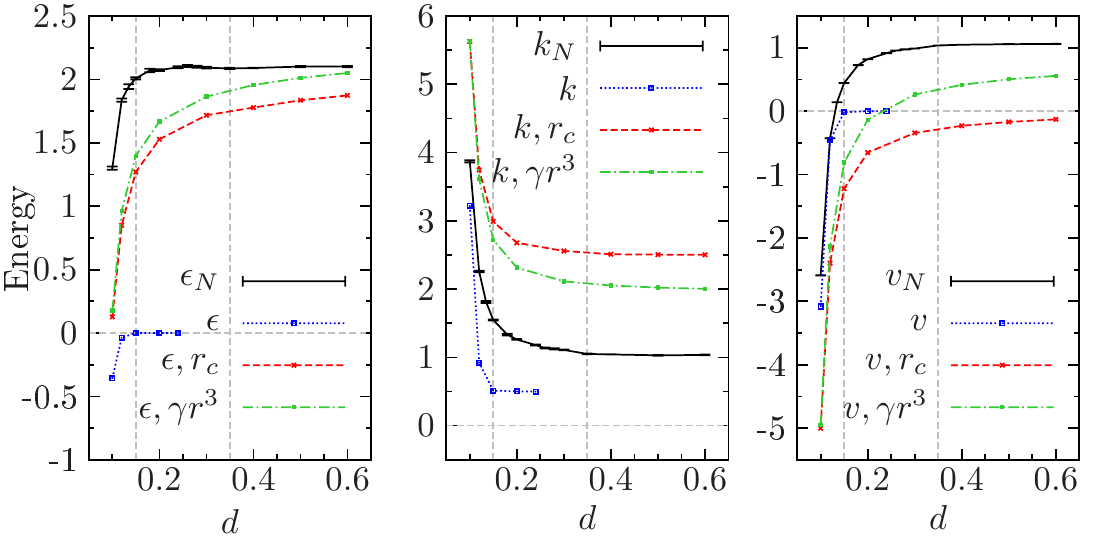}
%{../../../quasi2d/New/d55ed40/data/2d/m1new/d01/Ebind/energym1d01b1}
\end{center}
 \vspace{-0.7cm} 
\caption{The $d$-dependence of total, kinetic and potential energies, $\epsilon_N, k_N, v_N$ (per particle), from many-body simulations at $T=1$. For comparison the solution of a single dimer problem with the boundary conditions $\{r_c,\gamma r^3\}$ (see the text) is presented along with a free dimer case $\{\epsilon, v,k\}$.}
\label{fig:D01m1-energy}
 \end{figure}

{\em Thermodynamic properties.} 
Next we analyze the $d$-dependence of the total, kinetic and potential energies presented in Fig.~\ref{fig:D01m1-energy}. The dimer solution with the modified boundary conditions $\{r_c,\gamma r^3\}$, captures main features of the many-body result $\{\epsilon_N, k_N, v_N\}$. Both results predict qualitative changes below $d\sim 0.2$. This new regime can be identified as a transition from weakly to strongly bound states, when the energy scale specified by the dimer energy $\epsilon_d$ starts to dominate over the intralayer correlation energy. Similar to $D=1(5.5)$, we observe is a fast increase of the kinetic energy, and the build up of the dimerization peak $g_{12}(0)$, see Fig.~\ref{fig:grsk-D01m1}d. 

The energy characteristics, $\{\epsilon, k, v\}$ and $\{\epsilon_N, k_N, v_N\}$, differ in the absolute value. The boundary conditions enhance the kinetic energy $k$ of a single dimer compared to the many-body result $k_N$. The effect is present for all $d$ and is the largest for the hard wall potential, $\{k,r_c\}$. The many-body result, $k_N$, for $d> 0.35$ saturates slightly above the thermal kinetic energy $k_BT=1$.
This shift is due the many-body interactions, and gets larger for stronger coupling $D=1(5.5)$.

Some noticeable changes in the total energy $\epsilon_N$ (Fig.~\ref{fig:D01m1-energy}, first panel) are observed below $d \approx 0.2$. This point can be identified as the onset of dimerization. In comparison, the kinetic and potential energies ($k_N$, $v_N$) already show some weak $d$-dependence at a larger spacing, $d\sim 0.3$. Their contributions to $\epsilon_N$ mutually compensate in the range $0.2 \lesssim d\lesssim 0.4$, and produce a nearly flat curve for $\epsilon_N$. The internal kinetic energy fast increases below $d \approx 0.2$. The similar behavior is reproduced with the boundary conditions $\{r_c,\gamma r^3\}$, but is absent in the free dimer case (see dotted blue curves $\{\epsilon,k,v\}$ in Fig.~\ref{fig:D01m1-energy}). The earlier onset of dimerization, compared to a single free dimer, becomes possible due to the stabilizing effect of a many-body environment.

 \begin{figure}
 \begin{center} 
\hspace{-.0cm}\includegraphics[width=0.52\textwidth]{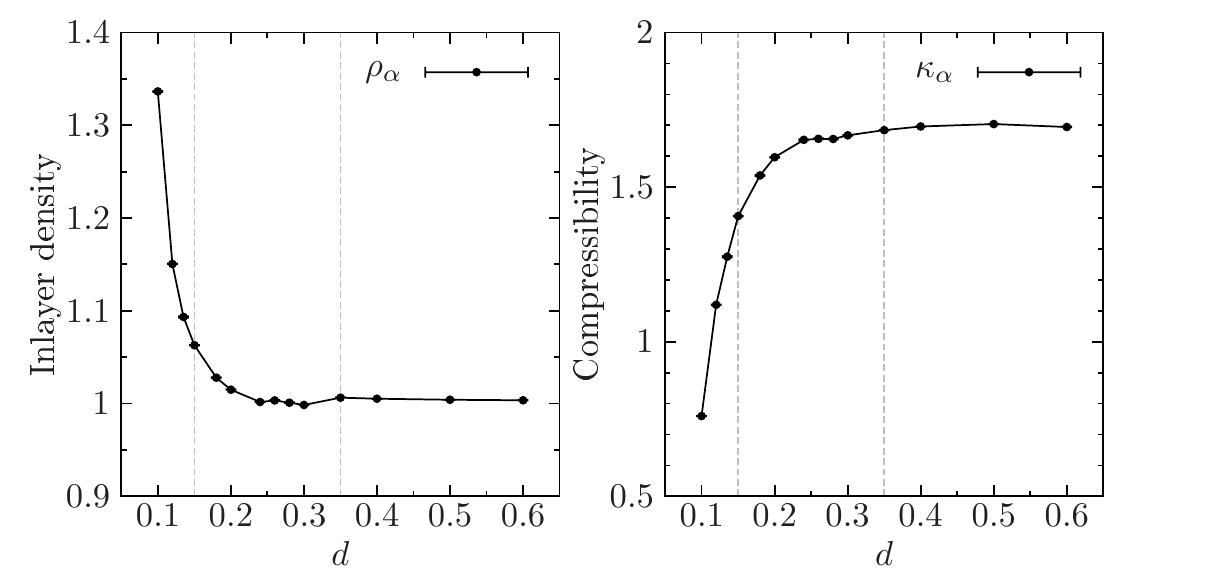}
%{../../../quasi2d/New/d55ed40/data/2d/m1new/d01/Ebind/densNd01}
\end{center}
 \vspace{-0.7cm} 
\caption{The $d$-dependence of inlayer density $\rho_\al$ and compressibility $\kappa_\al [10]$ for $D=0.1$.}
\label{fig:D01m10-sound}
 \end{figure}
 
The $d$-dependence of the inlayer density and the compressibility is analyzed in Fig.~\ref{fig:D01m10-sound}. The effect of the interlayer coupling on both quantities is well pronounced. The interval $0.15\leq d\leq 0.3$ can be considered a transient region, 
whereas a new $d$-dependent slope sets in for $d\lesssim 0.15$. It can be explained by the formation of dimer states. This is validated by $\Delta g(r)$ in Fig.~\ref{fig:D01m1-rho12} and $g_{12}(r)$ in Fig.~\ref{fig:grsk-D01m1}d.

The $d$-dependence of the compressibility $\kappa_\al$ follows the trend observed in 
$\{\epsilon_N,k_N,v_N\}$. The particle number fluctuations in both layers are suppressed by the energy penalty of the order of a dimer binding energy.

In summary, we demonstrated how the dimerization transition can be identified via the static properties and energy characteristics.

\section{Excitation spectrum of collective modes}\label{exc_sec}

In this section we analyze dispersion relations of collective modes. The generalization of the Feynman ansatz by the two-mode solution allows to distinguish the behavior of the spectral density at low and high frequencies. In the low frequency domain weakly damped collective modes (quasiparticles with a specific dispersion relation) provide a dominant contribution to $S(q,\omega)$. At high frequencies -- combinations of multiparticle excitations due their interaction and decay processes. 

We start from the diagonalization of the density response matrix. In this case spectral analyses significantly simplify.

\subsection{Diagonalization of density response matrix}
 
For a two-component system the matrix elements of the density-density correlation function in the imaginary time ($0\le \tau \leq \be$) are defined by~(\ref{g2})
and related with the density response function via the FDT~(\ref{fdt}).

We can introduce symmetric and antisymmetric density operators 
\begin{align}
 &\hat{n}_{\vec q +}(\tau)=\frac{1}{\sqrt{2}}\left[\hat{\rho}_{\vec q 1}(\tau)+\hat{\rho}_{\vec q 2}(\tau)\right],\nonumber\\
 &\hat{n}_{\vec q -}(\tau)=\frac{1}{\sqrt{2}}\left[\hat{\rho}_{\vec q 1}(\tau)-\hat{\rho}_{\vec q 2}(\tau)\right],
 \label{d3}
\end{align}
and switch to a new representation, where the matrix of the density-density correlation function, $G_{\al \be}(\vec q,\tau)$ with ($\al,\be=\pm$), becomes diagonal. 
Using (\ref{fdt}),(\ref{g2}) the diagonalization applies also to $S_{\al \be}(\vec q,\omega)$ and $\Im  \chi_{\al \be}(\vec q,\omega)$. The problem reduces to the spectral analysis of  the {\em in-phase} (symmetric) and {\em out-of-phase} (antisymmetric) mode. With~(\ref{d3}) the corresponding spectral densities can be written in terms of the partial dynamic structure factors
\begin{align}
 &S_{+}(\vec q,\omega)=S_{11}(\vec q,\omega)+S_{12}(\vec q,\omega),\nonumber\\
 &S_{-}(\vec q,\omega)=S_{11}(\vec q,\omega)-S_{12}(\vec q,\omega).
 \label{d4}
\end{align}
As a next step we introduce the symmetrized frequency power moments
\begin{align}
 \Avr{\omega^k}_{+}=\Avr{\omega^k}_{11}+\Avr{\omega^k}_{12},\nonumber\\
 \Avr{\omega^k}_{-}=\Avr{\omega^k}_{11}-\Avr{\omega^k}_{12},
 \label{d5}
\end{align}
which can be evaluated via the partial moments introduced in Sec.~\ref{twocomp}. 
In all expressions we have explicitly used the symmetry relations: $G_{11}(\vec q,\tau)=G_{22}(\vec q,\tau)$, $S_{11}(\vec q,\omega)=S_{22}(\vec q,\omega)$ and $\Avr{\omega^k}_{11}=\Avr{\omega^k}_{22}$.

\subsection{Moderate coupling: $D=1.0$}

{\em Two-mode solution.} 
The two-mode ansatz for $S_{\pm}(q,\omega)$ can obtained by a self-consistent treatment of Eqs.~(\ref{wH-eq})-(\ref{sH-sL}), using as an input the symmetrized moments~(\ref{d5}) determined via the partial frequency power moments~(\ref{wab})-(\ref{w3ab}).

\begin{figure}
 \begin{center} 
\hspace{-0.0cm}\includegraphics[width=0.51\textwidth]{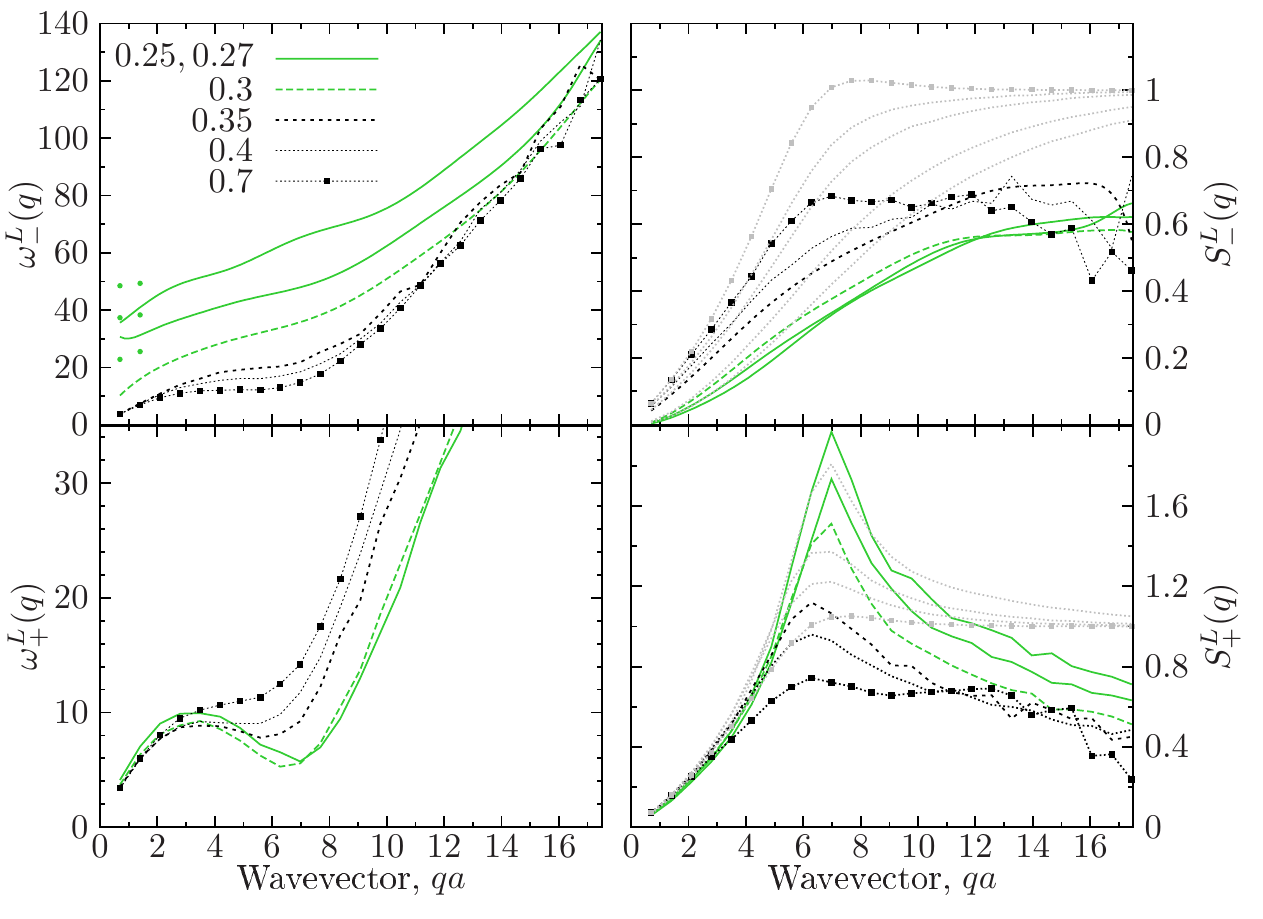}
%{../../../quasi2d/New/d55ed40/data/2d/m1new/d1/twomodepm}
 \end{center}
\vspace{-0.7cm} 
\caption{Left: The low-frequency symmetric and antisymmetric mode $\omega^L_{\pm}(q)$ from the two-mode ansatz. Right: Spectral weights $S^L_{\pm}(q)$. Coupling  $D=1$. Both solutions demonstrate an acoustic behavior in the long wavelength limit for $d\geq 0.35$. An optical branch in $\omega^L_{-}(q)$ is observed for $d \leq 0.30$. For $qa < 2$ the gap value is better described by the Feynman ansatz $\omega^f_-(q)$ (denoted by filled dots for $d=0.25,0.27,0.30$).  The legend indicates the layer spacing. Green/black color is used to distinguish a normal/superfluid phase.}
 \label{fig:two-m1D1}
 \end{figure}

The results are presented in Fig.~\ref{fig:two-m1D1} for different spacing $d$. Shown is the low-frequency branch, $\omega^L_{\pm}(q)$. The second solution, $\omega^H_{\pm}(q)$, is omitted. Typically, it does not describe a well defined dispersion relation, but characterizes some average weighted frequency of a broad multiexcitation continuum.

The generalized Feynman ansatz has several advantages over other approximations, like Singwi-Tosi-Land-Sjolander (STLS)~\cite{neilson96} and quasilocalized charge approximation (QLCA).~\cite{qlca} It predicts: i) spectral weights of collective modes; ii) the sum-rules~(\ref{wab})-(\ref{w3ab}) are exactly satisfied;  iii) sharp quasiparticle resonances can be distinguished from the multiexcitation continuum. 

The left and right panels in Fig.~\ref{fig:two-m1D1} show the wavenumber- and the $d$-dependence of the low-energy branch, $\omega_{\pm}^L(q)$, and its spectral weight, $S_{\pm}^L(q)$. For comparison the full spectral weight (normalization condition), specified by the symmetrized static structure factor $S_{\pm}(q)=S_{\pm}^L(q)+S_{\pm}^H(q)$, is also shown by dotted gray lines  monotonically increasing(decreasing) with $d$ for the symmetric(antisymmetric) mode.

% ------------------------ D=1 m=1 -------------

Several drastic changes in the dispersion relation are observed with variation of the layer spacing $d$.
 
First, for $d\geq 0.35$ and low wavenumbers the dispersion relation is acoustic both for the symmetric and antisymmetric mode. At these layer spacing, the system has a finite superfluid response, see Fig.~\ref{fig:D1m1-superf}. However, once the superfluid fraction drops to zero for $d\leq 0.3$, a finite energy gap develops in the spectrum of the antisymmetric (out-of-phase) mode $\omega^L_-(q)$. As was explained in Sec.~\ref{statD1}, suppression of the superfluidity at $d\leq 0.3$ is due to formation of strongly localized dimer states. Simultaneously, a deep roton minimum develops in the spectrum of the symmetric mode, $\omega^L_+(q)$. The roton wavenumber shifts continuously to larger momenta by lowering $d$ and saturates in the normal phase at the inverse inlayer interparticle spacing, $q \sim 2\pi/a$. The similar behavior demonstrates the roton gap. It has a strong $d$-dependence in the superfluid phase and saturates in the normal phase for $d\leq 0.3$. 

The difference in the resonance frequencies of the symmetric and antisymmetric mode increases at low $d$, both modes become well separated. With the formation of the optical gap, the dispersion $\omega^L_-(q)$ shifts to higher frequencies, while the symmetric mode $\omega^L_+(q)$ to lower frequencies. This has an effect on the $d$-dependence of their spectral weights. By lowering $d$, the spectral weight $S^L_+(q)$ continuously increases, while  $S^L_-(q)$ decreases, see Fig.~\ref{fig:two-m1D1}(right panel).

We conclude that with the formation of dimers, the in-phase density excitations have the largest spectral weight in the partial dynamic structure factor~(\ref{d4}), $S_{11}(q,\omega)=\frac{1}{2}[S_{-}(q,\omega)+S_{+}(q,\omega)]$, and dominate in the roton part of the spectrum. They are responsible for the corresponding peak in the static structure factor $S_{11}(q)$. In contrast, the spectral weight of the out-of-phase mode $S_{-}^L(q)$ shows only a monotonic increase with the wavenumber.

The in-phase excitations probe a collective behavior of a dimer gas and a strength of the dimer-dimer interaction. For the symmetric mode the system can be thought of as a single layer of composite bosons with a new dipole coupling $D^{\star}\sim 8D$. In contrast, the antisymmetric mode for $d\leq 0.3$ probes  intrinsic properties of dimer states. The out-of-phase oscillations act against a spatial localization in a bound state. At low $d$ the dimer binding energy and the interlayer coupling increases, see Fig.~\ref{fig:D1m1-energy}.  As a result the energy gap $\omega^L_-(0)$ gets larger. 

For classical systems presence of a gapped mode for two(multi)-component systems has been predicted by QLCA.~\cite{qlca_gap} However, as shows our analysis in Fig.~\ref{fig:two-m1D1} the spectral weight $S^L_{-}(q)$ of the gapped mode vanishes as $q\rightarrow 0$. Its experimental detection in the long wavelength can be difficult. The use of finite wavenumbers ($qa \gtrsim 1$) is more preferable.

The effect of the interlayer coupling is not restricted to the phonon-roton region, but extends also to large momenta. A fit to $\omega^L_{+}(q)$ for $qa> 8$ with the free-particle dispersion, $\epsilon_q(q;m^{\star},\epsilon_0)=[q^2/2 m^{\star}-\epsilon_0]$, ($\epsilon_0$ is used a fit parameter) results in a new effective mass, $m^\star > m$, which can be explained by the interlayer dimerization. 

For $D=1$ and $d \leq 0.25$, in the regime of strongly bound dimers (with the binding energy $\abs{\epsilon_d}\gtrsim 25$), the fit with $\epsilon_q(q)$ results in $m^{\star}\approx 2m$. The $q$-dependence of $\omega^L_{+}(q)$ is reproduced quite well up to the maximum considered wavenumber $qa \sim 17.4$. 
At larger layer spacing the excitation energies can significantly exceed the dimer binding energy, $\hbar \omega^L_{+}(q) > \abs{\epsilon_d}$, and as a result there exist an upper bound for the wavenumber to observe an effective mass $m^{\star}$ different from a bare particle mass $m$. 

In particular, for $d=0.27$ around $\tilde q a \sim 14.5$ we observe a smooth transition 
from $\epsilon_q(q;2m,\epsilon_0)$ to the free-particle dispersion with a bare mass, $\epsilon_q(q;m,\epsilon_0^{\prime})$. 
For larger $d$ this transition occurs earlier, and, typically, for $qa \gtrsim 12.5$ we recover $m^{\star}\approx m$. The dimer dispersion $\epsilon_q(q;m^{\star}=2m,\epsilon_0)$ is recovered just beyond the roton feature for $8 \lesssim qa \lesssim 12$. The transition to the dispersion $\epsilon_q(q;m,\epsilon_0^{\prime})$ starts around the wavenumber $\tilde qa$ corresponding to the excitation energy comparable to the dimer binding energy at a given layer spacing
\begin{align}
 \hbar \omega^L_{+}(\tilde qa) \sim g \cdot \abs{\epsilon_d(d)},\label{wl} 
\end{align}
with the scaling factor $g \lesssim 2$. 

For stronger coupling $D=5.5$ (see Fig.~\ref{fig:two-m1D55}) our observations are similar but reveal some new feature. Our present analyses are limited by the layer spacing $d\geq 0.6$ and the dimer binding energy $|\epsilon_d(d)| \leq 20$.

First, we observe (Fig.~\ref{fig:two-m1D55}) the roton feature (around $qa\sim 6.5$) with a slope specified by the ``roton mass''. Next, for $8 \lesssim qa \lesssim 10$ we recover the dimer dispersion $\epsilon_q(q;2m,\epsilon_0)$. This interval is followed by the transient region $10 \lesssim qa \lesssim 12.5$ where the dispersion exhibits a slight bending (flattens) once it becomes comparable to the dimer binding energy, Eq.~(\ref{wl}) with $g \sim 1$. Finally, for $qa \gtrsim 12.5$ the dispersion converges fast to a slope specified by a bare mass $\epsilon_q(q;m,\epsilon_0^{\prime})$.

In conclusion, the effective mass $m^{\star} > m$ due to the interlayer coupling can be observed beyond the roton feature with the upper bound for the momentum specified by Eq.~(\ref{wl}).

%The increase of the effective mass for $d\leq 0.3$ is limited by $m^\star \sim 2 m$ due to the dimer formation.

In Fig.~\ref{fig:two-m1D1}, some unsmooth behavior of the dispersion relations (and their weights) at large $q$ can be noted. This is a numerical artefact due to statistical errors in the input values of the frequency power moments~(\ref{wab})-(\ref{w3ab}). The errors, typically, increase with the wavenumber. Possible solutions for the set of equations~(\ref{wH-eq})-(\ref{sH-sL}) are found to be sensitive to any source of numerical uncertainties. Still in a wide $q$-range we reproduce quite smooth dependencies, and resolve a continuous evolution of the dispersion relations with the layer spacing.

Next, we discuss the observed transformation of the acoustic branch in $\omega^L_-(q)$ into the optical one. This occurs during the superfluid-normal fluid phase transition in the interval $0.3\leq d\leq 0.6$ (Fig.~\ref{fig:D1m1-superf}). It was explicitly shown by Gavoret and Nozi{\`e}res,~\cite{noz} that Bose condensation leads to hybridization of a single particle and collective density excitations.~\cite{griffinbook,glydebook} In the long wavelength and zero temperature limit, both spectral densities share a common pole -- the compressional sound. Presence of a gapless mode in the single particle spectra (but necessarily exhausted by this mode) has been theoretically proven. Hence, a linear dispersion should be present in $S_{\pm}(q,\omega)$ as lowest quasiparticle excitation (in addition to other energy resonances), if there exists off-diagonal long-range order. This fact can explain the presence of the acoustic branch in $\omega^L_-(q)$ in the superfluid phase. When the spatial coherence disappears due to the dimerization, the acoustic dispersion is substituted by the gapped mode. The density excitations are dominated by the interlayer correlations of the dimer states. The final conclusions can be drawn after we analyze the full spectral densities $S_{\pm}(q,\omega)$ and exclude a possibility that the current results are a  numerical artefact of the two-mode ansatz.

\begin{figure}
 \begin{center} 
 \hspace{-0.0cm}\includegraphics[width=0.5\textwidth]{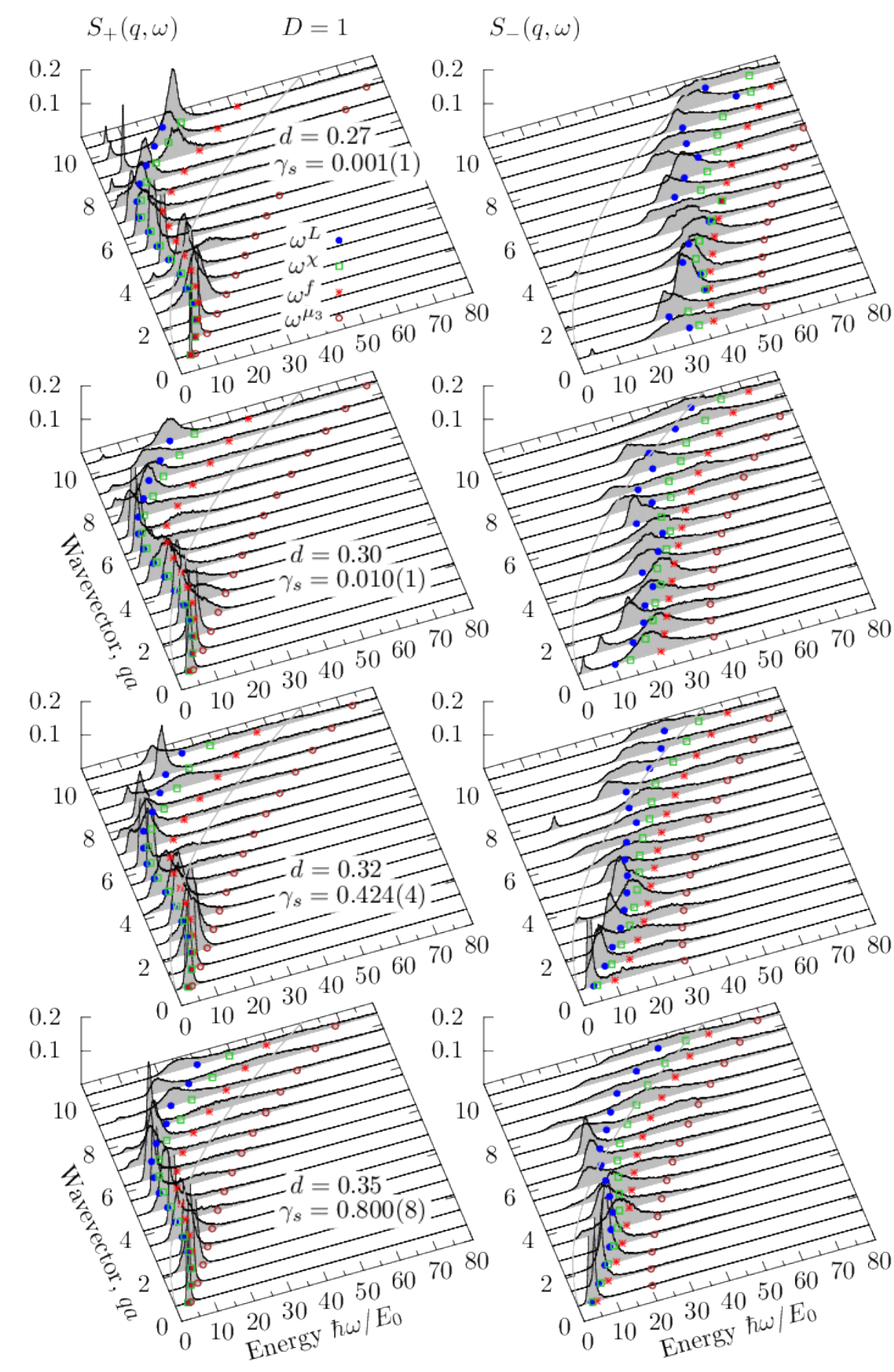}
 %{../../../quasi2d/New/d55ed40/data/2d/m1new/d1/stopmpaper/spec3dm1l027l035}     
 \end{center}
\vspace{-0.6cm} 
 \caption{Rescaled dynamic structure factor for the symmetric (left panel), $S_{+}(q,\omega)/S_+(q)$, and antisymmetric (right panel), $S_{-}(q,\omega)/S_-(q)$, modes. Coupling $D=1$. The legend indicates the layer spacing, $0.25 \leq d\leq 0.35$, and the superfluid fraction $\gamma_s$. For comparison several upper bounds for the dispersion relation are shown: $\omega_\pm(q) \leq \omega^\chi_\pm(q) \leq \omega^f_\pm(q) \leq \omega^{\mu_3}_\pm(q)$, see Eq.~(\ref{3upbound}). The ansatz $\omega^L_{\pm}$ (indicated by blue symbols) provides best agreement with the low-frequency resonances in $S_{\pm}(q,\omega)$. The solid gray line denotes the free-particle dispersion $\epsilon_q=q^2/2m$.}
 \label{fig:m1d1a}
 \end{figure}

{\em Dynamic structure factor.}
The dynamic structure factors~(\ref{d4}) are reconstructed from the density-density correlation functions~(\ref{g2}). The details of the method are provided in Ref.~\cite{fil2012} and are shortly reviewed in Appendix~\ref{SO}. In general, any reconstruction procedure is strongly influenced by the statistical noise present in the input data.~\cite{fil2012} However, as it is demonstrated in Appendix~\ref{SO}, the use of known frequency power moments can significantly reduce this dependence, and our present results demonstrate a continuous and systematic evolution of peak positions and their halfwidth with the layer spacing.  

In Fig.~\ref{fig:m1d1a} the spectral density is presented for a set of $d$-values. The legend indicates $d$ and the superfluid fraction $\gamma_s$. Corresponding changes in the static characteristics can be followed in Fig.~\ref{fig:grsk-D1m1}. At $d=0.3$ we are in the regime of a strong interlayer coupling. The dispersion $\omega^L_{+}(q)$ shows a well pronounced roton minimum, see Fig.~\ref{fig:two-m1D1}. For comparison, $d \sim 0.4$ corresponds to the onset of dimerization, and $d=0.7$ to a weak interlayer coupling. In the later case, the spectral density approaches the result for a single layer.~\cite{fil2012}

In Fig.~\ref{fig:m1d1a} the low-frequency resonances in $S_{\pm}(q,\omega)$ are compared with several upper bounds for dispersion relation (indicated by different symbols). At low temperatures (and $T=0$) they  satisfy the known inequality~\cite{lipp}
\begin{align}
 \omega_\pm(q)\leq \frac{\Avr{\omega^0}_\pm}{\Avr{\omega^{-1}}_\pm}\leq \frac{\Avr{\omega^1}_\pm}{\Avr{\omega^{0}}_\pm} \leq \sqrt{\frac{\Avr{\omega^3}_\pm}{\Avr{\omega^{1}}_\pm}},
 \label{w_ineq}
\end{align}
and have been introduced in Sec.~\ref{twomode} as the key ingredients for the two-mode solution $\omega^L_{\pm}(q)$. We observe that the upper bounds, indeed, form the correct sequence, but the best agreement with the peak positions in $S_{\pm}$ is provided by $\omega^L_{\pm}(q)$. This proves the advantage of the two-mode ansatz over other approximations, where the high-energy spectral features are not treated explicitly. A typical situation when our methods can fail is presented by $S_{-}(q,\omega)$ at $d=0.30$. The system has a small superfluid response ($\gamma_s=0.01$). At low $q$ the spectrum splits into one optical (gapped) and one acoustic mode. In this regime $\omega^L_{-}(q)$ characterizes some averaged weighted frequency which is not relevant for any of the modes. In contrast, the Feynman upper bound, $\omega^f_-(q)$, derived from the $f$-sum rule, is more sensitive to high-frequency spectral features. Therefore, it correctly predicts the gap value and the optical mode up to $qa \lesssim 2$. The predictions based on $\omega^{\mu_3}(q)$ are less reliable. The corresponding dispersion is shifted to high frequencies for all wavenumbers. Based on the third moment $\mu_3$,  it gets a main contribution from a slow decaying high-frequency tail of $S_{\pm}(q,\omega)$. Its applicability is limited to the acoustic range in $S_{+}(q,\omega)$, where only phonon resonances are present and the corresponding spectral density fast decays to zero at higher frequencies. As expected, in this case all upper bounds~(\ref{w_ineq}) converge to a single phonon dispersion, being the lowest energy mode for the in-phase density excitations.

Similar observations hold also for $d=0.32$. Now the optical mode is shifted to lower frequencies and demonstrates a significant broadening due to the overlap with the acoustic mode. The $\omega^L_{-}$-ansatz predicts an acoustic branch, while the Feynman mode -- an optical branch. Note, that compared to $d=0.30$, now the acoustic branch carries a significant spectral weight and the superfluid fraction increases to $\gamma_s=0.424$. This suggests that the presence of the gapless mode is related with the superfluid component. This mode completely dominates the spectrum  when $\rhos$ increases further, see $S_{-}(q,\omega)$ for $d=0.35$.

We assume that the split of the spectra into two modes is due to the hybridization of a single particle and collective density excitations, as was discussed above. The single particle spectra should be gapless in the long wavelength limit,~\cite{noz} and by sharing a common pole with the density response function is responsible for the low frequency acoustic branch. Below we investigate this result more in detail.

\begin{figure}
 \begin{center} 
 \hspace{-0.0cm}\includegraphics[width=0.5\textwidth]{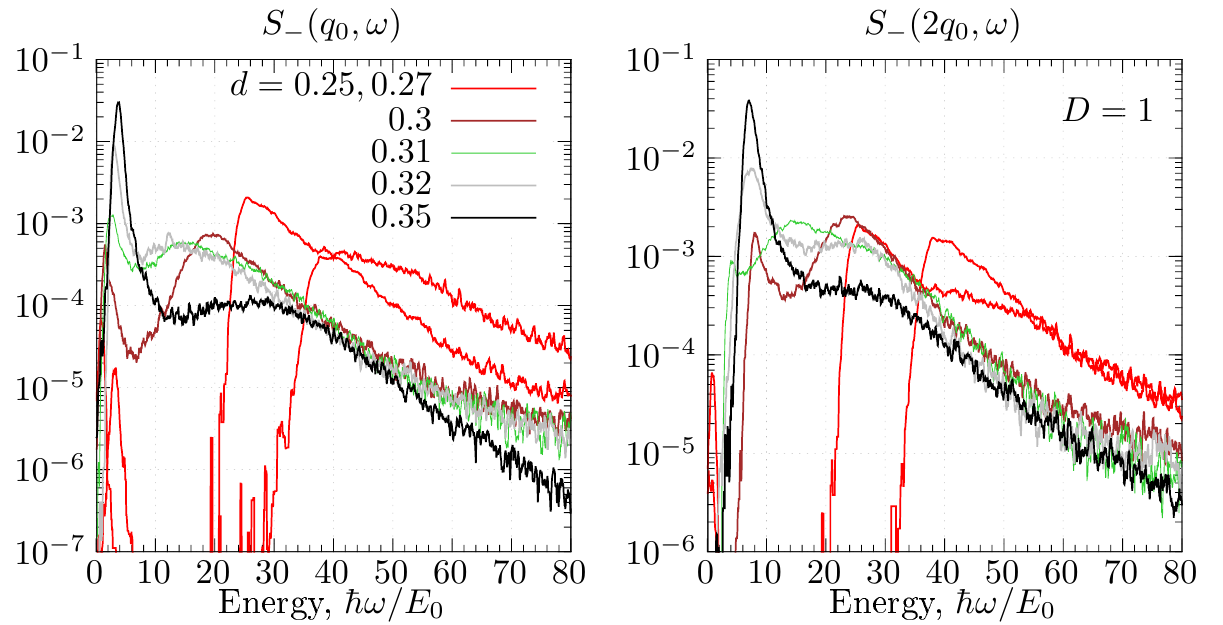}
 %{../../../quasi2d/New/d55ed40/data/2d/m1new/d1/stopmpaper/specSmk1}     
 \end{center}
\vspace{-0.6cm} 
 \caption{The $d$-dependence of $S_{-}(q,\omega)$ for $D=1$ at two wavenumbers $q=q_0 n$ ($q_0=2 \pi/L$): $n=1$ (left) and $n=2$ (right). The legend indicates the layer spacing, $0.25 \leq d\leq 0.35$.}
 \label{fig:Specmk1}
 \end{figure} 

Now we concentrate on the range of layer spacings $d$ where the superfluid fraction increases from zero to a finite value, and  $S_{-}(q,\omega)$ splits into two branches. Fig.~\ref{fig:Specmk1} allows to trace how relative contributions (the spectral weights) of the optical and the acoustic branch in $S_{-}(q,\omega)$ changes with the spacing $d$. Two smallest wavenumbers, where the splitting of the modes is more prominent, are chosen. The second peak (optical branch) completely merges with a first peak (acoustic branch) or drops out from the spectrum at $d=0.35$, when the superfluid fraction increases to $80\%$. In the opposite limit, at $d=0.25,0.27$ only the optical branch is observed. Simultaneously, the system undergoes a complete dimerization (see discussion in Sec.~\ref{D1coup}) and $\gamma_s$ drops below $0.1\%$.

\begin{figure}
 \begin{center} 
 \hspace{-0.0cm}\includegraphics[width=0.5\textwidth]{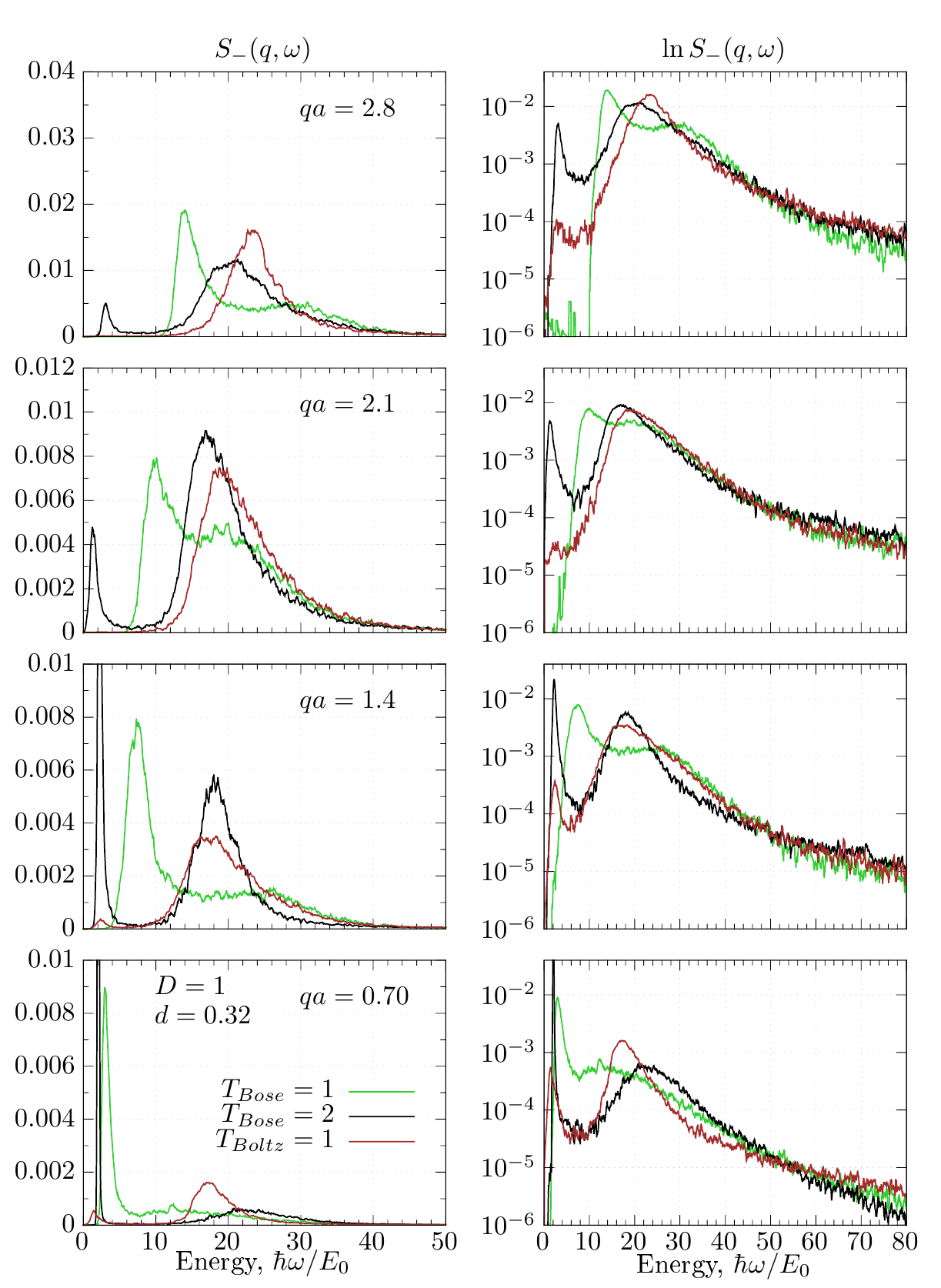}
 %{../../../quasi2d/New/d55ed40/data/2d/m1new/d1/stopmpaper/specd1m1l032boseboltz}     
 \end{center}
\vspace{-0.6cm} 
 \caption{Comparison of the antisymmetric mode spectra for Bose bilayers at $T=1$ ($\rhos=0.42$), $T=2$ ($\rhos=0$)  and the bilayer with Boltzmann statistics at $T=1$ ($\rhos=0$). Layer spacing $d=0.32$ and $D=1$. The right panels show the results on the log-scale. The legend indicates the wavenumbers $qa$. In the partially superfluid phase ($\rhos=0.42$) the spectra is dominated by the acoustic branch. The optical branch is  suppressed and strongly overlaps with the acoustic one. In contrast, in the simulations with $\rhos=0$ the spectral weight is carried by the optical branch. At low frequencies some additional resonance (but with a much smaller spectral weight) is observed for $T_{\text{Boltz}}=1$. The same resonance (but significantly enhanced) is observed also for $T_{\text{Bose}}=2$. This additional dispersionless branch is formed in the wide range of wavenumbers ($0<qa < 9$) and, can be interpreted as intrinsic excitations of the dimer states.}
 \label{fig:specl032}
 \end{figure}

To confirm that the presence of the acoustic branch is due to a superfluid component, we have performed additional calculations for $d=0.32$ (same values of temperature and chemical potential) but ``switched off'' the Bose statistics (PIMC simulations for distinguishable particles). The comparison of $S_{-}(q,\omega)$ for Bose and Boltzmann particles is presented in Fig.~\ref{fig:specl032}. One clearly observes the effect induced by the Boltzmann statistics: the spectrum of ``boltzmannons'' is composed of an optical branch. Such a clear distinction of the excitation spectra in the normal and superfluid phases can be used for practical applications, e.g. to distinguish both phases in experiments on ultra cold gases.

To further support our argument we have repeated simulations at a higher temperature, $T_{\text{Bose}}=2$, when the Bose system is non-degenerate with the global superfluid density $\rho_s=0$. The Bose statistics plays only a minor role, is limited to few-particle exchanges, and does not lead to a global spatial coherence. Similar to  the Boltzmann case ($T_{\text{Boltz}}=1$) the main peak position is practically $q$-independent and shows a gap in the long wavelength limit, see Fig.~\ref{fig:specl032}. The energy resonances demonstrate some thermal broadening and are slightly shifted to lower frequencies compared to $T_{\text{Boltz}}=1$. No sign of an additional acoustic branch is observed. Instead, by the reconstruction we recover a new low-frequency dispersionless mode, which can be related with the intrinsic excitations of the dimer states. Interestingly, that in the superfluid phase this mode is not observed, as the system behavior is dominated by the collective modes. When temperature is increased, one expects to observe a decay of the collective modes into combinations of two and more quasiparticles. As a result in the lower-frequency region the dimer mode is populated. The $q$-independence of this mode validates that it is of a single particle nature.

\begin{figure}
 \begin{center} 
 \hspace{-0.35cm}\includegraphics[width=0.5\textwidth]{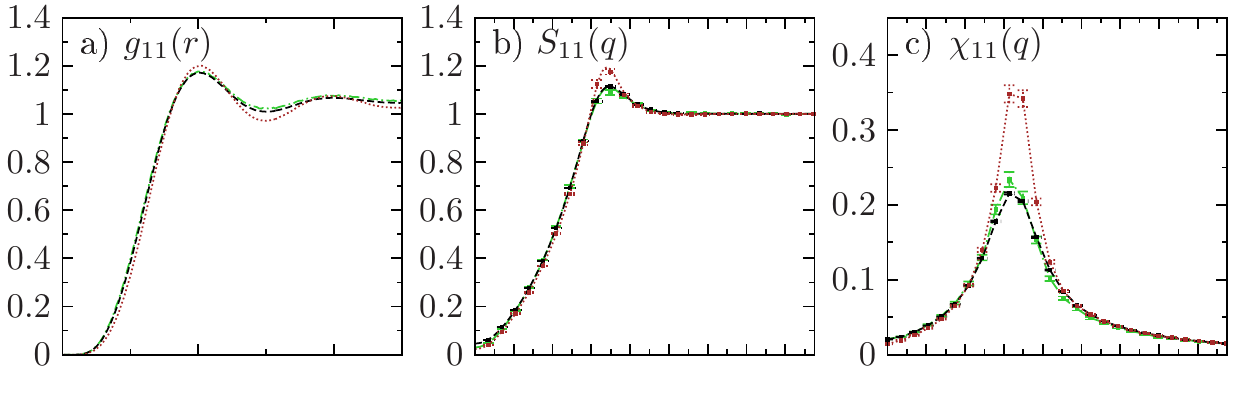}\\
 %{../../../quasi2d/New/d55ed40/data/2d/m1new/d1/gr11sk11chi11noxboltz}\\
 \vspace{-0.25cm}
 \hspace{-0.35cm}\includegraphics[width=0.5\textwidth]{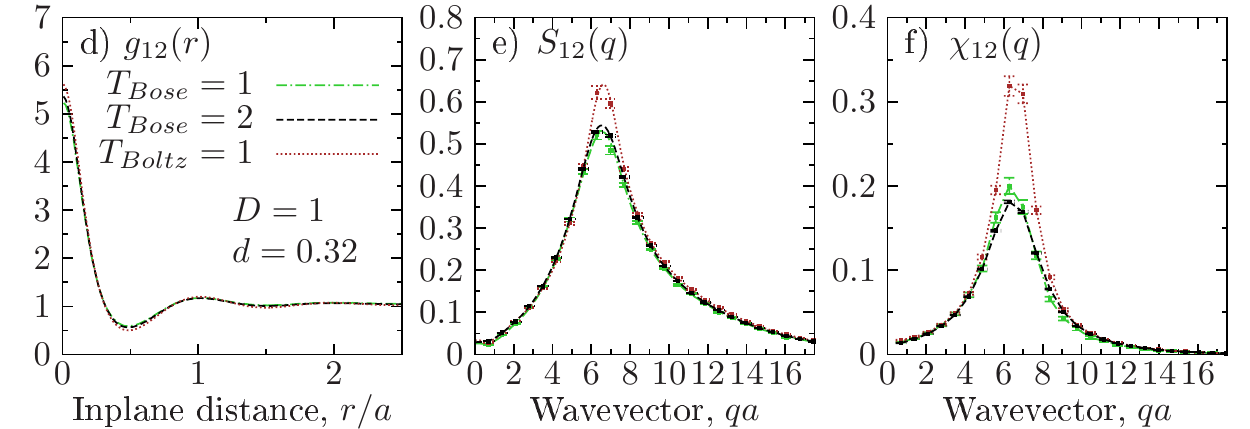}\\
 %{../../../quasi2d/New/d55ed40/data/2d/m1new/d1/gr12sk12chi12boltz}\\
 \vspace{-0.05cm}
 \hspace{-0.35cm}\includegraphics[width=0.5\textwidth]{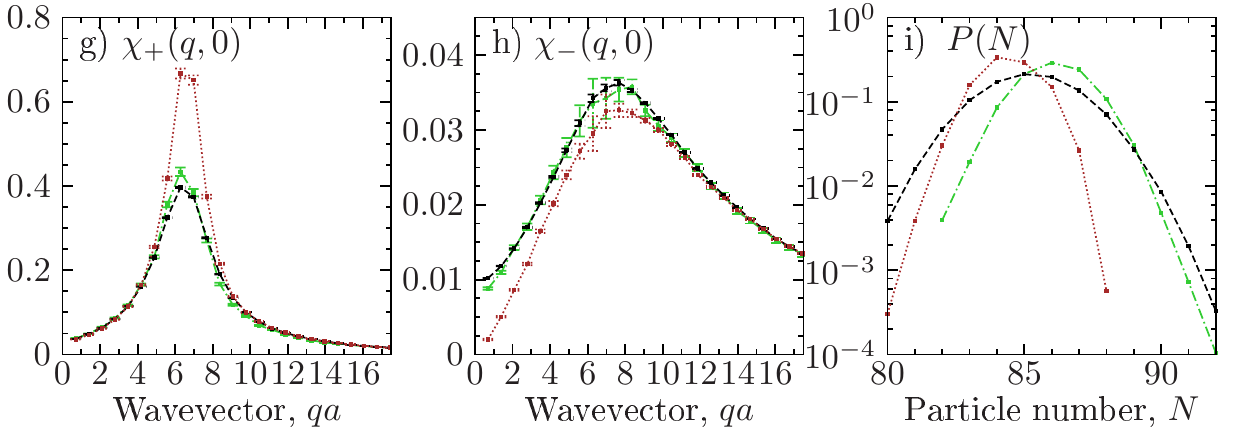}
 %{../../../quasi2d/New/d55ed40/data/2d/m1new/d1/chi0diagboltz}
 \end{center}
\vspace{-0.60cm} 
 \caption{Comparison of the static properties for $d=0.32$ and $D=1$. Compared are Bose bilayers at $T=1,2$ and the bilayer of boltzmannons at $T=1$.}
 \label{fig:specl032stat}
 \end{figure}

In Fig.~\ref{fig:specl032stat} we compare the static characteristics for the three cases discussed above. First, there is a difference in the average density, see panel (i) with the particle number distribution. The highest density corresponds to the superfluid phase. Second, among the three cases the static structure factor $S_{11(12)}(q)$ reaches its maximum value (around $qa\sim 7$) for $T_{\text{Boltz}}=1$, i.e. highest spatial correlations are reached in the non-superfluid phase at low temperatures. At the same time the static response function $\chi_{-}(q,0)$ converges nearly to zero as $q\rightarrow 0$. This quantity provides information of the interlayer particle number fluctuations. Using Eqs.~(\ref{Snm2})-(\ref{chinm}) and the symmetry relation ($\rho_{11}=\rho_{12}=\rho$) we can write
\begin{align}
&\frac{k_BT}{2} \abs{\Re \chi_{-}(0,\omega=0)}\nonumber \\
&=\rho [S_{11}(0)-S_{12}(0)]=\frac{\Avr{N_1^2} -\Avr{N_1 N_2}}{V}.
\label{fn12}
\end{align}
Thus, for boltzmannons at low temperatures the instantaneous particle number in both layers $N_\al$ ($\al=1,2$) are strongly correlated. This becomes possible due the interlayer dimerization when particles from different layers are pairwise coupled. 

Similar behavior is observed for Bose statistics at $d\leq 0.30$. The superfluid density drops to zero being a clear sign of the dimerization. The many-body exchange effects are suppressed due to a strong reduction of the mean dimer size, see Fig.~\ref{fig:D1m1-energu0}. The value $\chi_-(0,0)$ reduces nearly to zero, similar to the case $T_{\text{Boltz}}=1$.

For a non-zero value of $\Re \chi_{\pm}(0,0)$, the excitation spectrum should necessarily have a gapless mode in the long wavelength limit. For the symmetric mode this is always the case: the acoustic branch is present for all layer spacing, see $S_+(q,\omega)$ in Fig.~\ref{fig:m1d1a}. A finite value of $\Re \chi_+(0,0)$ is recovered independent on the quantum statistics and temperature, see Fig.~\ref{fig:specl032stat}g.

For the antisymmetric mode, a finite value of $\Re \chi_-(0,0)$ also depends on a gapless mode, and is observed in two cases. In the superfluid phase ($T_{\text{Bose}}=1$) and in the normal phase ($T_{\text{Bose}}=2$). In the first case, the acoustic branch is an intrinsic property of a superfluid. For $T_{\text{Bose}}=2$, it is due to a new low-frequency dispersionless mode, see Fig.~\ref{fig:specl032}.

Finally, we can conclude that the long wavelength limit of $\Re \chi_{-}(q,0)$ allows to identify either a gapless or a gapped mode.

{\em Model for dynamic structure factor.}
The static characteristics presented in Fig.~\ref{fig:specl032stat} are frequently used in different approximates for the density response function. Typically, they are included in the local-field corrector, $G(q,\omega)$, treated in the static approximation ($\omega=0$). The effects of quantum statistics, as shows the above comparison, can be equally important. A correct model of $G(q,\omega)$ should be able to reproduce the observed splitting into acoustic and optical branches in a partially superfluid phase. Our example demonstrates importance to include dynamical correlations into correct analyses of collective excitations in superfluids. Surprisingly, the information on these correlations is present and can be successfully recovered from the imaginary time dynamics of the density operator.

Our next goal is to find a relation between the spectral weight of the acoustic mode, intrinsic to a superfluid phase, and the superfluid density. The results in Fig.~\ref{fig:Specmk1} clearly demonstrate that such a relation should exist. To split a contribution of the acoustic and the optical branch we define the following procedure. Certainly, our treatment is approximate and its range of applicability is limited by the wavenumbers where both energy resonances do not overlap significantly. 

In the first approach, the spectral density, in vicinity of a second maximum (corresponding to the optical branch), is fitted with the equation of a damped harmonic oscillator (DHO)~\cite{dho} 
\begin{align}
 -\frac{\Im \chi_{\text{DHO}}(q,\omega)}{\pi \rho}=&Z(q) \frac{8\, \omega \, \omega(q) \Gamma(q)}{[\omega^2 -\omega(q)^2]^2+4 \omega^2 \Gamma(q)^2},\\
 S_{\text{DHO}}(q,\omega)=&-\frac{\Im \chi_{\text{DHO}}(q,\omega)}{\pi\rho (1-e^{-\beta \omega})}.
 \label{dho-an}
\end{align}
The fit parameters are the width $\Gamma(q)$ (which defines the damping), the dispersion $\omega(q)$ and the spectral weight $Z(q)$. Their $T$-dependence, as a second argument, is omitted. The spectral density of the optical branch is defined by $S_{\text{DHO}}(q,\omega)$. For us most important is a behavior of $S_{\text{DHO}}(q,\omega)$ in the region where the two modes overlap, producing a local minimum between two resonances. Here their contribution to the  dynamic structure factor $S_\pm(q,\omega)$ should be distinguished. Using results of the fit, we  can define the spectral density of the acoustic branch as
\begin{align}
S_{\text{A,DHO}}(q,\omega)=S(q,\omega)-S_{\text{DHO}}(q,\omega).\label{adho}
\end{align}
By the integration over frequency we get the spectral weight 
\begin{align}
S_{\text{A,DHO}}(q)=\int \limits_{0}^{\omega^\star} \db \omega \, S_{\text{A,DHO}}(q,\omega) \, (1+e^{-\beta \omega}).
\end{align}
The integration is performed up to the frequency $\omega^\star$, where $S_{\text{A}}(q,\omega)$ drops to zero or becomes negative due to the used approximation~(\ref{adho}).

Alternatively, the optical mode can be approximated by the ansatz from the method of moments~\cite{Ark2010} (MM)
\begin{align}
 -\frac{\Im \chi_{\text{M}}(q,\omega)}{\pi \rho\, \omega}=&\frac{1}{\pi} \frac{ \tilde\mu_2 (\omega_2^2-\omega_1^2) \Gamma(q)}{\omega^2(\omega^2-\omega_2^2)^2+\Gamma(q)^2 (\omega^2-\omega_1^2)^2},\label{MM1}\\
 S_{\text{M}}(q,\omega)=&-\frac{\Im \chi_{\text{M}}(q,\omega)}{\pi\rho (1-e^{-\beta \omega})}
 \label{MM2},
\end{align}
with the fit parameters $\omega_{1(2)}(q)$, $\tilde\mu_2(q)$ and $\Gamma(q)$. By its construction the density response function~({\ref{MM1}}) exactly satisfies five frequency power moments $\{\tilde\mu_0,0,\tilde\mu_2,0,\tilde\mu_4,0\}$
\begin{align}
\tilde\mu_k=\int\limits_{-\infty}^{\infty}\db \omega\, \omega^k\, \left[-\frac{\Im \chi_{\text{M}}(q,\omega)}{\pi \rho\, \omega}\right].
\end{align}
A resonance position is constrained to  $\omega_1(q) \leq \omega(q) \leq \omega_2(q)$ and depends on the frequencies $\omega_1^2=\tilde\mu_2/\tilde\mu_0$ and $\omega_2^2=\tilde\mu_4/\tilde\mu_2$. The role of the damping $\Gamma(q)$, or more general the Nevanlinna parameter, is to shift a resonance position within the interval $[\omega_1,\omega_2]$, rescale the spectral weight defined the frequency integral of~(\ref{MM2}), and to define a halfwidth of the resonance peak, also influenced by the width of the interval $[\omega_1,\omega_2]$.

\begin{figure}
 \begin{center} 
 \hspace{-0.0cm}\includegraphics[width=0.5\textwidth]{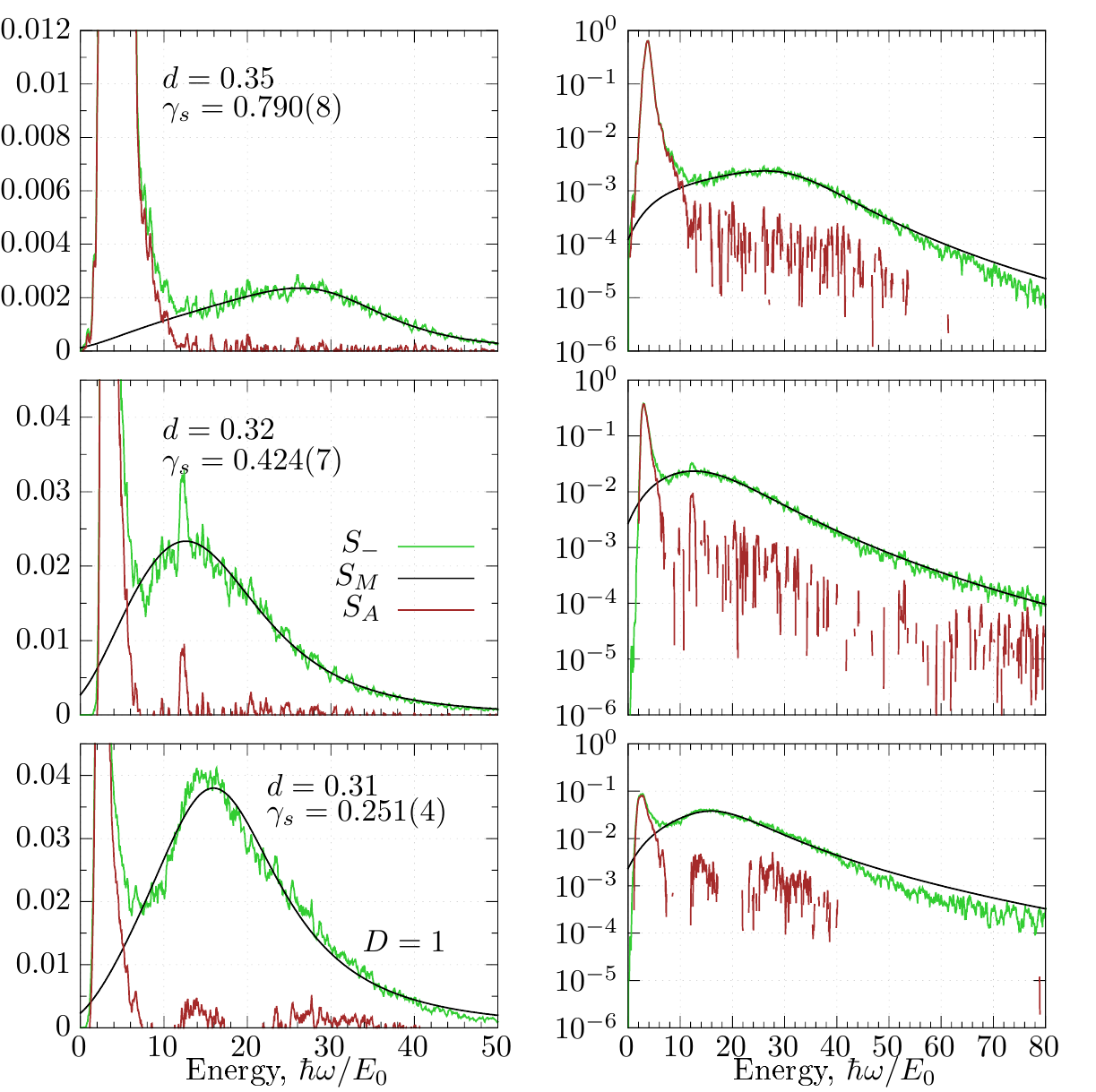}
 %{../../../quasi2d/New/d55ed40/data/2d/m1new/d1/stopmpaper/twomodefit}     
 \end{center}
\vspace{-0.6cm} 
  \caption{Decomposition of the renormalized dynamic structure factor, $\tilde{S}_{-}(q,\omega)= S_{-}(q,\omega)/S_{-}(q)$ at $qa=2\pi/L$, into the acoustic and the optical branch: $\tilde S_{-}(q,\omega)=S_{\text{A}}(q,\omega)+S_{\text{O}}(q,\omega)$. The high-frequency optical branch is fitted with the MM-ansatz: $S_{\text{O}}=S_{\text{M}}$, Eq.~(\ref{MM2}). The  spectral weight of the acoustic mode $S_{\text{A}}$ is compared in Tab.~\ref{tab1} with the superfluid fraction $\gamma_s$. The legend indicates $\gamma_s$ at different layer spacing $d$.}
 \label{fig:twomodeAB}
 \end{figure}

In our fit procedure, as a first approximation for the power moments in~(\ref{MM1}), we use the results obtained from the DHO: $S_{\text{DHO}}$ is set to zero below the acoustic resonance at $\omega_{\text{A}}(q)$
\begin{align}
 & S_{\text{A}}(q,\omega)=\begin{cases} 
                            S_{-}(q,\omega), \hspace{2.4cm} \omega \leq \omega_{\text{A}(q)} \\
                            S_{-}(q,\omega)-S_{\text{DHO}}(q,\omega), \quad \omega > \omega_{\text{A}(q)}                           
                           \end{cases}\label{sab}\\
 &\tilde\mu_0/2=\Avr{\omega^{-1}}_{S_-}-\Avr{\omega^{-1}}_{S_{\text{A,DHO}}},\\ &\tilde\mu_2/2=\Avr{\omega^{1}}_{S_-}-\Avr{\omega^{1}}_{S_{\text{A,DHO}}},\\
 &\tilde\mu_4/2=\Avr{\omega^{3}}_{S_-}-\Avr{\omega^{3}}_{S_{\text{A,DHO}}},
\end{align}
where $S_{-}(q,\omega)$ is the reconstructed spectral density with the stochastic optimization (Appendix~\ref{SO}), $S_{\text{A,DHO}}(q,\omega)$ is the ansatz for the acoustic mode obtained with the DHO, and $\Avr{\omega^k}_{S_{\text{A,DHO}}}$ is the contribution of the acoustic mode to different sum rules. Once $\{\omega_{1(2)}\}$ are fixed by this procedure (and correspondingly the frequency interval for the resonance of the optical branch), we proceed to a numerical fit using $\Gamma(q)$ and $\tilde\mu_2$ as free parameters. As a next step, $\{\Gamma(q),\tilde\mu_2\}$ are kept fixed, and the frequencies $\{\omega_{1(2)}\}$ are varied. In the final iteration all parameters are allowed to vary, but the convergence is fast, as the fit parameters are already near their optimal values.

Once the MM-fit is constructed, we reevaluate the spectral density of the acoustic branch by replacing $S_{\text{DHO}}$ with $S_{\text{M}}$ in Eq.~(\ref{sab}).

The efficiency of this fit procedure for the high-frequency part of the spectrum is demonstrated in Fig.~\ref{fig:twomodeAB}. In all cases the MM-ansatz quite accurately reproduces an asymptotic high-frequency decay of $S_{-}(q,\omega)$. This is guaranteed by the fulfillment of the sum rule, $\tilde\mu_4/2 \approx \Avr{\omega^3}_{-}$. A contribution of the acoustic branch in this moment is small.

Finally, in Tab.~\ref{tab1} we compare a spectral weight of the acoustic branch, $S_A(q)$,  with the superfluid fraction $\gamma_s$. Both $S_{A}(q)$ and the static structure factor $S_{-}(q)$ increase progressively with the superfluid fraction $\gamma_s$. The used fit parameters for the high-frequency part of the spectrum~(\ref{MM1}) are also included. 

\begin{figure}
 \begin{center} 
 \hspace{-0.0cm}\includegraphics[width=0.5\textwidth]{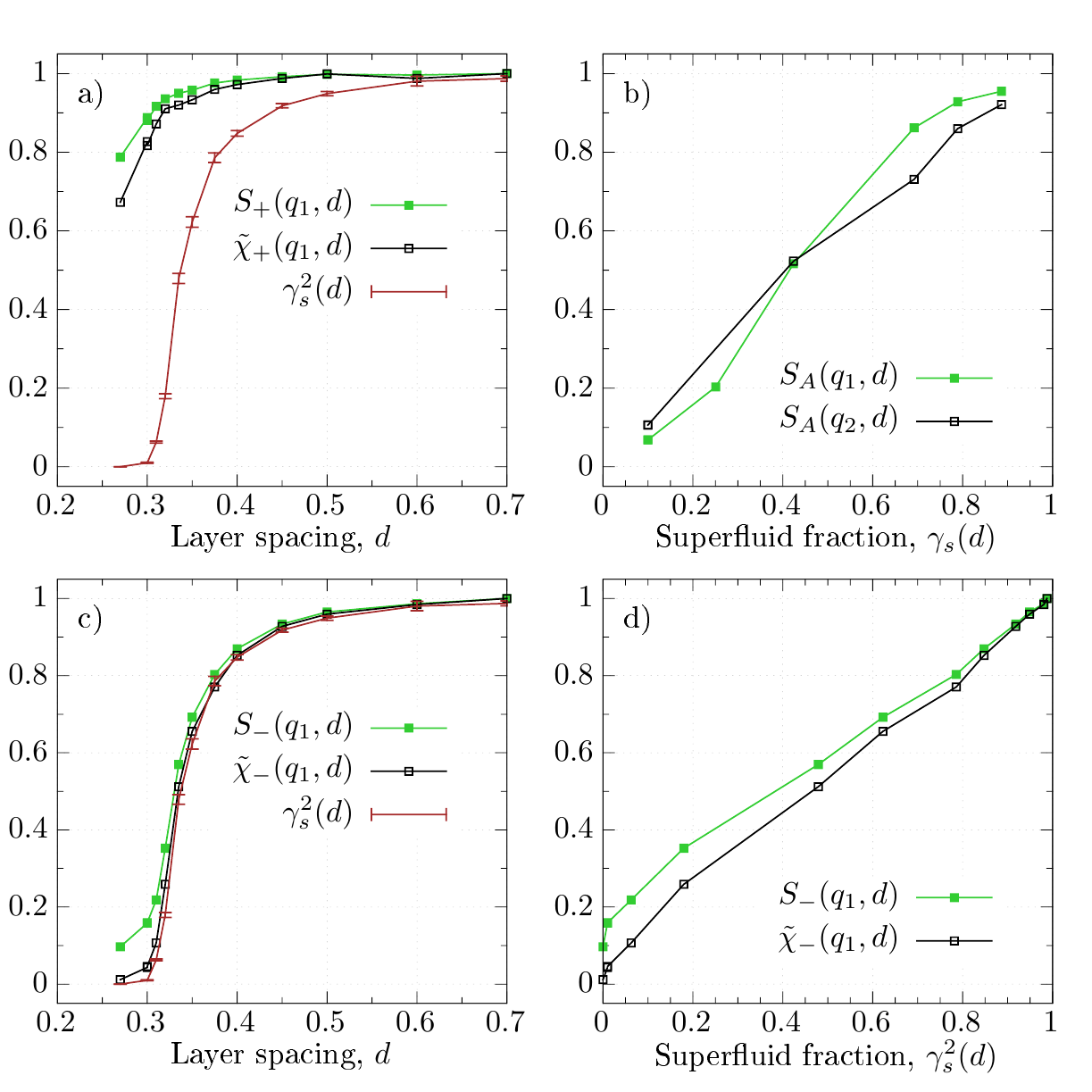}
 %{../../../quasi2d/New/d55ed40/data/2d/m1new/d1/stopmpaper/rhosweight}     
 \end{center}
\vspace{-0.6cm} 
  \caption{a),c) The $d$-dependence of the renormalized static structure factor, $S_{\pm}(q_1,d)=S_{\pm}(q_1,d)/S_{\pm}(q_1,d=0.7)$, the static response function $\tilde\chi_{\pm}(q_1,d)=\Re \chi_{\pm}(q_1,d)/ \Re \chi_{\pm}(q_1,d=0.7)$ for the (anti)symmetric mode, and of the square of the superfluid fraction $\gamma^2_s(d)$. The normalization factors: $S_{\pm}(q_1,d=0.7)=0.07663(0.06741)$ and $\frac{1}{2}\Re \chi_{\pm}(q_1,d=0.7)=0.02112(0.01704)$. b) The integrated  spectral weight of the acoustic branch, $S_{\text{A}}(q_n,d)$, versus $\gamma_s$. The considered wavenumbers: $q_n= 2 \pi n/L$ ($n=1,2$). The system size $L=9$. d) Same as in the panel (c) plotted versus $\gamma^2_s(d)$ to demonstrate a linear dependence.}
 \label{fig:rhosweight}
 \end{figure} 

In Fig.~\ref{fig:rhosweight} we analyze our results more in detail. The panels a),c) show the $d$-dependence of the static density response function $\Re \chi_{\pm}(q_1,0)$ and $S_{\pm}(q_1)$ taken at the wavenumber $q_1=2\pi/L$ ($L=9$). For comparison we also plot the square of the superfluid fraction $\gamma^2_s(d)$.  

For the symmetric mode, we can easily identify two regions characterized by a strong and weak $d$-dependence. For  $d\lesssim 0.32$, the superfluid fraction fast drops to zero from $50\%$. As was discussed in Sec.~\ref{D1coup}, in this regime thermodynamic properties are dominated by properties of a single dimer state. In contrast, for $d>0.32$ a weak $d$-dependence is observed. The interlayer coupling effects are screened by a homogeneous superfluid phase within each layer. 

More, interesting features are observed for the asymmetric mode, see Fig.~\ref{fig:rhosweight}c. The system response to compressibility modes with a phase shift quite closely reproduces the $d$-dependence of the superfluid response. After rescaling the plotted quantities converge to a single curve with the $d$-dependent slope which reproduces the square of the superfluid fraction $\gamma^2_s(d)$. The similarity of $S_{-}(q)$ and $\chi_{-}(q)$ is not surprising, as both values at the smallest wavenumber $q_1$ can not differ significantly from $S_{-}(0)$ and $\Re\chi_{-}(0,0)$ related by the compressibility sum-rule~(\ref{Snm2})-(\ref{chinm}), independent on the many-body correlation effects and quantum statistics. This dependence is demonstrated more explicitly in the panel (d), where a nearly  linear dependence on $\gamma_s^2(d)$ is reported in a wide range of layer spacing. The plotted data corresponds to $0.27\leq d\leq 0.7$.

The excitation spectrum of the antisymmetric mode is also influenced by the superfluidity. In a partially superfluid phase it splits into one acoustic and one optical branch. The relative spectral weight of the acoustic mode $S_A$ (taken at two smallest wavenumbers, see Tab.~\ref{tab1}) is plotted in Fig.~\ref{fig:rhosweight}b versus $\gamma_s(d)$. Almost a linear dependence on $\gamma_s$ is observed, validating that presence of the acoustic branch is directly related with the superfluid density. The acoustic branch can either dominate the excitation spectra, $S_{-}(q,\omega)$, when $\gamma_s \rightarrow 1$, or completely  vanish in the opposite limit, when it is substituted by the spectrum of the normal component. In the later case, a finite energy gap develops in the spectrum in the long wavelength limit, see Fig.~\ref{fig:m1d1a}.

Our results allow us to conclude that the out-phase density excitations, once experimentally measured, can be efficiently used as a probe for the inlayer superfluidity. 

\begin{table}[h]
\caption{Dipole coupling $D=1.0$. The $d$-dependence of $\gamma_s$ and the relative spectral weight of the acoustic mode $S_{\text{A}}(q_n) =\int_{-\infty}^{\infty} S_{\text{A}}(q_n,\omega)\db \omega /S_{-}(q) $ at $q_n=2\pi n/L$ ($n=1,2$). The last four columns represent the fit parameters for the optical branch, Eq.~(\ref{MM1})-(\ref{MM2}). The superfluid fraction has a statistical error $\delta \gamma_s=0.007$. The second moment $\tilde\mu_2$ enters as a fit parameter in~(\ref{MM1}) and is limited by the upper bound, $\tilde\mu_2(q) < \hbar^2 q^2/m S_{-}(q)$, if $S_{\text{MM}}(q,\omega) \leq S_{-}(q,\omega), \, \forall \omega$. The static structure factor is rescaled as  $\tilde S_-(q)=10\cdot S_-(q)$.}
  \label{tab1}
 \begin{tabular}{c|c c c c c c c|}
 \hline
 \hline
  $d$ & $\gamma_s$ & $S_{A}(q_1)$ & $S_{A}(q_2)$ & $\omega_1 (\omega_2)$ & $\tilde\mu_{2}$ &$\Gamma$ & $\tilde S_{-}(q_1)$\\
 \hline
 \hline
 0.30 & 0.101  & 0.0684 & 0.106  & 20.51 (76.74) & 42.96 & 558.8 &  0.107(6) \\
 0.31 & 0.251 & 0.203 & -- &  17.01 (46.28) & 35.08 & 105.9 & 0.147(7) \\
 0.32 & 0.424 & 0.516 & 0.523 & 14.06 (36.54)  & 17.88 & 64.86 & 0.238(5) \\
 0.335 & 0.692& 0.862 & 0.731  & 20.36 (38.64) & 6.55 & 64.64 & 0.384(5) \\
 0.35 &  0.790 & 0.928& 0.860  & 21.95 (40.39) & 3.67 & 47.31 & 0.466(4) \\
 0.375 & 0.886 & 0.955 & 0.921  & 17.35 (39.60) & 1.90  & 40.00  & 0.541(3) \\
 \hline
 \hline
 \end{tabular}
 \end{table}

\subsection{Strong coupling $D=5.5$}

% ------------------------ D=55 m=1 -------------

{\em Dynamic structure factor.}
Now we discuss the case of strong dipole coupling. It can be realized  either by increase of the dipole moment, the particle mass or the inlayer density. Many features in the excitation spectrum observed for $D=1$ are also reproduced here.  A key difference is the onset of the inlayer crystallization below $d\sim 0.7$, see Fig.~\ref{fig:grsk-D55m1}. A sign of a triangular Wigner lattice can be observed in the density snapshots in Fig.~\ref{fig:D55m1-superf}, however, the temperature is not low enough and many structural defects are present. 

\begin{figure}
 \begin{center} 
\hspace{-0.0cm}\includegraphics[width=0.51\textwidth]{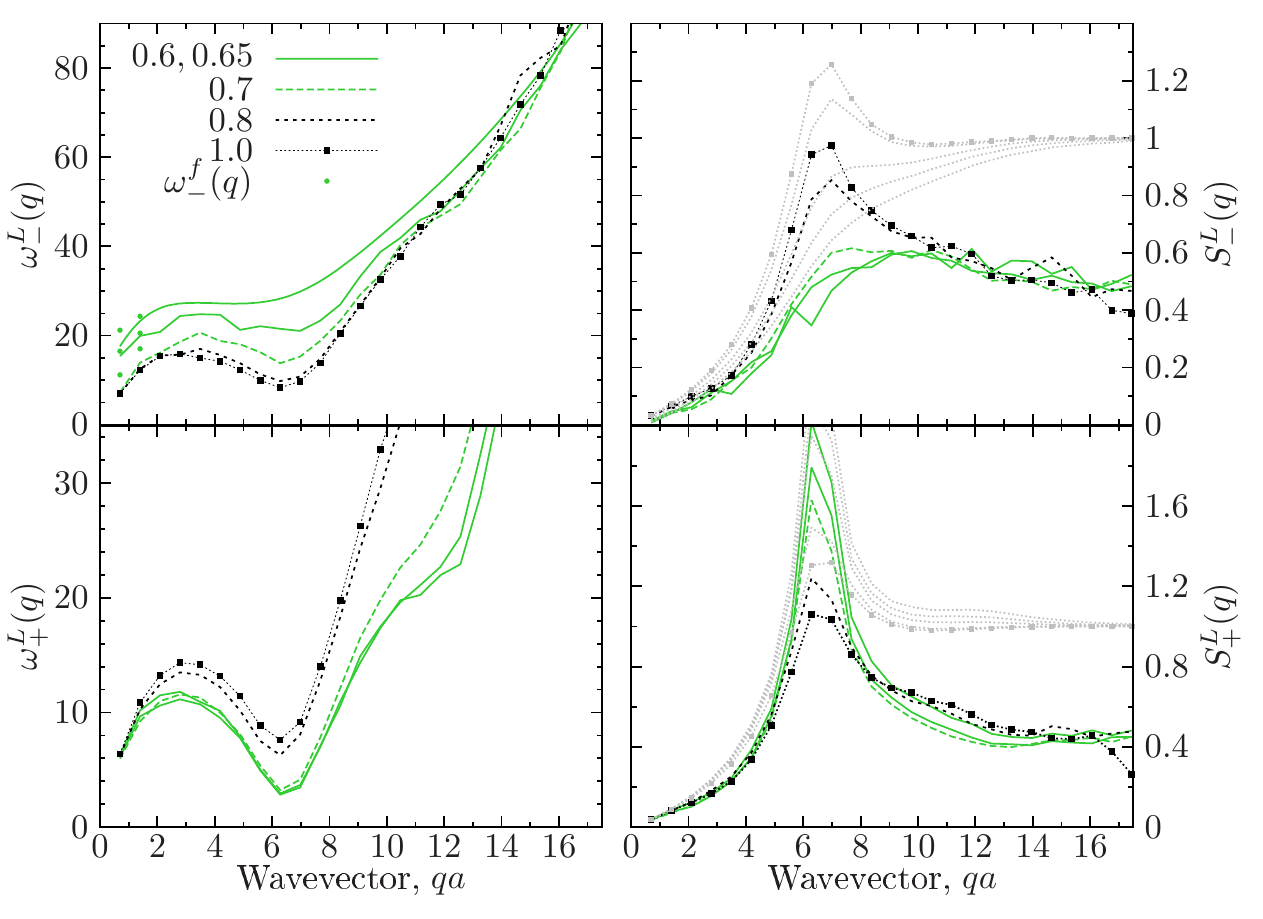}
%{../../../quasi2d/New/d55ed40/data/2d/m1new/d55/twomodepm}
 \end{center}
\vspace{-0.7cm} 
\caption{The low-frequency modes, $\omega^L_{\pm}(q)$, and their spectral weights, $S^L_{\pm}(q,\omega)$, for $D=5.5$. Both solutions demonstrate an acoustic behavior in the long wavelength limit for $d> 0.75$. An optical branch is present in the out-of-phase mode, $\omega^L_{-}(q)$, for $d \leq 0.75$. Its behavior for $qa < 2$ and the gap value is better described by the Feynman ansatz $\omega^f_-(q)$ (denoted by filled dots for $d=0.6,0.65,0.7$).  The legend indicates the layer spacing. Different colors are used to distinguish a superfluid ($\gamma_s \geq 0.9$ and $d\geq 0.8$) and a normal phase ($\gamma_s=0$ and $d\leq 0.65$).}
 \label{fig:two-m1D55}
 \end{figure}

\begin{figure}
 \begin{center} 
 \hspace{-0.0cm}\includegraphics[width=0.5\textwidth]{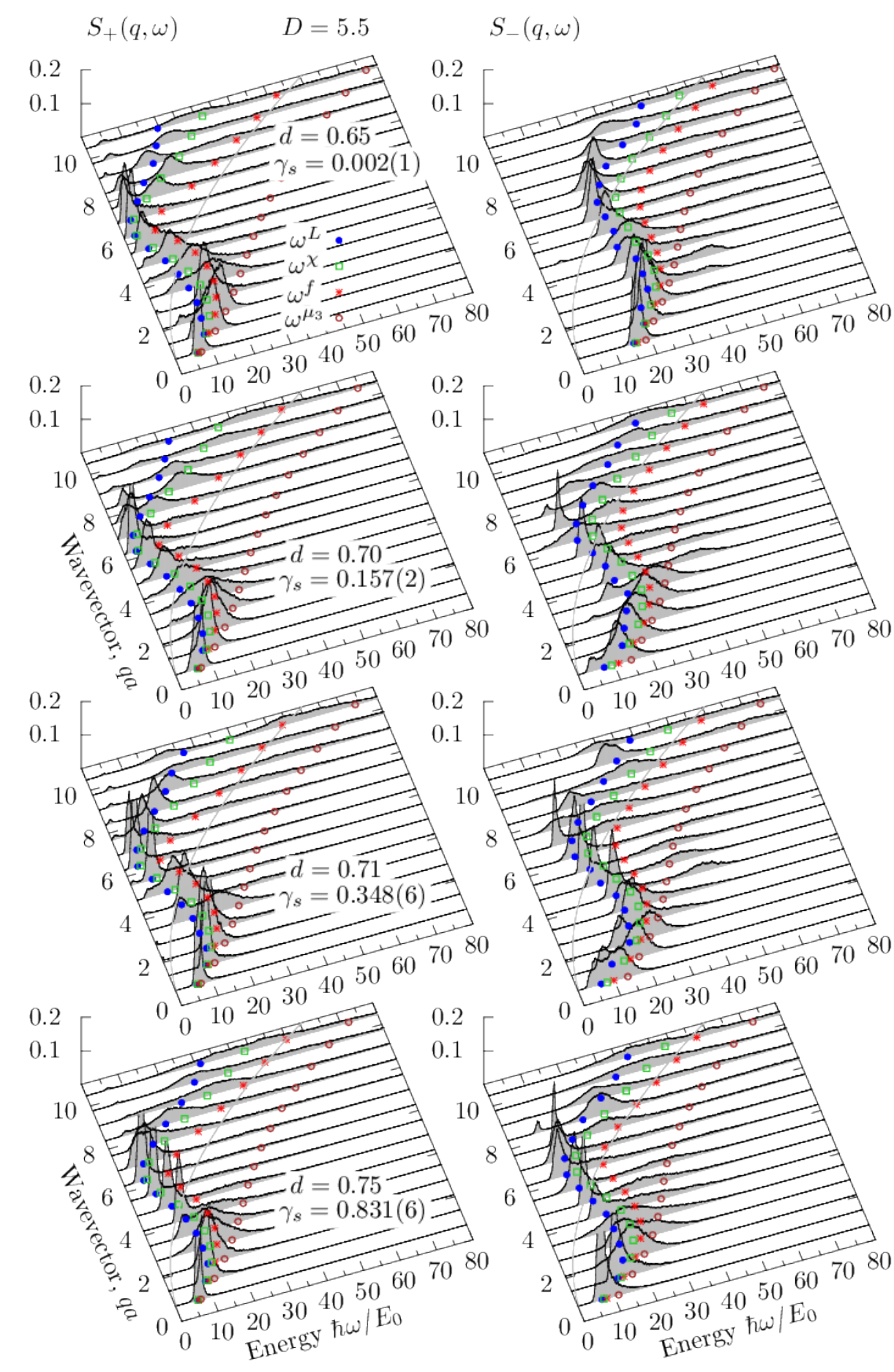}
 %{../../../quasi2d/New/d55ed40/data/2d/m1new/d55/stopmpaper/spec3dm1l065l075}     
 \end{center}
\vspace{-0.6cm} 
 \caption{The same as in Fig.~\ref{fig:m1d1a} for $D=5.5$. The legend indicates the layer spacing, $0.65 \leq d \leq 0.75$, and the superfluid fraction $\gamma_s$. The symbols show three upper bounds $\{\omega_\pm^\chi,\omega_\pm^f,\omega_\pm^{\mu_3}\}$ and the low-energy mode $\omega^L_{\pm}$ indicated by the blue symbols. The solid gray line denotes the free-particle dispersion $\epsilon_q=q^2/2m$.}
 \label{fig:m1d55a}
 \end{figure}

The  dispersion relations for the symmetric and antisymmetric density modes from the two-mode  ansatz are compared in Fig.~\ref{fig:two-m1D55}. For $d> 0.75$ both solutions predict the acoustic dispersion for $qa \leq 2$ and a roton feature around $qa \sim 2\pi$. When the interlayer spacing reduces below $d\approx 0.8$, the roton gap is also reduced and saturates for $d < 0.7$. This behavior is different from the case $D=1$, where a continuous evolution of the roton parameters with $d$ was observed, and the existence of the roton was attributed to formation of bound dimer states. In the present case, the roton is present also at large $d$ (e.g. $d=1$), hence, its origin is the intralayer correlations which are enhanced at low $d$. Indeed, the pair distribution function $g_{11}(r)$ in Fig.~\ref{fig:grsk-D55m1} shows formation of a quasi long-range order at $d\leq 0.7$, being a precursor for crystallization. The spatial ordering is also reflected in the increased spectral weight around the roton wavenumber, see $S_+^L(q)$ in Fig.~\ref{fig:two-m1D55}.

\begin{figure}
 \begin{center} 
 \hspace{-0.0cm}\includegraphics[width=0.5\textwidth]{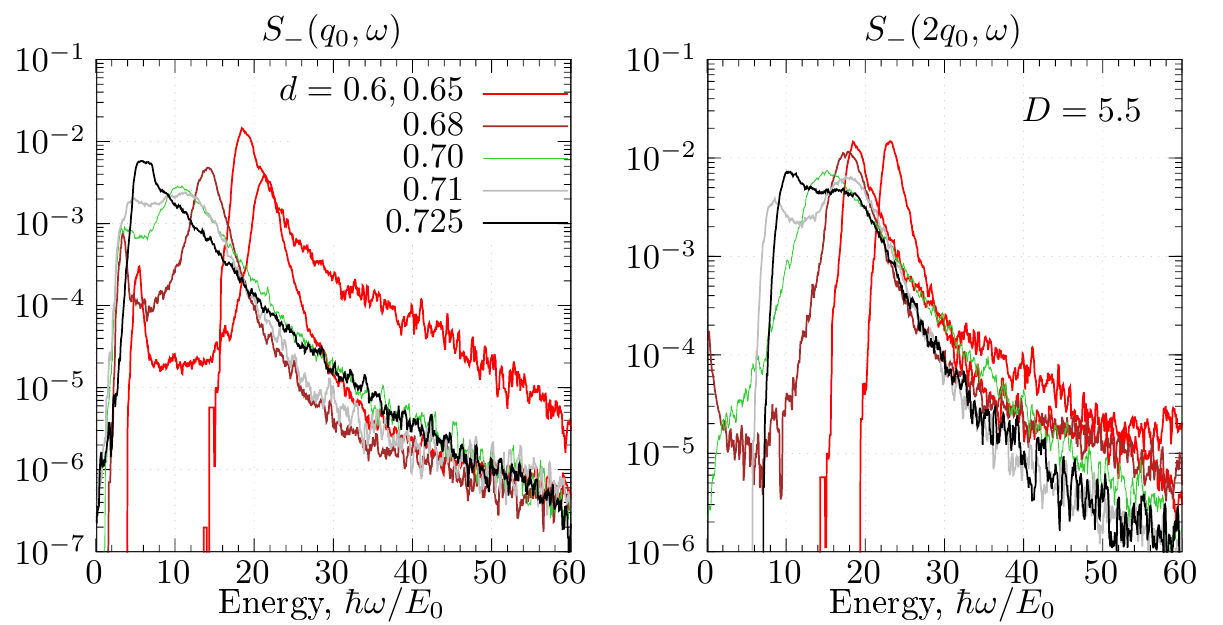}
 %{../../../quasi2d/New/d55ed40/data/2d/m1new/d55/stopmpaper/specSmk1}     
 \end{center}
\vspace{-0.6cm} 
 \caption{The $d$-dependence of $S_{-}(q,\omega)$ for the asymmetric mode spectra at two wavenumbers $q=q_0 n$: $n=1$ (left) and $n=2$ (right). The legend indicates the layer spacings: $0.60\leq d\leq 0.725$. Dipole coupling $D=5.5$.}
 \label{fig:Specmk1D55}
 \end{figure}

In Fig.~\ref{fig:m1d55a}, similar to Fig.~\ref{fig:m1d1a}, we perform comparison of the reconstructed dynamic structure factor $S_{\pm}(q,\omega)$ with the upper bounds~(\ref{w_ineq}). The best agreement again is given by the two-mode ansatz $\omega^L_{\pm}(q)$. At low wavelengths the symmetric mode remains acoustic independent on $d$ and the superfluid fraction. The spectrum of antisymmetric mode shows similar features as for $D=1$. In the superfluid phase with $\gamma_s > 0.8$ only one acoustic branch is present. In the partially superfluid phase, see $d=0.71,0.70$ in Fig.~\ref{fig:m1d55a}, both an acoustic and  an optical branch are observed simultaneously. When $\gamma_s$ is reduced, the spectral weight continuously transfers to the optical branch. In the normal phase with $\gamma_s=0$  only the optical branch remains, see $d=0.65$ in Fig.~\ref{fig:m1d55a}. Here, the same scenario applies as for $D=1$. With a lost of spatial coherence, the antisymmetric mode probes the intrinsic properties of dimer states. The reduction of $d$ increases the binding energy and the interlayer coupling. As a result the gap value is increased. See the long wavelength limit of $\omega^L_-(q)$  in Fig.~\ref{fig:two-m1D55} and, in more detail, the $d$-dependence of $S_{-}(q,\omega)$ in Fig.~\ref{fig:Specmk1D55}

\begin{figure}
 \begin{center} 
 \hspace{-0.0cm}\includegraphics[width=0.5\textwidth]{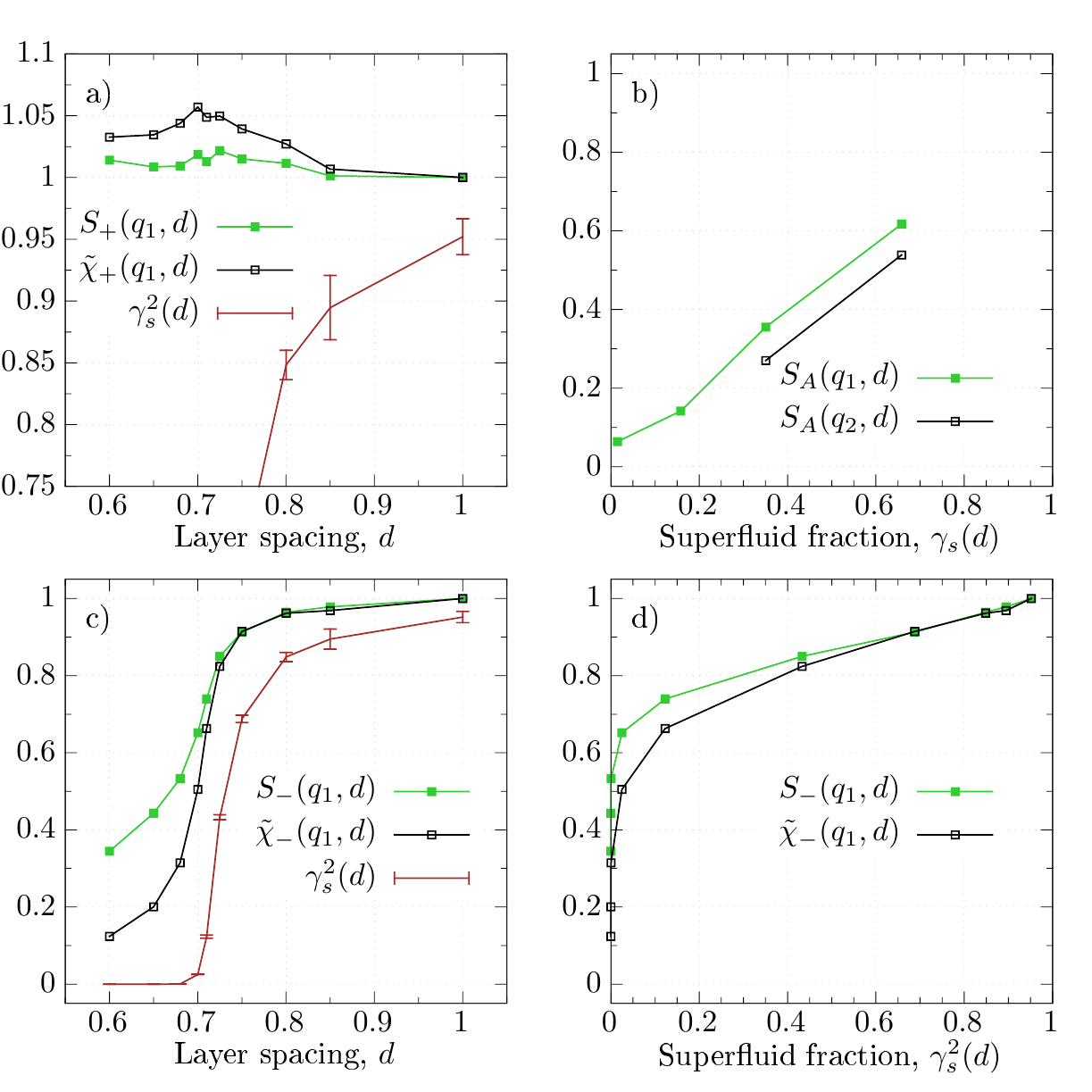}
 %{../../../quasi2d/New/d55ed40/data/2d/m1new/d55/stopmpaper/rhosweight}     
 \end{center}
\vspace{-0.6cm} 
  \caption{a),c) The $d$-dependence of the renormalized static structure factor, $S_{\pm}(q_1,d)=S_{\pm}(q_1,d)/S_{\pm}(q_1,d=0.7)$, the static response function $\tilde\chi_{\pm}(q_1,d)=\Re \chi_{\pm}(q_1,d)/ \Re \chi_{\pm}(q_1,d=0.7)$ for the in-phase and out-of-phase modes, and the square of the superfluid fraction $\gamma^2_s(d)$. b) The integrated  spectral weight of $S_{\text{A}}(q_1,d)$ vs. $\gamma_s$. d) Same as in panel (c) plotted vs. $\gamma^2_s(d)$ to demonstrate a linear dependence. The wavenumber $q_1$ corresponds to the smallest wavenumber in the simulation $2\pi/L$ ($L=9$). The normalization values: $S_{\pm}(q_1,d=1.0)=0.03775(0.03334)$ and $\frac{1}{2}\Re \chi_{\pm}(q_1,d=1.0)=0.5874\cdot 10^{-2} (0.4629 \cdot 10^{-2})$.}
 \label{fig:rhosweightD55}
 \end{figure}

Next, we repeat our analyses to find a relation between the superfluid response and the spectral weight of the acoustic mode $S_A$. The high-frequency optical mode is fitted with the DHO~(\ref{dho-an}) and the MM-ansatz~(\ref{MM2}) following the same procedure as for $D=1$. The range of layer spacing and the dynamic structure factor used for these analyses is illustrated by Fig.~\ref{fig:Specmk1D55}. The results are presented in Tab.\ref{tab2} and Fig.~\ref{fig:rhosweightD55}b. We confirm a linear dependence between $\gamma_s(d)$ and $S_A(q_1,d)$ [and also $S_A(q_2,d)$ used as an independent test].

For a non-vanishing superfluid response ($\gamma_s>0.1$), a linear dependence, but now on $\gamma_s^2(d)$, is observed for the static structure factor $S_-(q_1)$ and the density response function $\chi_-(q_1)$. Both are taken at the smallest wavenumber $q_1$, when they are mutually related via the particle number fluctuations~(\ref{fn12}). In the normal phase at $d< 0.68$, the interlayer particle number fluctuations are significantly reduced due to the pairwise coupling in the dimer states. Hence, in a partially superfluid phase the density fluctuations are mainly due to the superfluid component. This explains the observed dependencies in Fig.~\ref{fig:rhosweightD55}c,d. 

Similar results for the symmetric mode are presented in Fig.~\ref{fig:rhosweightD55}a. In contrast, they capture only the collective properties of the dimer states  and the inlayer density fluctuations, which are not much sensitive to either the dimer states are weakly or strongly bound. Some non-monotonic behavior observed in the range $0.68 \leq d \leq 0.85$ is related to the phase transition from a superfluid to a normal phase, and the onset of formation of a Wigner-type lattice with defects.

 \begin{table}
\caption{Same as in Tab.~\ref{tab1} for the dipole coupling $D=5.5$. The static structure factor is rescaled as  $\tilde S_-(q)=10\cdot S_-(q)$.}
  \label{tab2}
 \begin{tabular}{c|c c c c c c c|}
 \hline
 \hline
  $d$ & $\gamma_s$ & $S_{A}(q_1)$ & $S_{A}(q_2)$ & $\omega_1 (\omega_2)$ & $\mu_{2}$ &$\Gamma$ & $\tilde S_{-}(q_1)$\\
 \hline
 \hline
 0.68 & 0.0165  & 0.064 & --  & 13.66 (16.71) & 29.53 & 33.21 &  0.178(2) \\
 0.70 & 0.157 & 0.142 & 0.135 &  10.07 (16.08) & 22.30 & 25.99 & 0.217(1) \\
 0.71 & 0.348 & 0.355 & 0.269 & 9.81 (15.87)  & 19.35 & 21.91 & 0.247(2) \\
 0.725 & 0.652& 0.618 & 0.539  & 8.5473 (14.28) & 10.89 & 20.80 & 0.283(1) \\
 \hline
 \hline
 \end{tabular}
 \end{table}

\begin{figure}
 \begin{center} 
 \hspace{-0.0cm}\includegraphics[width=0.5\textwidth]{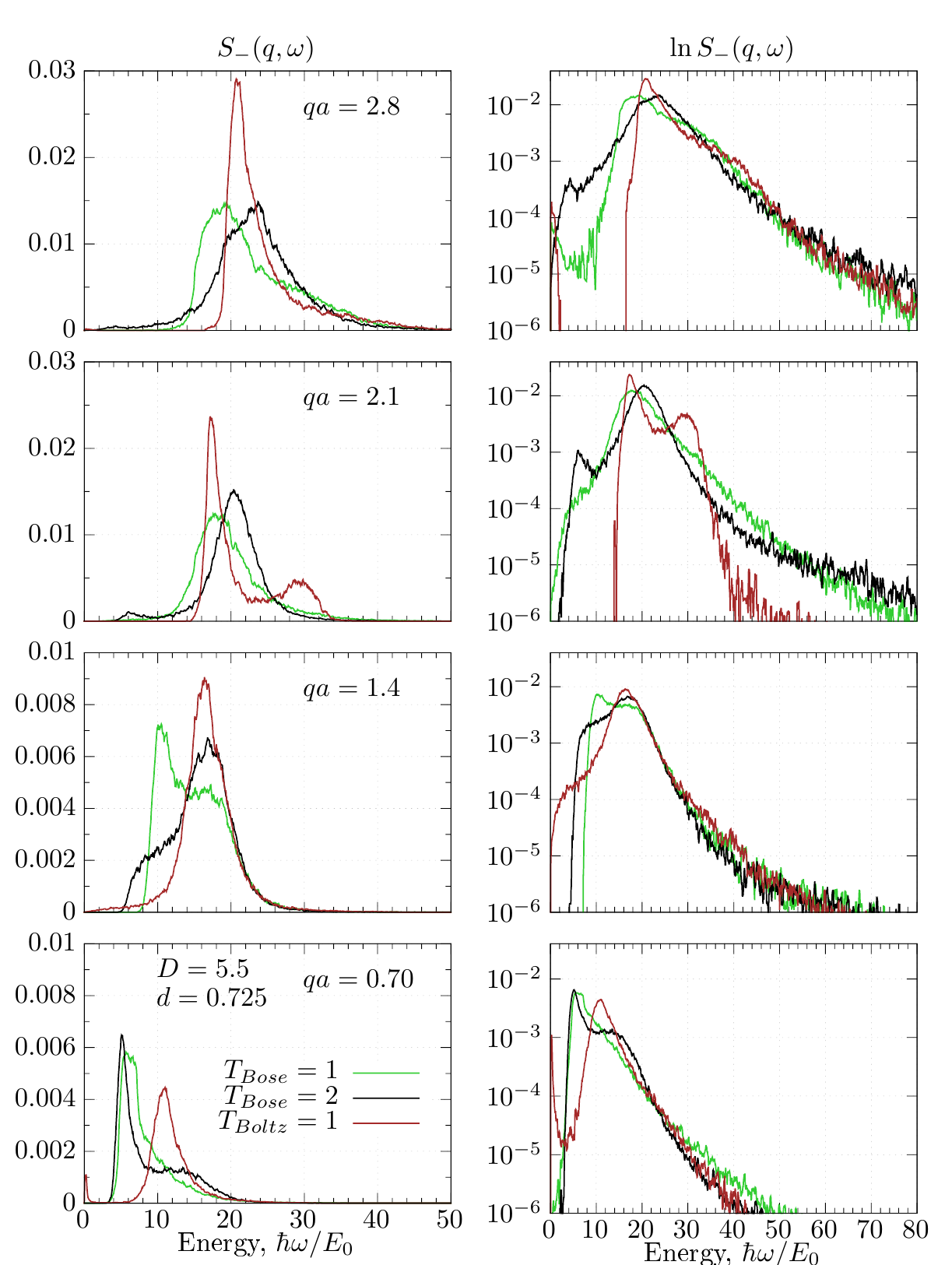}
 %{../../../quasi2d/New/d55ed40/data/2d/m1new/d55/stopmpaper/specd55m1l0725boseboltz}     
 \end{center}
\vspace{-0.6cm} 
 \caption{Comparison of the antisymmetric mode spectra for a bosonic bilayer at $T_{\text{Bose}}=1$ ($\rhos=0.625(5)$), $T_{\text{Bose}}=2$ ($\rhos=0$)  and a bilayer with Boltzmann statistics at $T_{\text{Boltz}}=1$ ($\rhos=0$). Layer spacing $d=0.725$ and $D=5.5$. The right panels show the results on the log-scale. The legend indicates the wavenumbers $qa$. In the superfluid system ($T_{\text{Bose}}=1$) the spectra is dominated by the acoustic branch.  For the Boltzmann case ($\rhos=0$) the spectral weight is carried by the optical branch. For $T_{\text{Bose}}=2$ a new dispersionless branch is formed  around $\hbar\omega /E_0\sim 6$, most probable, due to intrinsic excitations of the dimer states.}
 \label{fig:specl0725}
 \end{figure}

\begin{figure}
 \begin{center} 
 \hspace{-0.35cm}\includegraphics[width=0.5\textwidth]{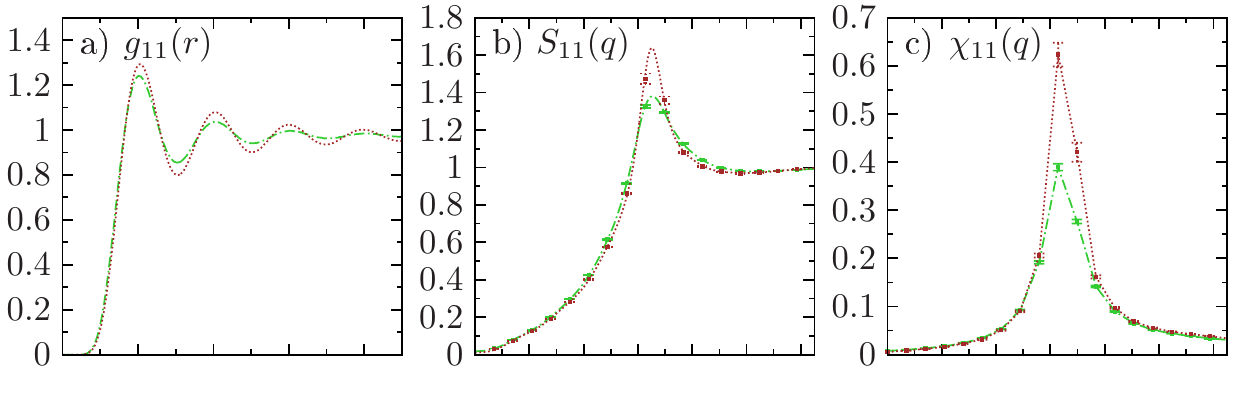}\\
 %{../../../quasi2d/New/d55ed40/data/2d/m1new/d55/gr11sk11chi11noxboltz}\\
 \vspace{-0.25cm}
 \hspace{-0.35cm}\includegraphics[width=0.5\textwidth]{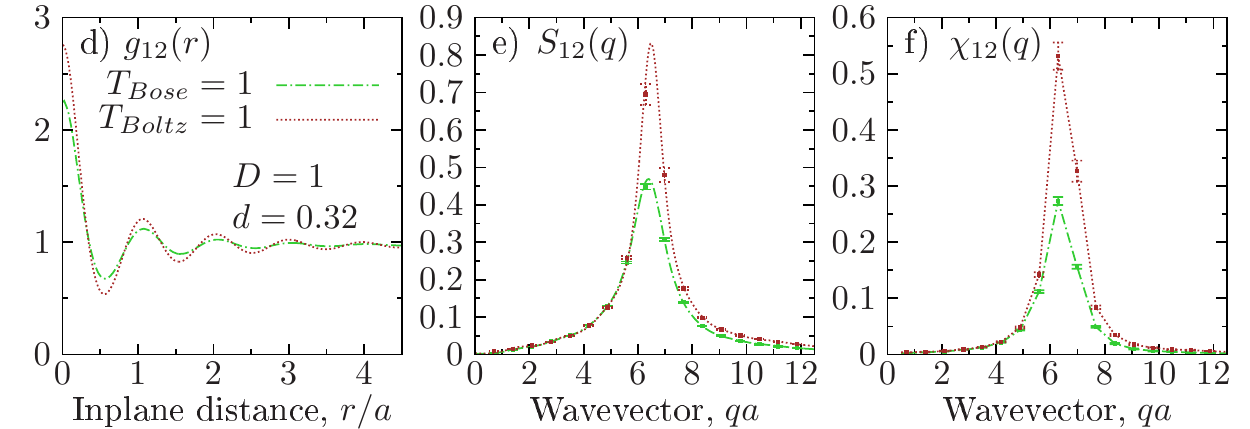}\\
 %{../../../quasi2d/New/d55ed40/data/2d/m1new/d55/gr12sk12chi12boltz}\\
 \vspace{-0.05cm}
 \hspace{-0.35cm}\includegraphics[width=0.5\textwidth]{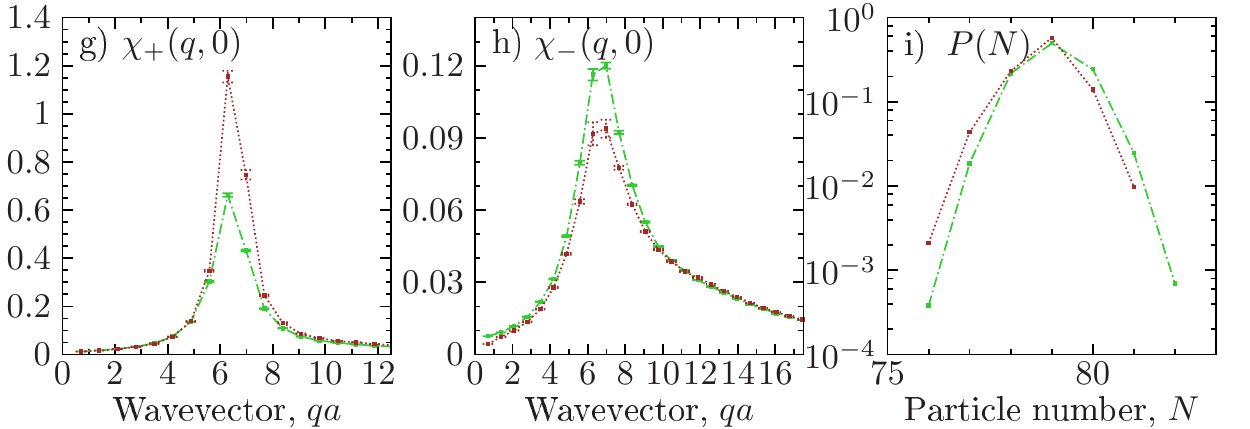}
 %{../../../quasi2d/New/d55ed40/data/2d/m1new/d55/chi0diagboltz}
 \end{center}
\vspace{-0.60cm} 
 \caption{Comparison of the static properties for $d=0.725$ and $D=5.5$. Compared are the bilayers at $T=1$ with Bose and Boltzmann statistics.}
 \label{fig:specl0725stat}
 \end{figure}

Similar to the case $D=1$, we repeat the test with the distinguishable ``boltzmannons'' to prove that the origin of the acoustic branch is a superfluid component. Both spectra are taken at the same temperature and layer spacing ($d=0.725$), and are compared in Fig.~\ref{fig:specl0725}. The Boltzmann case shows an optical branch with a finite gap in the long wavelength limit. Difference in the quantum statistics is also reflected in the static properties, see Fig.~\ref{fig:specl0725stat}. The enhanced amplitude of the peaks in $g_{11(12)}(r)$, $S_{11(12)}(q)$, $\chi_{11(12)}(q)$ and $\chi_{+}(q)$ testifies that in the Boltzmann case particles positions are more correlated. This can be interpreted as an earlier onset of crystallization, which starts at a larger layer spacing compared to the bosonic case. 
 
\begin{figure}
 \begin{center} 
  \hspace{-0.70cm}\includegraphics[width=0.5\textwidth]{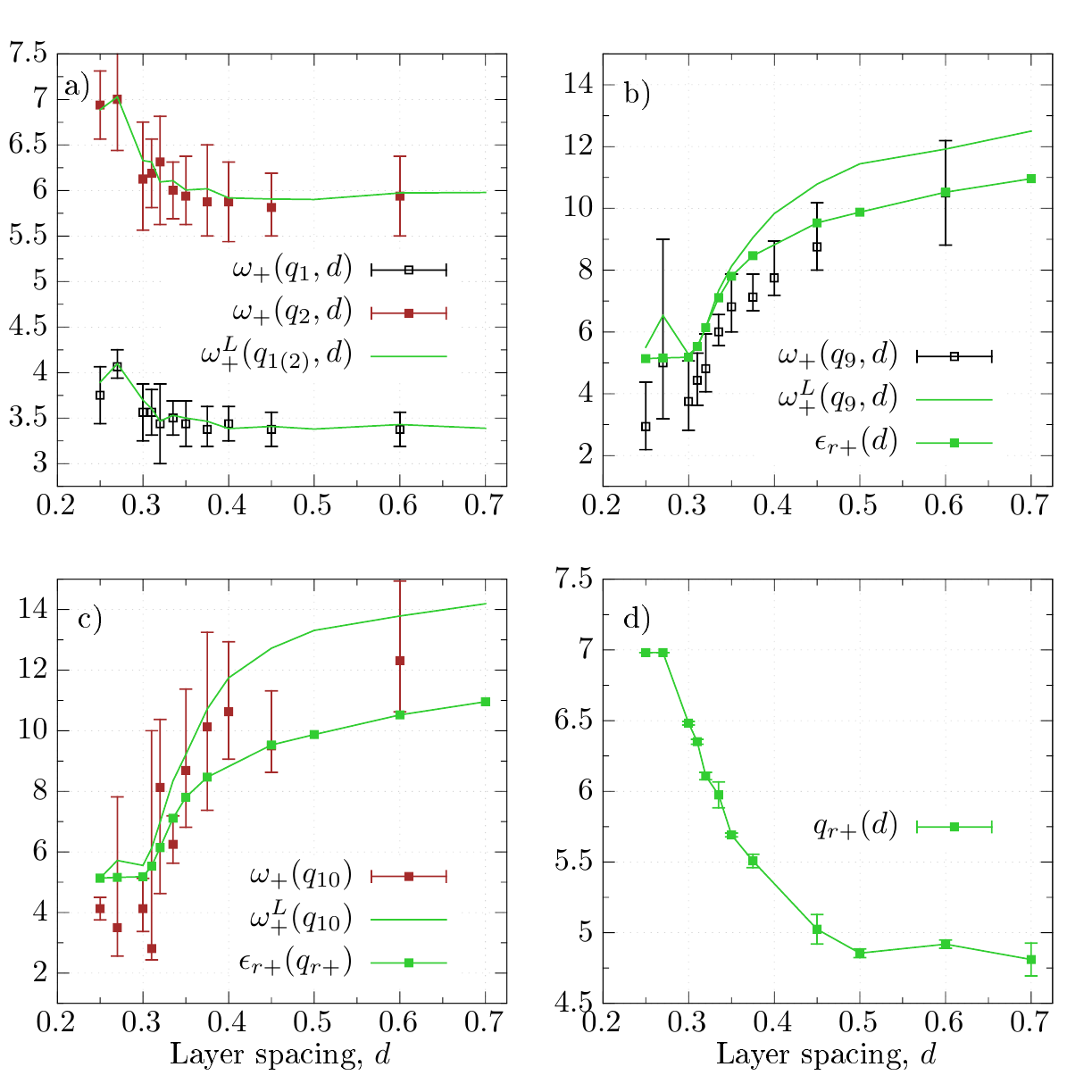}
  %{../../../quasi2d/New/d55ed40/data/2d/m1new/d1/stopmpaper/wpq1q2}    
 \end{center}
 \vspace{-0.6cm} 
\caption{The $d$-dependence of the phonon and roton parameters of the symmetric mode $S_{+}(q,\omega)$ for  $D=1.0$. a) The position of the maximum in $S_{+}$ with its halfwidth [shown as errorbars]. Solid line is the prediction from the two-mode ansatz $\omega^L_{+}$. b),c) The energy resonances and their halfwidth [symbols with errorbars] in the roton region, $q_n=2 \pi n/L$ ($n=9,10$). The solid line with bold symbols is a fit to $\omega^L_{+}(q)$ by the roton-ansatz  $\epsilon^r_{+}$~(\ref{rotonfit}). d) The $d$-dependence of the roton wavenumber $q^r_{+}$ from~(\ref{rotonfit}).}
 \label{fig:roton-D1}
 \end{figure}

\begin{figure}
 \begin{center} 
  \hspace{-0.70cm}\includegraphics[width=0.5\textwidth]{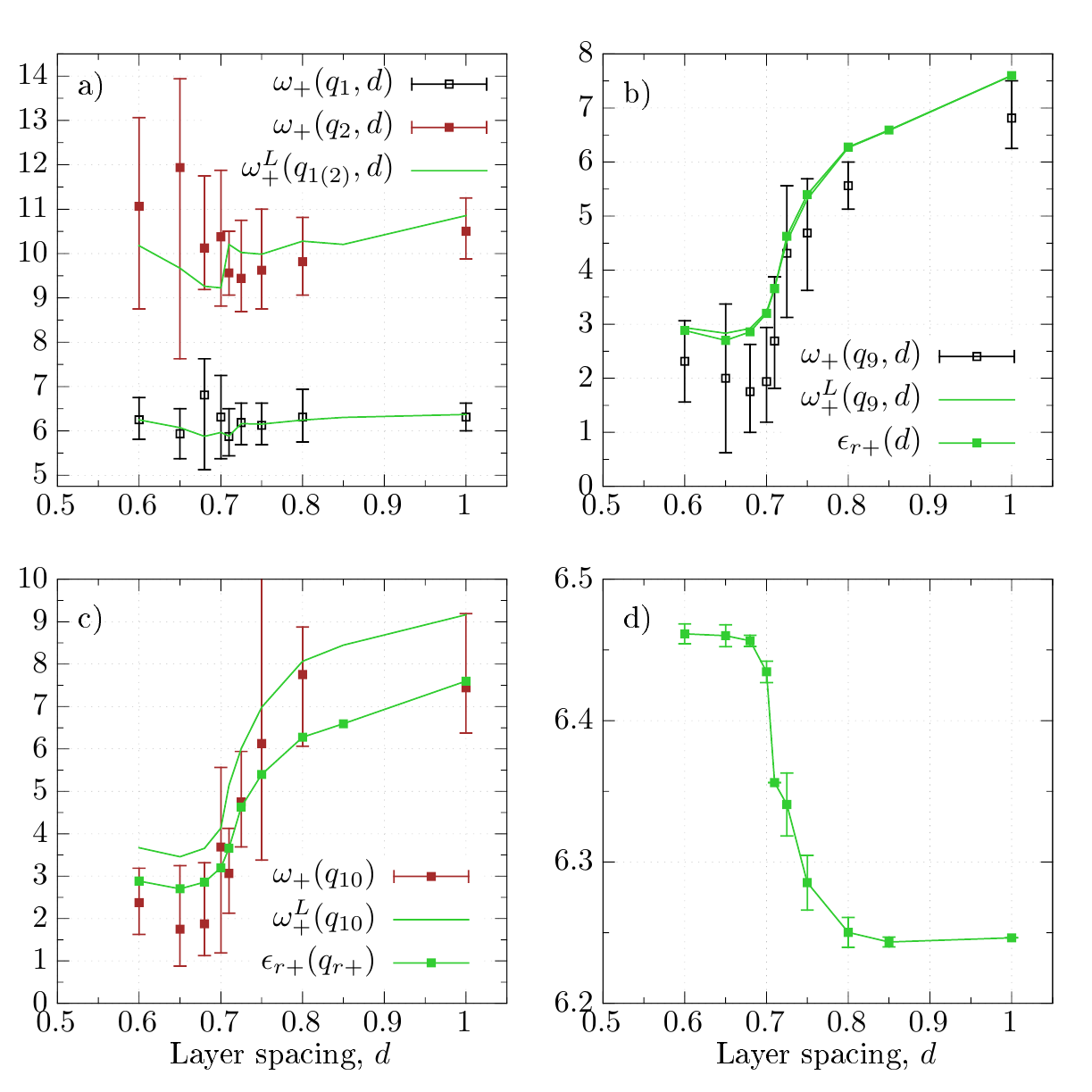}
  %{../../../quasi2d/New/d55ed40/data/2d/m1new/d55/stopmpaper/wpq1q2}    
 \end{center}
 \vspace{-0.6cm} 
\caption{Similar analyses as in Fig.~\ref{fig:roton-D1} for $D=5.5$.}
 \label{fig:roton-D55}
 \end{figure}

{\em Spectrum in the phonon and roton regions.} Now we analyze in more detail the $d$-dependence of the phonon and roton resonances. Here we combine the discussion of moderate ($D=1$) and strong ($D=5.5$) coupling, as they demonstrate similar trends. To extract  resonance positions we use the dispersion relation $\omega^L_{+}(q)$ derived from the sum-rules, and the full dynamic structure factor $S_{+}(q,\omega)$. The phonon and roton spectrum  is analyzed at the wavenumbers $q_n=2\pi n/L$ ($n=1,2$) and $q_n=2\pi n/L$ ($n=9,10$), correspondingly.  The results are presented in Fig.~\ref{fig:roton-D1} and Fig.~\ref{fig:roton-D55}.

In Fig.~\ref{fig:roton-D1}a positions of the phonon resonances and their halfwidth in $S_{+}(q,\omega)$ are indicated by the symbols with errorbars. There is a nice agreement with the $d$-dependence of the two-mode ansatz, $\omega^L_{+}(q,d)$, shown by a solid line. The increase of the acoustic sound  speed $c_+(d)$ for $d\leq 0.4$ is directly correlated with the reduction of the superfluid response $\gamma_s(d)$ and formation of bound dimer states. After a transition into a normal phase at $d\leq 0.28$, the value of the sound speed saturates. 

At strong coupling, $D=5.5$, the dimerization and the superfluid-normal phase transition does not show some pronounced $d$-dependence, see Fig.~\ref{fig:roton-D55}a. The sound speed of the symmetric mode shows a non-monotonic behavior with a local minimum around $d \sim 0.7$. The absolute value of $c_+$ changes within 15$\%$. Below $d \sim 0.7$ the system goes into the normal phase  and $\gamma_s$ reduces to zero. Note that $c_+(d)$ should reproduce the $d$-dependence of the static response function and the static structure factor shown in Fig.~\ref{fig:rhosweightD55}. This follows from the compressibility sum-rule
\begin{align}
 \lim\limits_{q \rightarrow 0} \frac{S_{\pm}(q)}{k_BT}=\lim\limits_{q \rightarrow 0} \frac{\abs{\Re \chi_{\pm}(q,0)}}{2 \rho}=\kappa_{\pm} \rho=\frac{1}{m c_{\pm}^2}.
 \label{cs}
\end{align}
The last equality is written in the assumption that as $q \rightarrow 0$ two upper bounds converge to the acoustic dispersion, i.e. $\omega^{\chi}_{\pm}(q)\approx  \omega^{f}_{\pm}(q)\approx c_{\pm} q$. Note, that once the spectrum contains an additional mode (as the antisymmetric mode with an optical branch), the estimate of the sound speed from Eq.~(\ref{cs}) will be wrong. For the symmetric mode the dispersion contains a single acoustic branch and Eq.~(\ref{cs}) remains valid. As a result, a local maximum in $S_+(q)$ and $\chi_+(q)$ around $d\sim 7$, as observed in Fig.~\ref{fig:rhosweightD55}, translates into a local minimum in $c_+(d)$. For $D=1$, by decreasing $d$ both $S_+(q)$ and $\chi_+(q)$ show a monotonic decrease, see Fig.~\ref{fig:rhosweight}. In this case, $c_+(d)$ monotonically increases being in agreement with~(\ref{cs}).

The acoustic sound speed for the antisymmetric mode can be determined from~(\ref{cs}) when the dispersion relation is linear and no optical branch is present. This holds in the superfluid phase with $\gamma_s \gtrsim 0.8$. For smaller $\gamma_s$ the spectrum splits into one acoustic and one optical branch, and Eq.~(\ref{cs}) is invalid.

Next, in Figs.~\ref{fig:roton-D1},~\ref{fig:roton-D55}b-d we show the $d$-dependence of the roton parameters: the roton frequency (value of the roton gap) and the roton wavenumber $q^r$. We compare the resonances $\omega_+(q)$ in $S_{+}(q,\omega)$ and their halfwidth (indicated by the errorbars) with the two-mode ansatz $\omega^L_+(q)$ in the roton region at the wavenumbers $q_n=2\pi n/L$ ($n=9,10$). In addition, we plot the curve $\epsilon^r(q^r)$ (solid lines with bold symbols) -- a fit to $\omega^L_+(q)$ around $q^r$
\begin{align}
\omega_+^L(q)=\epsilon^r_{+}(q^r_{+})+\frac{(q-q^r_{+})^2}{2 m_{+}^r}.
\label{rotonfit}
\end{align}
The numerical results for the roton frequency and the fit to the two-mode ansatz~(\ref{rotonfit}) are in a reasonable agreement.
We observe a systematic trend: the resonances in the reconstructed spectra are shifted to lower frequencies and predict a more deep roton minimum. Typically by $10\%$ in the superfluid phase, and by up to $50\%$ in the normal phase, when  strongly bound dimer states are formed: see $d\leq 0.3$ in Fig.~\ref{fig:roton-D1}b,c and $d\leq 0.7$ in Fig.~\ref{fig:roton-D55}b,c. A discrepancy between the resonances in $S_{+}(q,\omega)$ and $\omega^L_+(q)$ starts to increase quite rapidly beyond the roton region, see the left columns in Fig.~\ref{fig:m1d1a} and Fig.~\ref{fig:m1d55a}.

For both coupling strength we observe a systematic reduction of the roton gap at low $d$, which follows the reduction of the superfluid density. The roton energy saturates after   transition in the normal phase. For $D=1$ this happens below $d \sim 0.3$ and for $D=5.5$ below $d \sim 0.7$.

The roton wavenumber shows just an opposite trend, see Fig.~\ref{fig:roton-D1}d and~\ref{fig:roton-D55}d. It decreases during the transition from a normal to a superfluid phase.

\begin{figure}
 \begin{center}
 \hspace{-0.35cm}\includegraphics[width=0.51\textwidth]{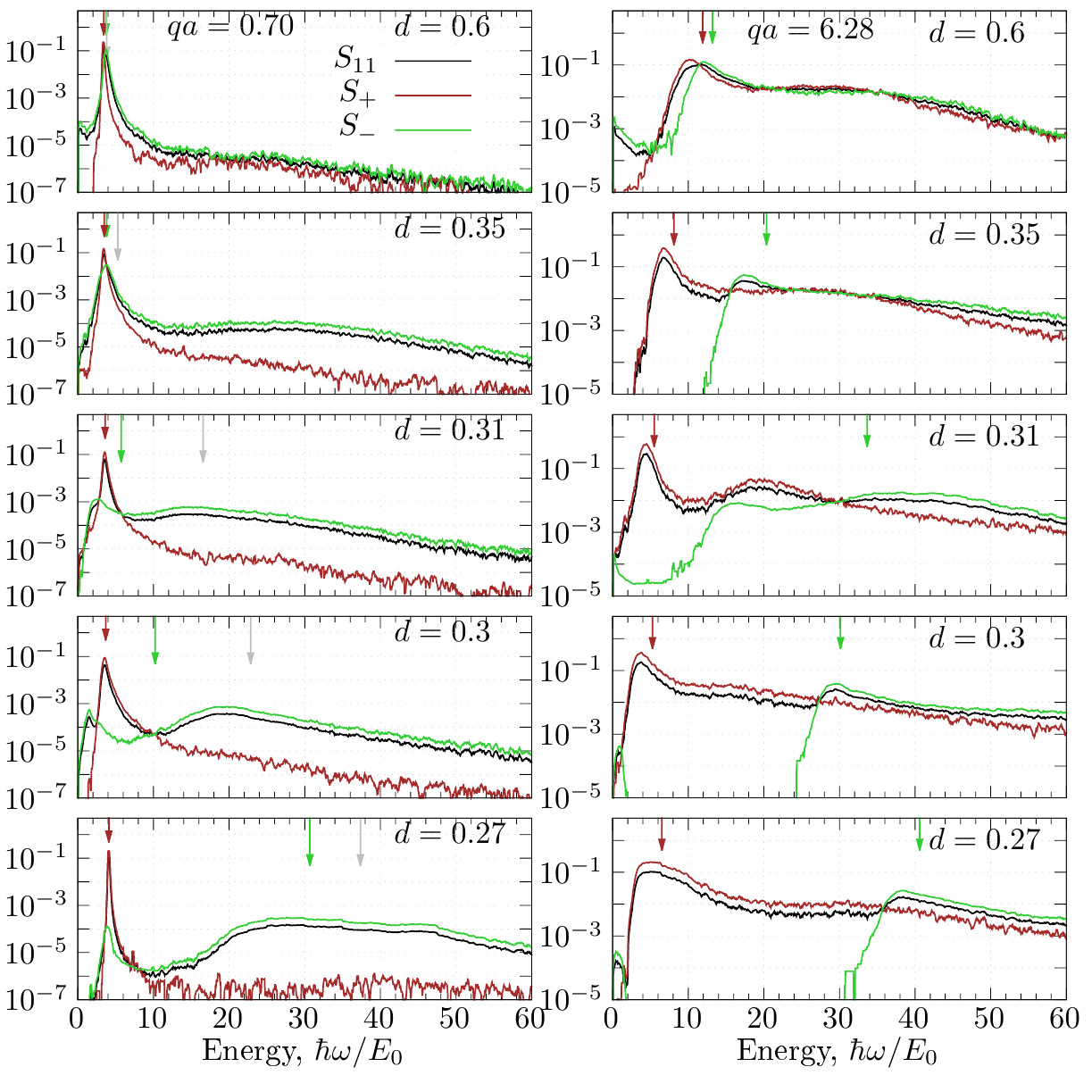}
%{../../../quasi2d/New/d55ed40/data/2d/m1new/d1/stopm/m1d1k1alld}
 \end{center}
\vspace{-0.8cm} 
\caption{Dynamic structure factors $S_{\pm}(q,\omega)$ and $S_{11}(q,\omega)=\frac{1}{2}[S_{+}(q,\omega)+S_{-}(q,\omega)]$ for $D=1$. Each spectral density is identified by a color as specified in the legend. Left and right column shows the result for the phonon ($qa=0.70$) and roton ($qa=6.28$) wavenumbers. Vertical arrows indicate position of the peaks predicted by the sum-rules analyses: $\omega^L_{+}(q)$ (brown), $\omega^L_{-}(q)$ (green) and $\omega^f_{-}(q)$ (gray, shown only for $qa=0.70$).}
 \label{fig:alld-m1D1}
 \end{figure}

\begin{figure}
 \begin{center} 
\hspace{-0.0cm}\includegraphics[width=0.51\textwidth]{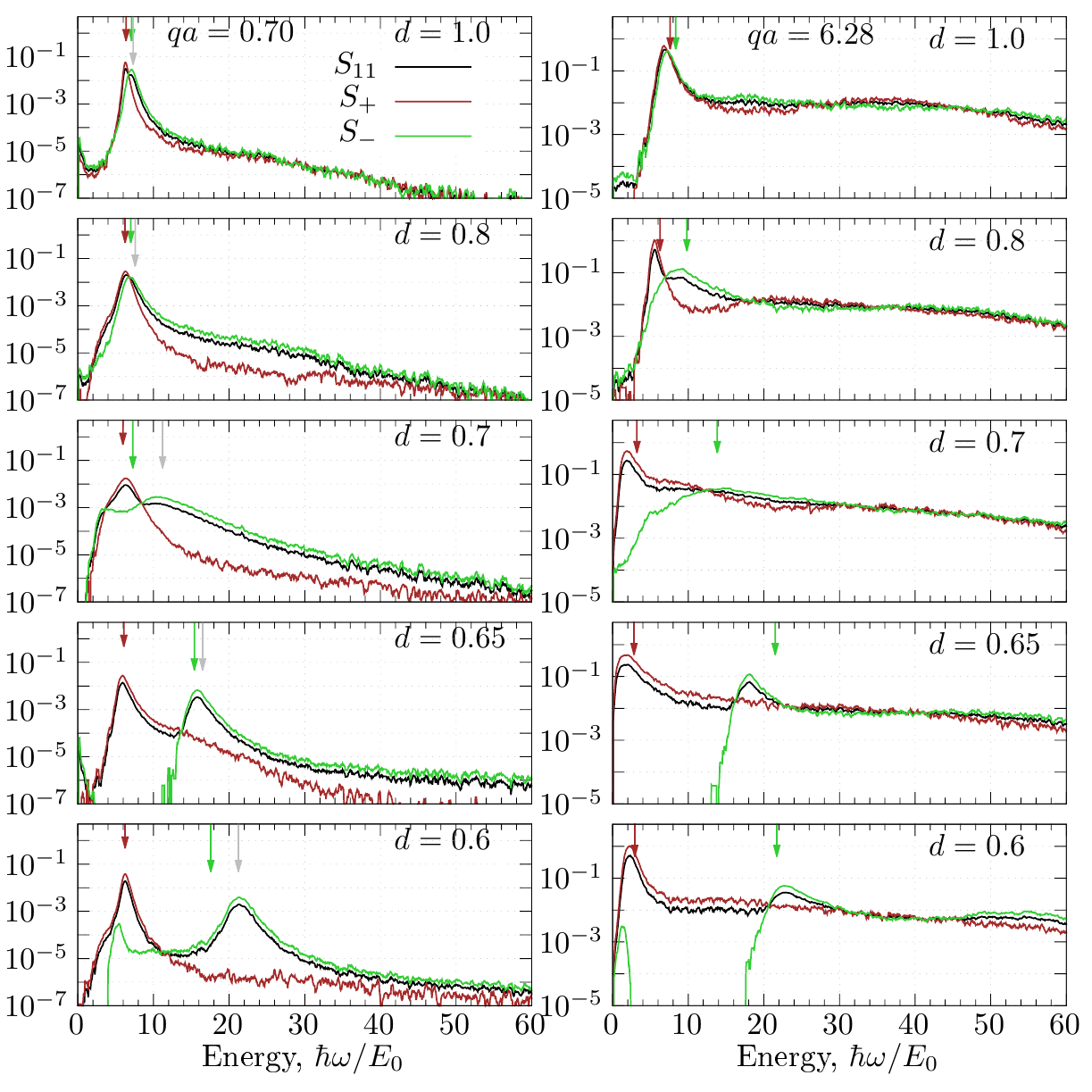}
%{../../../quasi2d/New/d55ed40/data/2d/m1new/d55/stopm/m1d55k1alld}
 \end{center}
\vspace{-0.8cm} 
\caption{Same as in Fig.~\ref{fig:alld-m1D1} for $D=5.5$.}
 \label{fig:alld-m1D55}
 \end{figure}

An excitation of the symmetric and antisymmetric mode requires a special experimental setup which induce the in-phase and the out-phase density fluctuations in both layers. More common is a direct probe of a single layer corresponding to the inlayer partial structure factor, $S_{11}(q,\omega)=\frac{1}{2}[S_{+}(q,\omega)+S_{-}(q,\omega)]$. The comparison of three spectral densities is presented in Fig.~\ref{fig:alld-m1D1} and~\ref{fig:alld-m1D55}. Several layer spacings are considered and two wavenumbers -- for acoustic phonons and a roton. The vertical arrows indicate the resonances  predicted by the two-mode ansatz, $\omega^L_{\pm}(q)$. For the antisymmetric mode at $qa=0.70$ we also show the Feynman ansatz, $\omega^f_-(q)$. It provides a better agreement with the resonances of the optical branch in $S_-(q,\omega)$ in the partially superfluid phase, see $d=\{0.3,0.31\}$ for $D=1$ and $d=\{0.6,0.65,0.70\}$ for $D=5.5$.

The presented comparison clearly shows how the resonances in $S_{11}(q,\omega)$ can be explained in terms of the symmetric and antisymmetric density excitations. For $d\leq 0.31$ ($D=1$) and $d\leq 0.7$ ($D=5.5$) the contribution of the symmetric and antisymmetric mode is well distinguished.
In particular, the observed high-frequency tail in $S_{11}(q,\omega)$ (in the phonon region) is due to the out-of-phase density excitations. In contrast, the sharply peaked acoustic resonances originate from the in-phase excitations and are present for all $d$. 
 
At large layer spacing, i.e. $d=0.6$ ($D=1$) and $d=1$ ($D=5.5$), a clear distinction of both modes in $S_{11}(q,\omega)$ is problematic. The corresponding spectral densities significantly overlap and the resonances are in a close vicinity. In this regime the interlayer dynamic structure factor vanishes, i.e. $S_{12}(q,\omega)\ll  S_{11}(q,\omega)$, and both spectra are similar, $S_{+}(q,\omega)\approx S_{-}(q,\omega)$. 

The layer spacing $d=0.35$ ($D=1$) and $d=0.8$ ($D=5.5$) represents an intermediate case. The acoustic resonances in $S_+$ and $S_-$ nearly coincide (see the left panel with $qa=0.70$), but both spectra are well distinguished in the roton region (see the right panel, $qa=6.28$). A similarity in the acoustic range is related, as was discussed above, with the hybridization of the collective density excitations with the single-particle spectra in a superfluid phase.

\subsection{Weak coupling $D=0.1$}

A weakly coupled regime is currently accessible with the experimental setups for ultra-cold dipolar systems.\cite{atomic1,atomic2,pfau}

First, we note an order of magnitude larger compressibility compared to the strong and moderate coupling, see Fig.~\ref{fig:D01m10-sound}. The inlayer density increases fast below $d\approx 0.2$ due to the dimerization identified by formation of a peak in $g_{12}(r)$, see Fig.~\ref{fig:D01m1-rho12}, and a slight reduction of the superfluid fraction, see Fig.~\ref{fig:D01m1-superf}.

% ------------------ m=1 D=0.1
 \begin{figure}
 \begin{center} 
\hspace{-0.0cm}\includegraphics[width=0.51\textwidth]{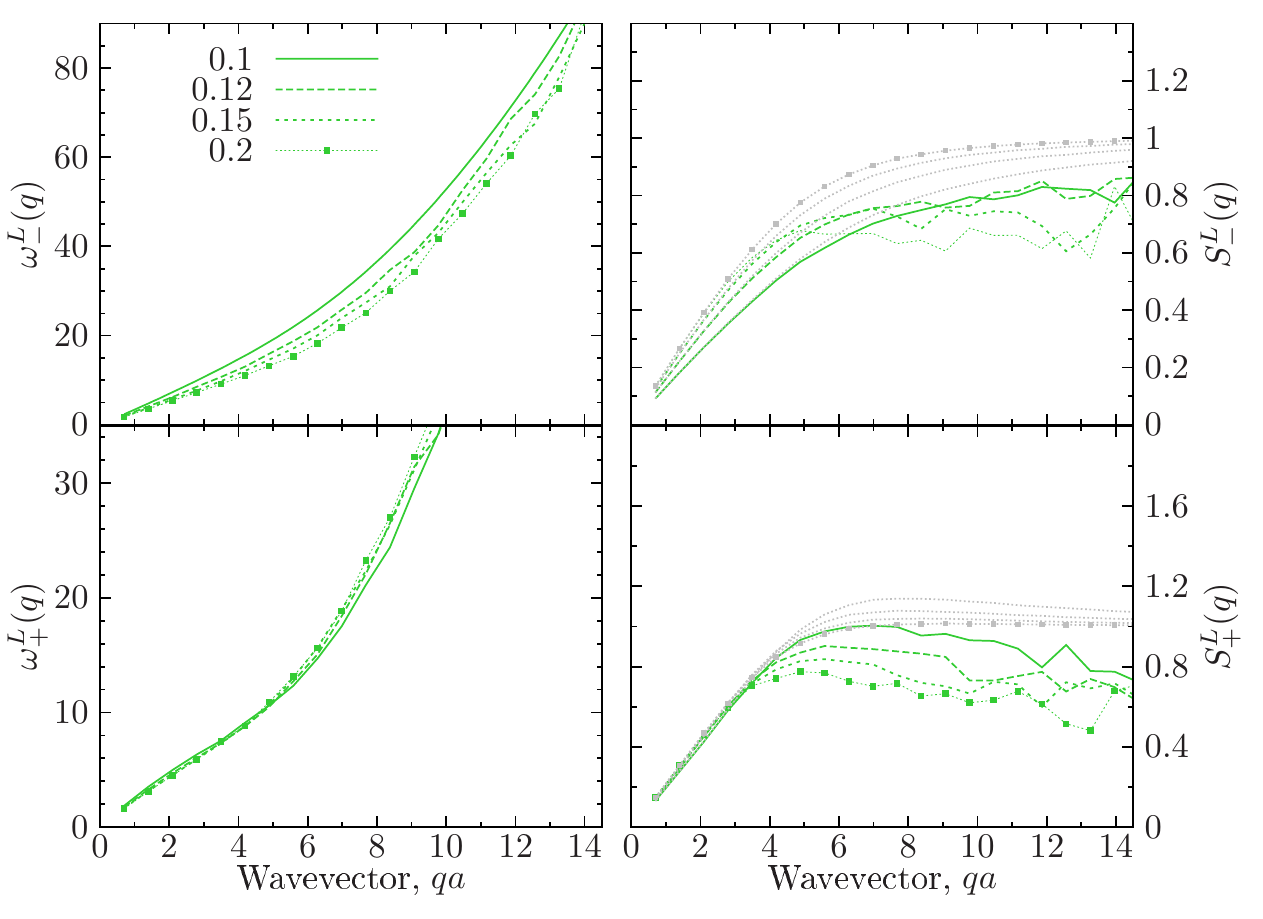}
%{../../../quasi2d/New/d55ed40/data/2d/m1new/d01/twomodepm}
 \end{center}
\vspace{-0.9cm} 
\caption{The low-frequency modes, $\omega^L_{\pm}(q)$, and their spectral weights, $S^L_{\pm}(q,\omega)$, for $D=0.1$. Both solutions demonstrate an acoustic behavior in the long wavelength limit for $d \geq 0.1$. Formation of an optical branch at these layer spacings is not observed due to a high superfluid fraction, $\gamma_s > 0.9$.}
 \label{fig:two-m1D01}
 \end{figure}
 
The dispersion relation predicted by the two-mode ansatz is presented in Fig.~\ref{fig:two-m1D01}. For all layer spacing ($d \geq 0.1$) we observe no sign of rotonization similar to $D=1(5.5)$. In the former case the roton feature was always accompanied by the oscillations in the pair distribution function $g_{11}(r)$. This is not the case here, see Fig.~\ref{fig:grsk-D01m1}.  We either do not observe an optical branch. The spectrum of the symmetric and antisymmetric mode remains acoustic for all considered $d$-values.  The amplitude of the dimerization peak in $g_{12}(r)$ can be relatively large, see $d=0.1$ and $d=0.12$ in Fig.~\ref{fig:D01m1-rho12}, however, the observed long decaying tail of the dimer distribution  $g_d$ results in a significant overlap of the adjacent dimer states. The net effect is that the inlayer spatial coherence is not perturbed and the system remains in the superfluid phase. Based on our previous analyses for $D=1(5.5)$, we expect that the acoustic branch  will completely dominate the spectrum of the antisymmetric mode once $\gamma_s \gtrsim 0.8$. In the present case, even for the smallest (considered) layer spacing $d=0.1$ the superfluid fraction does not drop below $0.93$. Hence, to observe an optical branch and probe intrinsic properties of dimer states, the layer spacing should be reduced further. According to the introduced coupling strength $U_0$ of a single dimer problem~(\ref{u01}) the interlayer spacing  $d=0.1$ and the inlayer coupling $D=0.1$ used in the many-body simulations corresponds to $U_0=1$. As shows Fig.~\ref{fig:D1m1-energu0}, in this regime the interlayer binding dimer energy is significantly reduced, whereas the dimer size diverges to several inlayer interparticle spacings.

Presence of a second layer has a largest effect of the antisymmetric mode. By lowering $d$ the dispersion relation $\omega^L_-(q)$ shifts to higher frequencies, and its tangent in the low-$q$ range and the acoustic sound speed increases, see Fig.~\ref{fig:two-m1D01}. The net effect is a systematic reduction of the spectral weight, see the $d$-dependence of $S_-^L(q)$ in Fig.~\ref{fig:two-m1D01}. In contrast, the interlayer coupling shows a minimal influence on the symmetric mode dispersion $\omega^L_+(q)$ and $S_+^L(q)$. Such a result is expected when the interlayer coupling does not enhance the inlayer correlations, being opposite to the case $D=1(5.5)$ when the reduction of $d$ has led to the oscillations in $g_{11}(r)$. 

\begin{figure}
 \begin{center} 
 \hspace{-0.0cm}\includegraphics[width=0.5\textwidth]{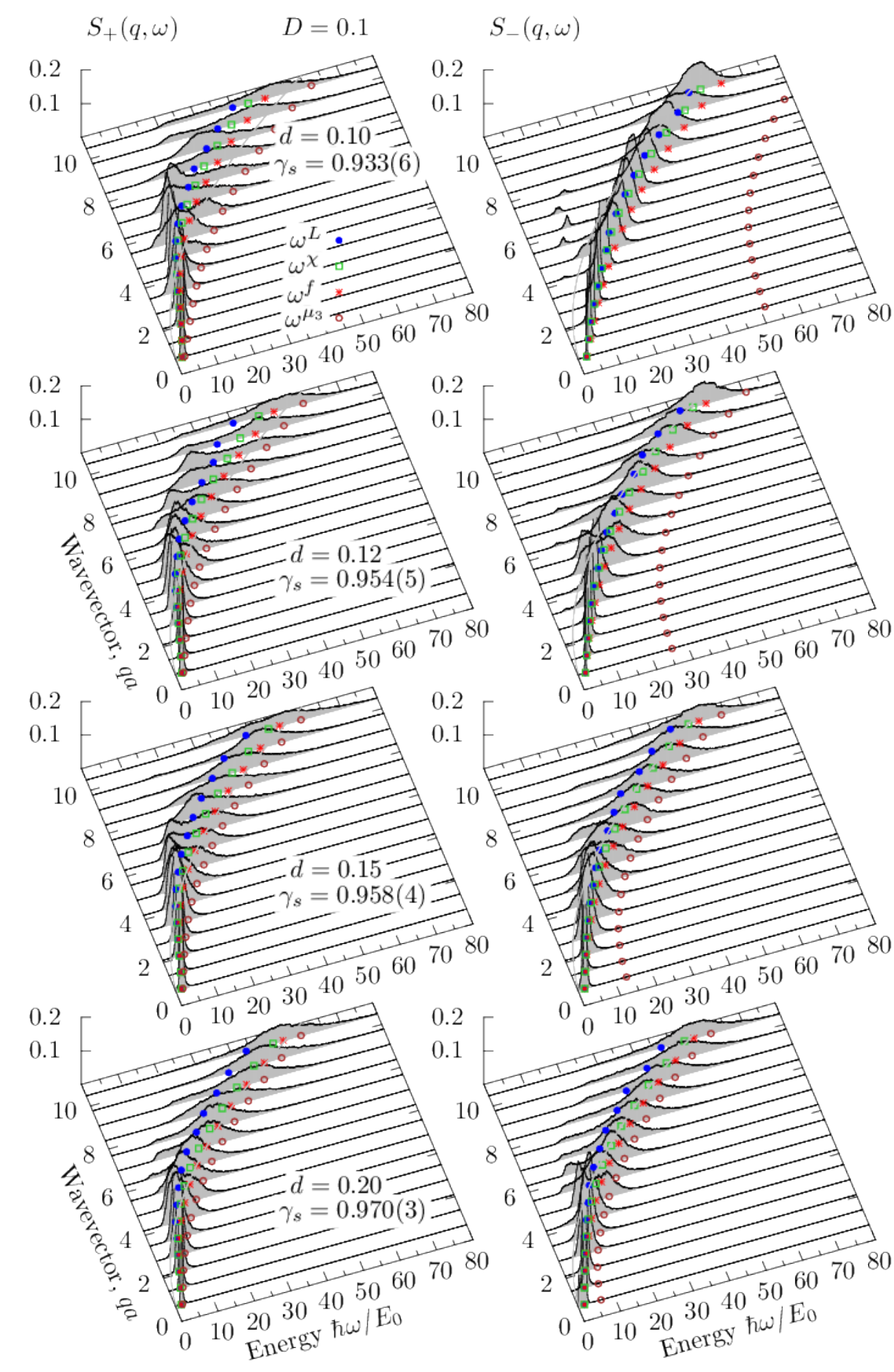}
 %{../../../quasi2d/New/d55ed40/data/2d/m1new/d01/stopm/spec3dm1l01l02}     
 \end{center}
\vspace{-0.6cm} 
 \caption{Rescaled dynamic structure factor for the symmetric (left panel), $S_{+}(q,\omega)/S_+(q)$, and antisymmetric (right panel), $S_{-}(q,\omega)/S_-(q)$, modes. Coupling $D=0.1$. The legend indicates the layer spacing, $0.1 \leq d\leq 0.2$, and the superfluid fraction $\gamma_s$. For comparison several upper bounds for the dispersion relation are shown: $\omega_\pm(q) \leq \omega^\chi_\pm(q) \leq \omega^f_\pm(q) \leq \omega^{\mu_3}_\pm(q)$, see Eq.~(\ref{3upbound}). The ansatz $\omega^L_{\pm}$ (indicated by blue symbols) provides best agreement with the low-frequency resonances in $S_{\pm}(q,\omega)$. The solid gray line denotes the free-particle dispersion $\epsilon_q=q^2/2m$.}
 \label{fig:m1d01a}
 \end{figure} 
 
In Fig.~\ref{fig:m1d01a} we perform comparison of the reconstructed dynamic structure factor $S_{\pm}(q,\omega)$ with the upper bounds for the dispersion relation. In the acoustic range $\omega^f_{\pm}(q)$ and $\omega^{\chi}_{\pm}(q)$ converge to a single dispersion relation with a linear $q$-dependence. Beyond the acoustic domain, similar to $D=1(5.5)$, the two-mode solution $\omega^L_{\pm}(q)$ provides best agreement with the resonances in $S_{\pm}(q,\omega)$. On the contrary, the largest deviations are found for $\omega^{\mu_3}_{\pm}(q)$, specifically for small $d$ and for the antisymmetric mode. As was discussed above, the main contribution to the $\Avr{\omega^3}$-sum rule is given the high-frequency behavior of the spectral density. At low $d$ (e.g. $d=0.10,0.12$) a slow decaying high-frequency asymptotic behavior of $S_-(q,\omega)$ can be resolved on the log-scale. 

\begin{figure}
 \begin{center} 
 \hspace{-0.0cm}\includegraphics[width=0.5\textwidth]{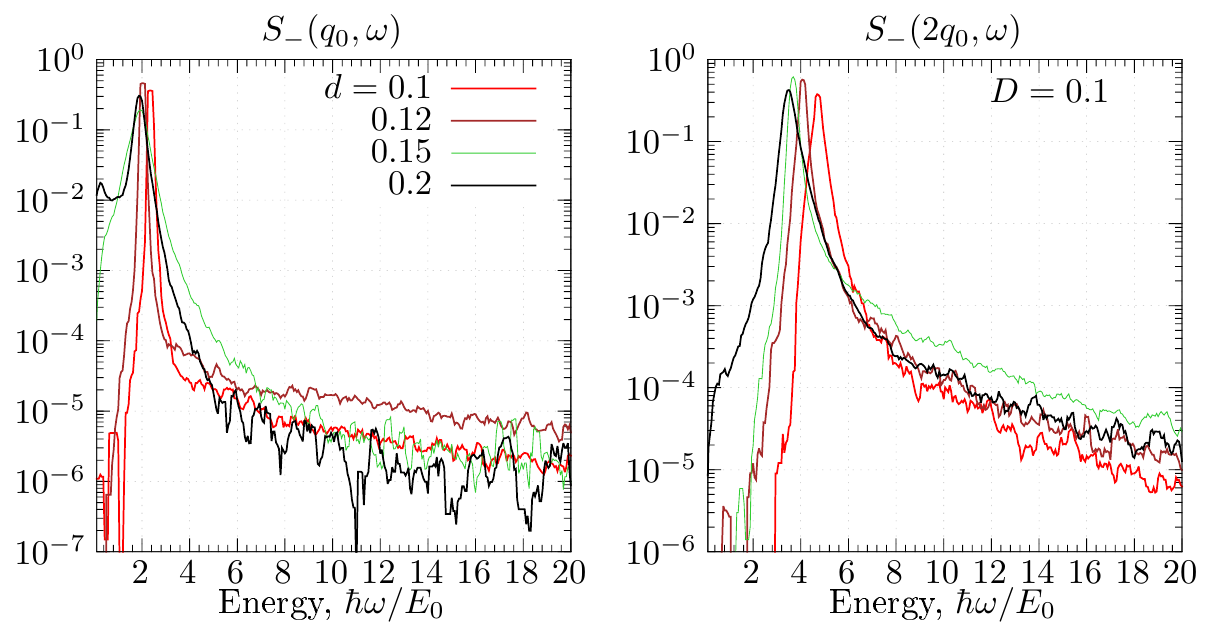}
 %{../../../quasi2d/New/d55ed40/data/2d/m1new/d01/stopaper/specSmk1}     
 \end{center}
\vspace{-0.6cm} 
 \caption{The $d$-dependence of $S_{-}(q,\omega)$ for $D=1$. Used wavenumbers $q=q_0 n$: $n=1$ (left) and $n=2$ (right). The legend indicates the layer spacing: $0.1\leq d \leq 0.2$.}
 \label{fig:Specmk1d01}
 \end{figure} 

For all $d$ the system remains in a superfluid phase with $\gamma_s > 0.9$ and, therefore, no sign of an optical branch is observed in $S_{-}(q,\omega)$. In more detail, the $d$-dependence, for the two smallest wavenumbers, is shown in Fig.~\ref{fig:Specmk1d01}. For low $d$ there is a systematic shift of the acoustic resonances to larger frequencies. This testifies an increase of the acoustic sound speed.

 \begin{figure}
 \begin{center} 
 \hspace{-0.0cm}\includegraphics[width=0.5\textwidth]{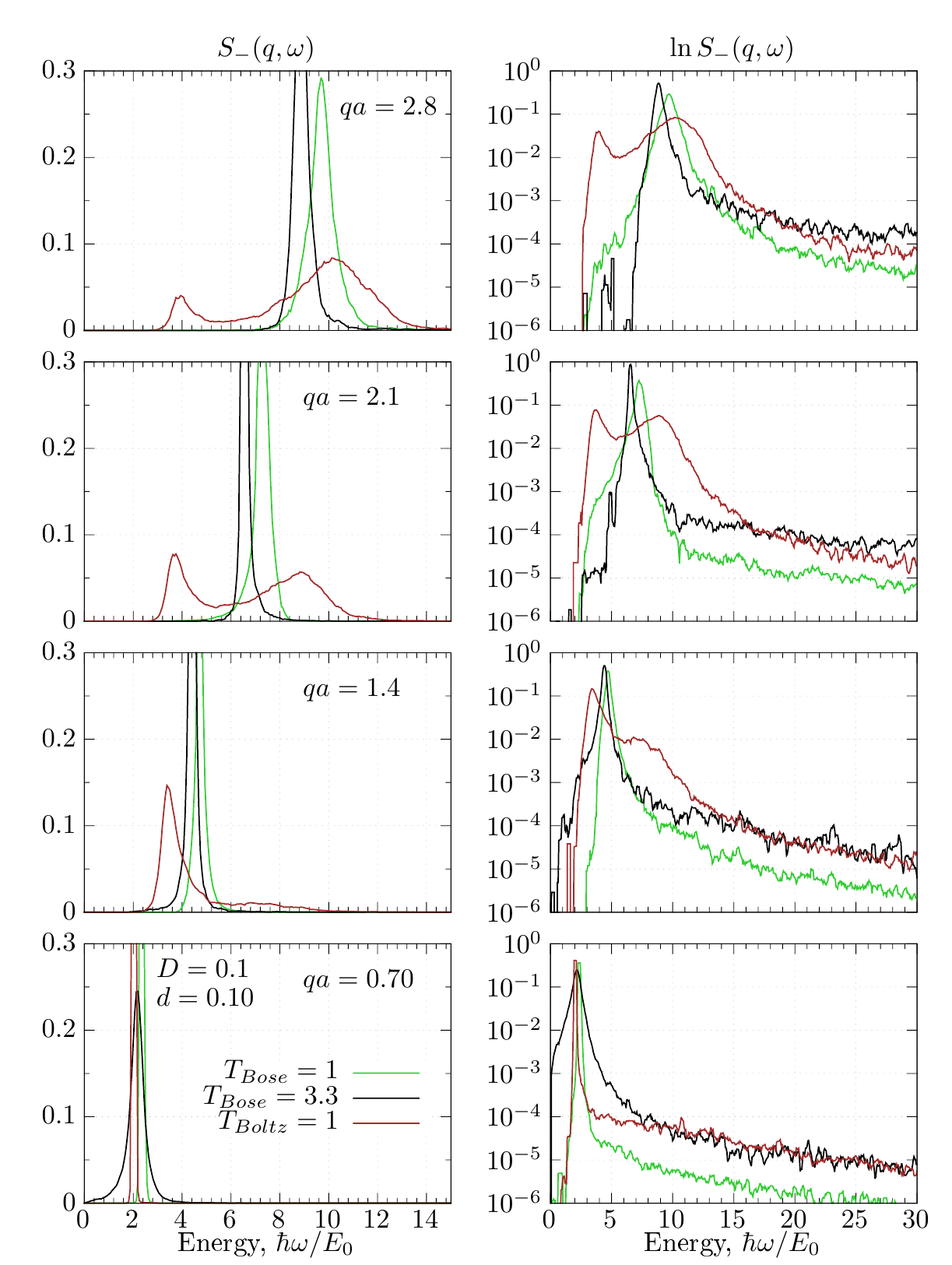}
 %{../../../quasi2d/New/d55ed40/data/2d/m1new/d01/stopaper/specd01m1l01boseboltz}     
 \end{center}
\vspace{-0.6cm} 
 \caption{Comparison of the antisymmetric mode spectra for a bosonic bilayer at $T_{\text{Bose}}=1$ ($\rhos=0.93$), $T_{\text{Bose}}=3.3$ ($\rhos=0$)  and a bilayer with Boltzmann statistics at $T_{\text{Boltz}}=1$ ($\rhos=0$). Layer spacing $d=0.1$ and $D=0.1$. The right panels show the results on the log-scale. The legend indicates the wavenumbers $qa$. For the Bose statistics only the acoustic branch is observed. Both the acoustic and the optical branch are present in the Boltzmann case.}
 \label{fig:specl01d01}
 \end{figure} 

 \begin{figure}
 \begin{center} 
 \hspace{-0.0cm}\includegraphics[width=0.5\textwidth]{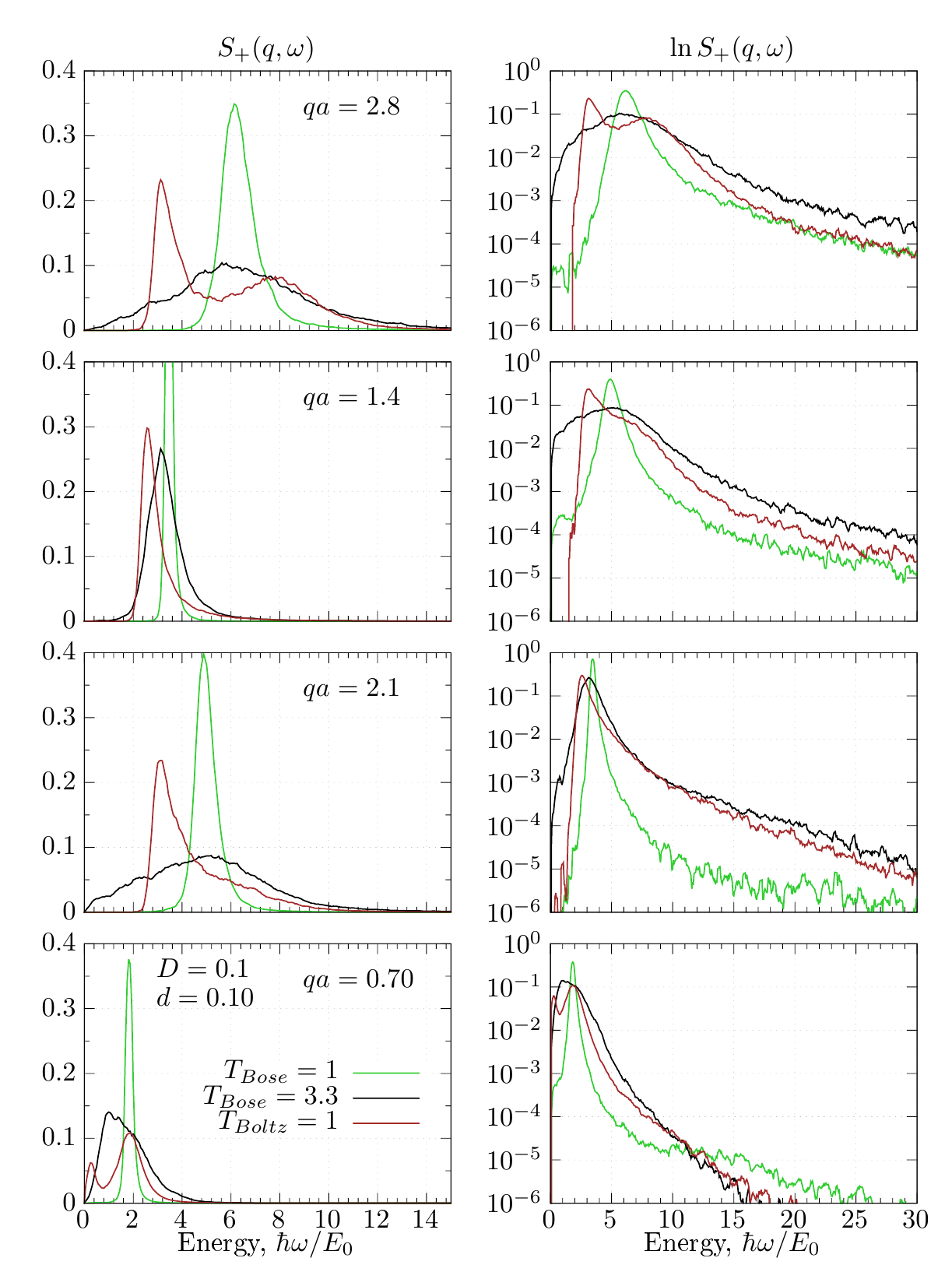}
 %{../../../quasi2d/New/d55ed40/data/2d/m1new/d01/stopaper/specd01m1l01boseboltzplus}     
 \end{center}
\vspace{-0.6cm} 
 \caption{Same as in Fig.~\ref{fig:specl01d01} for the symmetric mode.}
 \label{fig:specl01d01plus}
 \end{figure} 
  
We repeat our analyses for distinguishable ``boltzmannons'' to check a relation between an optical branch and a superfluidity. Fig.~\ref{fig:specl01d01} presents the spectrum for the antisymmetric mode. The simulations with the Bose statistics are performed at the two temperatures, $T_{\text{Bose}}=\{1,3.3\}$ and show a single acoustic branch. For $T_{\text{Bose}}=3.3$ we observe a systematic shift of the resonances to lower frequencies and reduction of the acoustic sound speed. 

On the contrary, for the Boltzmann case we observe a splitting into two branches for $qa \geq 1.4$. The lower branch is dispersionless with the energy $\hbar\omega/E_0 \sim 3.5$. A weak dependence of the wavenumber indicates its relation with intrinsic excitations of the interlayer dimers. The second high-frequency branch is acoustic with the sound speed which exceeds the one for the Bose system. A shift of the ``dimer mode'' to a low frequency domain can be explained by a significantly reduced dimer binding energy compared to $D=1(5.5)$, see $\epsilon_d$ in Fig.~\ref{fig:D1m1-energu0} for $U_0=1$. The dimer binding energy can be also estimated from $\epsilon^T(d)$ in Figs.~\ref{fig:D1m1-energy},~\ref{fig:D55m1-energy} and $\epsilon(d) (r_c,\gamma r^3)$ in Fig.~\ref{fig:D01m1-energy}.

For the boltzmannons a similar splitting into two branches (``dimer'' and acoustic), but not so pronounced, is also observed in the symmetric mode spectra, see Fig.~\ref{fig:specl01d01plus}.

In the Boltzmann case for $qa < 1.4$, both modes collapse into a single resonance, see $qa=0.70$ in Fig.~\ref{fig:specl01d01}. In contrast to $D=1(5.5)$, presence of an optical gap in the long wavelength limit here can not be clearly confirmed. At the smallest wavenumber ($qa=0.70$) only a single resonance is recovered in the symmetric and antisymmetric modes, see Figs.~\ref{fig:specl01d01} and~\ref{fig:specl01d01plus}. To check whether the spectrum splits again into two branches at lower $q$, the simulations with a significantly larger system size (to access smaller wavenumbers) are required.
 
\begin{figure}
 \begin{center} 
 \hspace{-0.35cm}\includegraphics[width=0.5\textwidth]{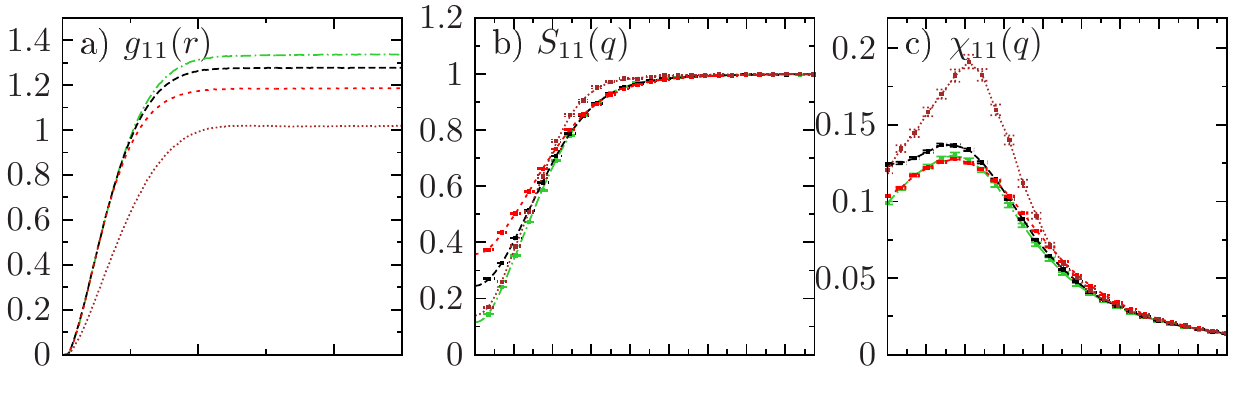}\\
 %{../../../quasi2d/New/d55ed40/data/2d/m1new/d01/gr11sk11chi11noxboltz}\\
 \vspace{-0.25cm}
 \hspace{-0.35cm}\includegraphics[width=0.5\textwidth]{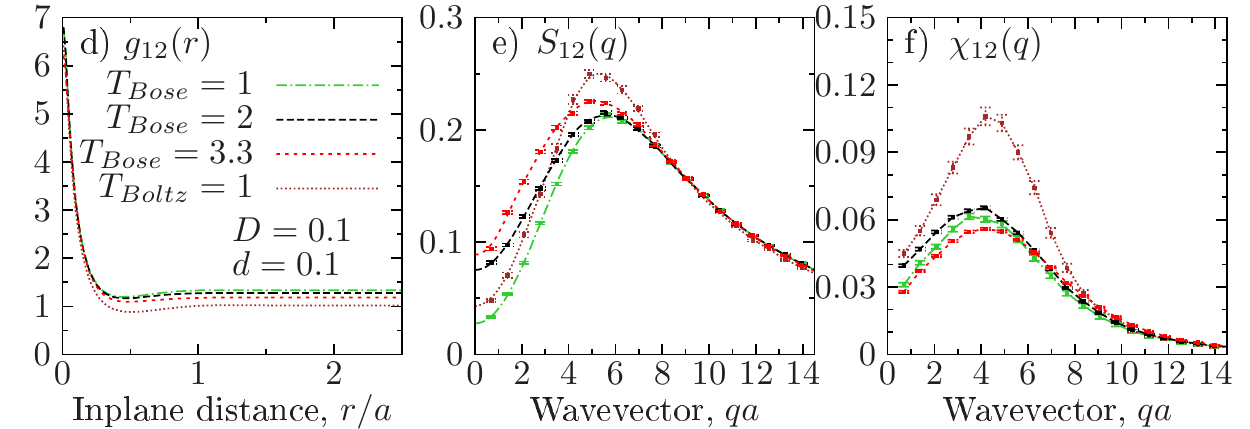}\\
 %{../../../quasi2d/New/d55ed40/data/2d/m1new/d01/gr12sk12chi12boltz}\\
 \vspace{-0.05cm}
 \hspace{-0.35cm}\includegraphics[width=0.5\textwidth]{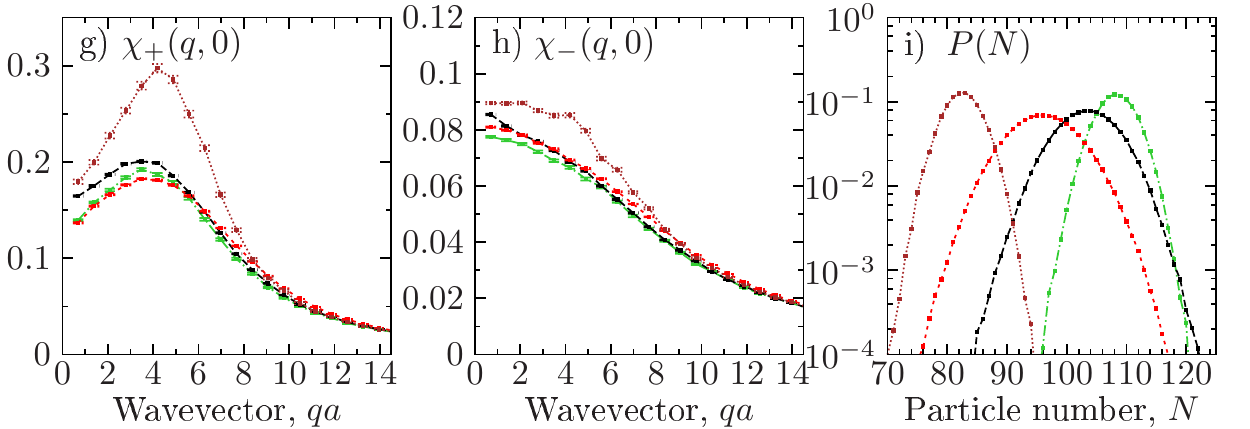}
 %{../../../quasi2d/New/d55ed40/data/2d/m1new/d01/chi0diagboltz}
 \end{center}
\vspace{-0.70cm} 
 \caption{Comparison of the static properties for $d=0.1$ and $D=0.1$. Compared are a bosonic bilayer at the three temperatures, $T_{\text{Bose}}=\{1,2,3.3\}$, and a bilayer with Boltzmann statistics at $T_{\text{Boltz}}=1$.}
 \label{fig:specl01statd01}
 \end{figure}

Presence of the low-frequency dispersionless mode for $T_{\text{Bolt}}=1$ is related with a strong interlayer coupling identified by a peak $g_{12}(0)$ in Fig.~\ref{fig:specl01statd01}d. In the superfluid phase we observe a similar value of $g_{12}(0)$, but the ``dimer mode'' is masked (or substituted) by the collective excitations of a superfluid component and acoustic phonons. 

The effect of quantum statistics on the static properties can be followed in detail in Fig.~\ref{fig:specl01statd01}. The Bose case is presented by three temperatures which cover the transition from the superfluid to the normal phase. 

First, in Fig.~\ref{fig:specl01statd01}a we note a significantly higher inlayer density for the Bose system, by $20\%-35\%$ compared to the boltzmannons [for the same chemical potential $\mu$]. The inlayer static characteristics, $g_{11}(r)$ and $S_{11}(q)$, remain structureless for all temperatures. No qualitative changes are observed during the superfluid-normal transition by variation of $T_{\text{Bose}}$. In contrast, in the Boltzmann case we observe an enhancement of the peak amplitude of  $\chi_{11(12)}(q)$, and their symmetrized counterparts, $\chi_{\pm}(q)$. The effect is more pronounced for the symmetric mode, and can be explained by the Kramers–Kronig relation which states the relation between the static limit of the density response function and the low-frequency behavior of the spectral density
\begin{align}
 -\frac{\Re \chi_{\pm}(q,0)}{2}=\int\limits_{-\infty}^{\infty}  \frac{1}{\omega}\, S_{\pm}(q,\omega) \, \db \omega.
\end{align}
The observed enhancement of $\chi_{\pm}(q)$ for the boltzmannons can be uniquely identified with an additional low-frequency dispersionless (``dimer'') branch in $S_\pm(q,\omega)$, which is not observed in the Bose system, see Fig.~\ref{fig:specl01d01} and~\ref{fig:specl01d01plus}. 

In both cases, up to the wavenumber $qa \sim 8$ the spectral weight of the ``dimer'' mode is comparable with the one of the main excitation branch, specified by the Bogolyubov-type dispersion $\omega_{\pm}(q) \sim \omega^L_\pm(q)$ (Fig.~\ref{fig:two-m1D01}). As a result a pronounced difference between Bose and Boltzmann statistics can be observed in $\chi_{\pm}(q)$  in Fig.~\ref{fig:specl01statd01}g,h.

{\em Spectrum in the phonon and roton region.} Now we analyze the $d$-dependence of the resonances in the range of phonon, $q_n=2\pi n/L$ ($n=1,2$), and ``roton'' wavenumbers. For $D=0.1$ the roton feature is absent, therefore, as the ``roton'' wavenumber we use similar values as for $D=1(5.5)$, i.e. $q_n=2\pi n/L$ ($n=9,10$).

\begin{figure}
 \begin{center} 
  \hspace{-0.70cm}\includegraphics[width=0.5\textwidth]{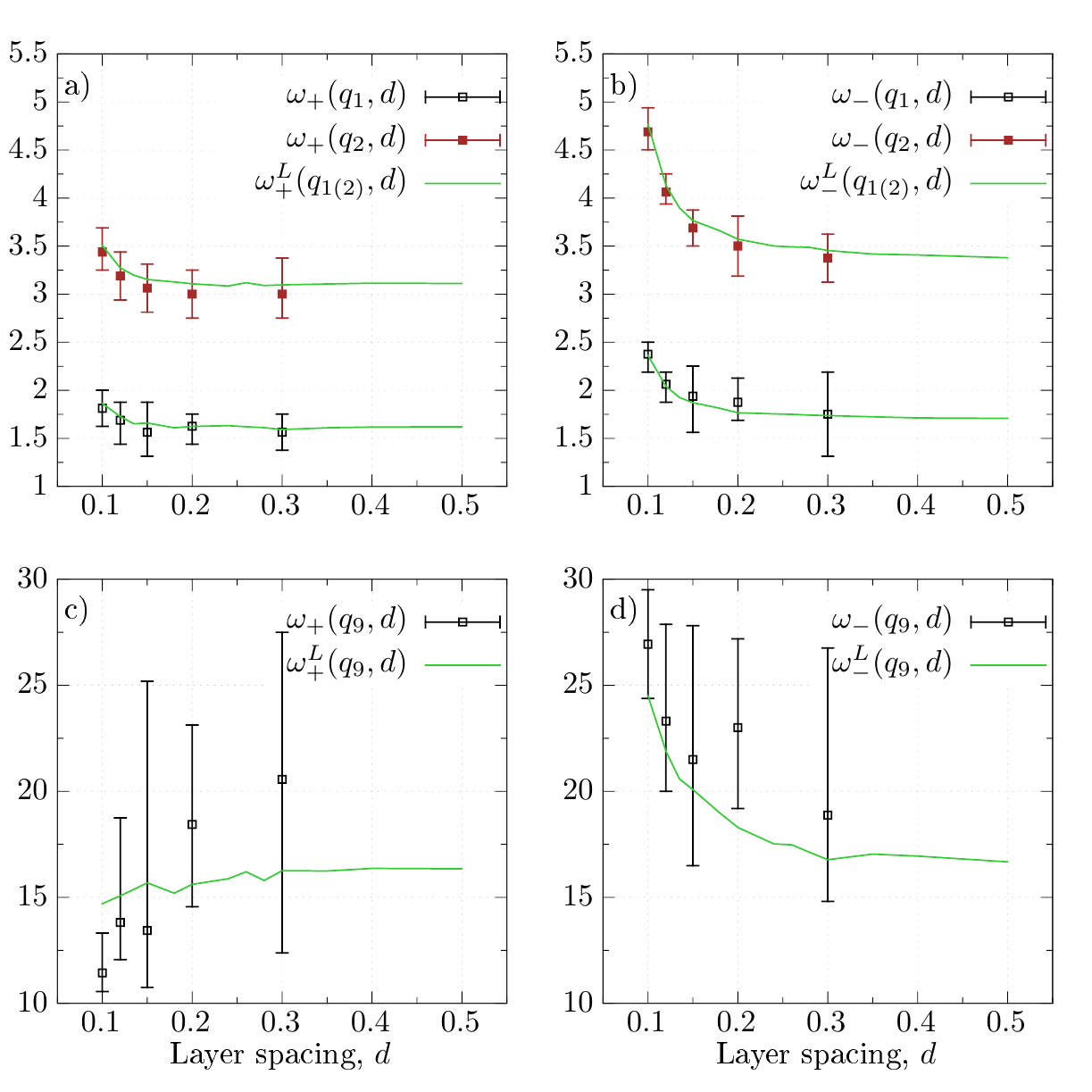}
  %{../../../quasi2d/New/d55ed40/data/2d/m1new/d01/stopaper/wpq1q2}    
 \end{center}
 \vspace{-0.6cm} 
\caption{The $d$-dependence of the low-frequency resonances $\omega_{\pm}(q)$ (maximum in $S_{\pm}(q,\omega)$ and their halfwidth for $D=0.1$. The wavenumbers $q_n=2 \pi n/L$ in the phonon ($n=1,2$) and ``roton'' ($n=9$) domain are considered. Predictions from the two-mode ansatz $\omega^L_{\pm}$ are indicated by solid lines.}
 \label{fig:roton-D01}
 \end{figure}

To extract resonance positions we use the dispersion $\omega^L_{\pm}(q)$ and the  dynamic structure factor $S_{\pm}(q,\omega)$. The results are presented in Fig.~\ref{fig:roton-D01}.

As shows Fig.~\ref{fig:roton-D01}a,b, when $d$ is reduced the phonon resonances are shifted to higher frequencies, i.e. the acoustic sound speed $c_\pm(d)$ increases. The effect is larger for the antisymmetric mode. The increase of $c_\pm(d)$ means an increase of the intralayer coupling. The latter originates from two effects. First, from the density increase (see Fig.~\ref{fig:D01m10-sound}), and, second, from the spatial localization of particles due to onset of the interlayer dimerization below $d\sim 0.2$ (see $g_{12}(r)$ in Fig.~\ref{fig:grsk-D01m1}). 

The phonon resonances $\omega_\pm(q)$ and their halfwidth recovered from $S_{\pm}(q,\omega)$ are indicated by  symbols with errorbars. The resolved $d$-dependence is found to be in a nice agreement with the two-mode ansatz, $\omega^L_{\pm}(q,d)$, shown by solid lines. A rapid increase of $c_-(d)$ for $d\leq 0.15$ is directly correlated with the reduction of the superfluid fraction $\gamma_s(d)$ and enhancement of the interlayer coupling. Formation of the dimerization peak  $g_{12}(0)$ is observed in Fig.~\ref{fig:specl01statd01}, however, in contrast to $D=1(5.5)$  the dimer states can not be resolved individually, as in each layer the system stays in a homogeneous superfluid phase. 

The interlayer coupling also influences the ``roton'' region, see Fig.~\ref{fig:roton-D01}c,d. The resonances of the antisymmetric mode shift to higher frequencies. There is a good agreement between $\omega_-(q_9)$ and $\omega_-^L(q_9)$. As shows Fig.~\ref{fig:m1d01a}(right panels),  $\omega_-^L(q)$ remains quite accurate, practically, for all $d$ and in a wide range of wavenumbers. In contrast, $\omega_+^L(q)$ agrees with the resonances in $S_{+}(q,\omega)$ only for $d \geq 0.2$. As shows  Fig.~\ref{fig:m1d01a}(left panels), at small $d$ around the ``roton'' wavenumber and beyond, the spectrum splits into two branches. In this case  $\omega_+^L(q)$ becomes an estimate of their average. The lower branch can be considered as a continuation of the acoustic dispersion. Its formation is due to decay processes of the quasiparticles with the quadratic dispersion $q^2/2 m^{\star}$. This splitting is the main reason for the discrepancy between $\omega_+^L(q)$ and $\omega_+(q)$ as observed in Fig.~\ref{fig:roton-D01}c for $d \leq 0.15$.

\begin{figure}
 \begin{center} 
 \vspace{-0.3cm}
 \hspace{-0.0cm}\includegraphics[width=0.5\textwidth]{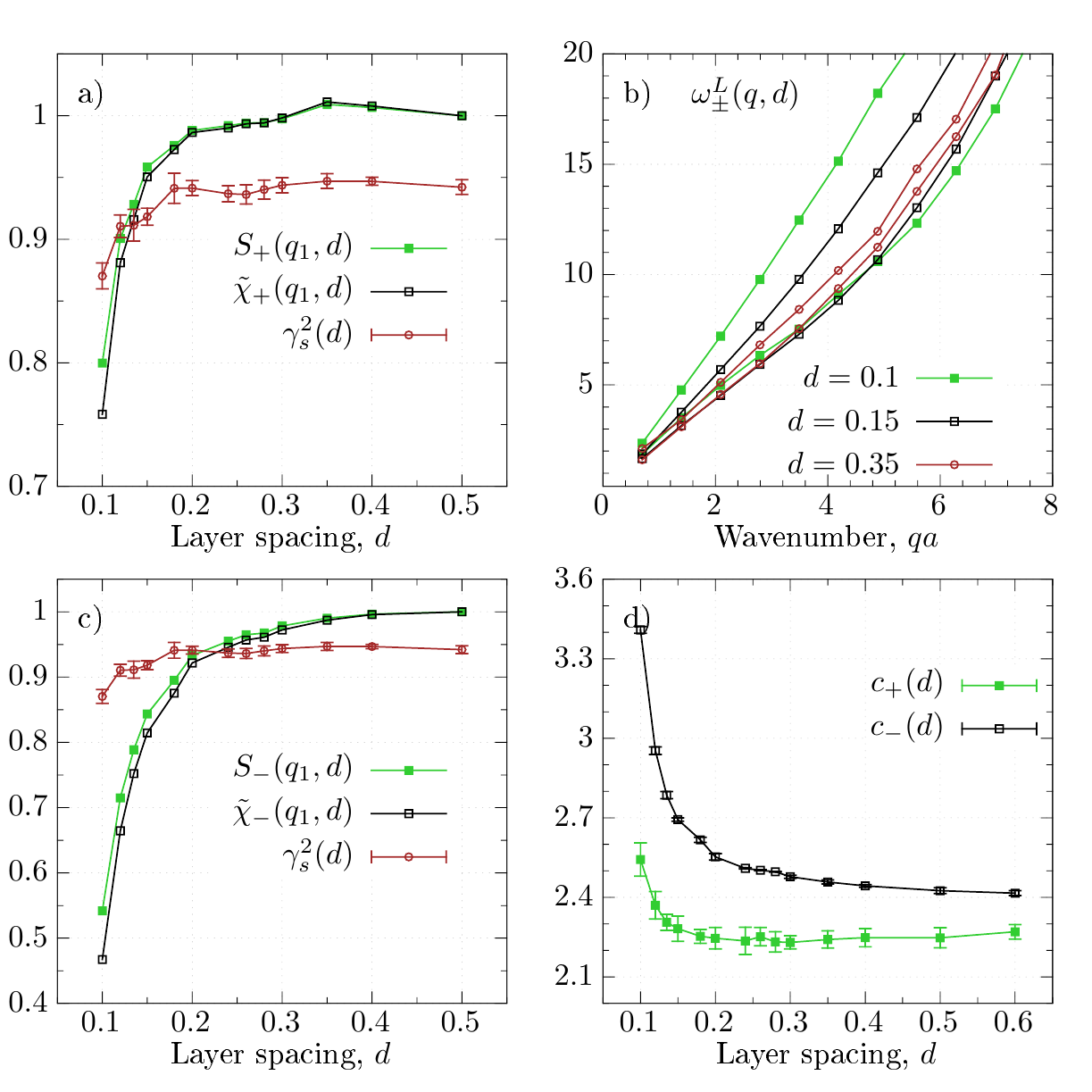}
 %{../../../quasi2d/New/d55ed40/data/2d/m1new/d01/stopaper/rhosweight}     
 \end{center}
\vspace{-0.8cm} 
  \caption{a), c) The $d$-dependence of the renormalized static structure factor, $S_{\pm}(q_1,d)=S_{\pm}(q_1,d)/S_{\pm}(q_1,d=0.5)$, the static response function $\tilde\chi_{\pm}(q_1,d)=\Re \chi_{\pm}(q_1,d)/ \Re \chi_{\pm}(q_1,d=0.5)$ for the symmetric and antisymmetric mode, and the square of the superfluid fraction $\gamma^2_s(d)$. b) The dispersion $\omega^L_{\pm}(q)$ from the two-mode ansatz at two layer spacings. d) The $d$-dependence of the acoustic sound speed obtained from the linear fit, $\omega^L_{\pm}(q,d)=c_{\pm}(d) q$, for $qa \leq 2$. The increase of $d$ has a larger effect on the antisymmetric mode (see also Fig.~\ref{fig:two-m1D01}). The normalization values: $S_{\pm}(q_1,d=0.5)=0.2239(0.2045)$ and $\Re \chi_{\pm}(q_1,d=0.5)/2=0.09209(0.08296)$.}
 \label{fig:rhosweightd01}
 \end{figure}

Next, in Fig.~\ref{fig:rhosweightd01}a,c we analyze the $d$-dependence of the static structure factor and the static response function in the long wavelength limit. Both the symmetric and antisymmetric spectrum shows only the acoustic branch, therefore, the compressibility sum-rule~(\ref{cs}) can be applied to determine the isothermal sound speed. The interlayer dimerization, as observed in $g_{12}(r)$  at $d\leq 0.2$, results in a strong reduction of both $S_{\pm}(q_1)$ and $\chi_{\pm}(q_1)$ at the smallest wavenumber $q_1$. In this case Eq.~(\ref{cs}) predicts a strong enhancement of $c_{\pm}$. This result is in a full agreement with the independent estimate obtained from the fit, $\omega^L_\pm(q)\approx c_\pm q$, see Fig.~\ref{fig:rhosweightd01}b,d. The effect is more pronounced for the antisymmetric mode.

The dispersion relation $\omega^L_{\pm}(q,d)$ for several layer spacing is presented in Fig.~\ref{fig:rhosweightd01}b. Two modes corresponding to the same $d$-value are indicated by the same color: $\omega^L_{-}$ (upper curve) and $\omega^L_{+}$ (lower curve). Again a more strong $d$-dependence  is demonstrated by $\omega^L_{-}(q)$.

We can conclude that while the interlayer dimerization has only a little effect the inlayer  superfluidity, it has a strong influence on both static and dynamic properties. We observe the pronounced effect in the $d$-dependence of the acoustic sound speed $c_\pm(d)$. This example can be complemented by the case $D=5.5$, where the interlayer dimerization, first, leads to formation to strongly bound dimers, and, second, as a result of spatial localization of these dimers and suppression of the inlayer superfluidity, to formation of a Wigner-like crystalline structure. The latter, in its turn, modifies all static and dynamic characteristics.

\begin{figure}
 \begin{center} 
\hspace{-0.0cm}\includegraphics[width=0.51\textwidth]{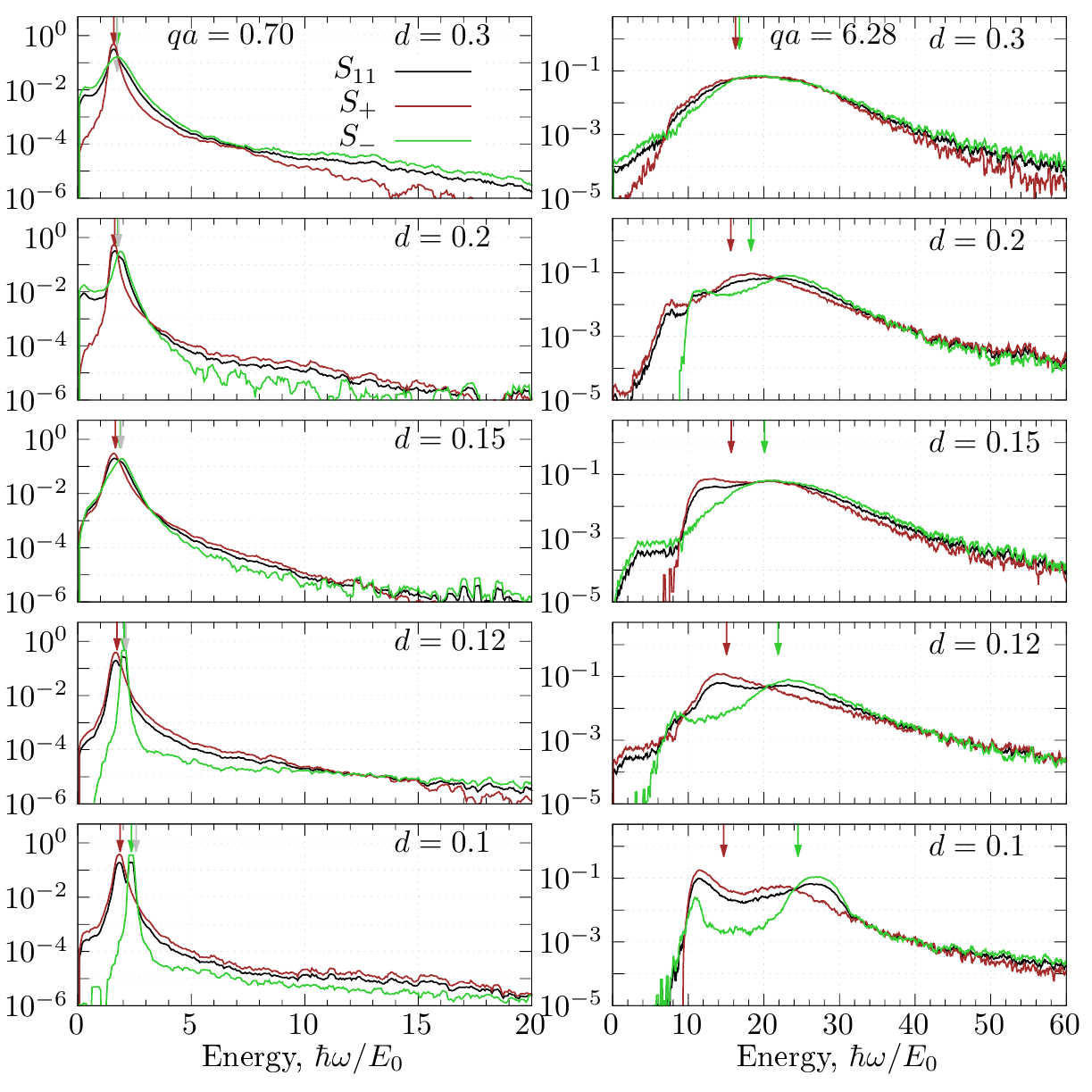}
%{../../../quasi2d/New/d55ed40/data/2d/m1new/d01/stopm/m1d01k1alld}
 \end{center}
\vspace{-0.9cm} 
\caption{Dynamic structure factors $S_{\pm}(q,\omega)$ and $S_{11}(q)$ for $D=0.1$. Each spectral density is identified by the color as specified by the legend. Left (right) column shows the result for the wavenumber $qa=0.70$ ($qa=6.28$). Vertical arrows indicate position of the peaks predicted from the sum-rules: $\omega^L_{+}(q)$, $\omega^L_{-}(q)$ and $\omega^f_{-}(q)$.}
 \label{fig:alld-m1D01}
 \end{figure}

Similar to $D=1(5.5)$, in Fig.~\ref{fig:alld-m1D01} we compare a relative contribution of the symmetric and antisymmetric modes in the partial dynamic structure factor $S_{11}(q,\omega)$. The results are shown for several $d$  in the region of acoustic phonons ($qa=0.70$) and the ``roton'' ($qa=6.28$). The predictions from the two-mode ansatz are indicated by vertical arrows. Both spectral densities, $S_{-}$ and $S_{+}$, can be well distinguished for  $d \leq 0.15$. The lower (higher) frequency resonance in $S_{11}(q,\omega)$  corresponds to the symmetric (antisymmetric) mode. In the ``roton'' domain both resonances can be well distinguished, whereas in the acoustic one they are quite close and the spectral densities strongly overlap. Certainly, the symmetrization of the density response function in the terms of its eigenmodes ``$\pm$'' significantly simplifies physical interpretation of the observed spectral features, and helps to understand evolution of the spectral density with variation of the layer spacing.

{\em Mean-field analyses.} 
At large $d$, we are in the regime of weak inlayer coupling and can compare our results with the mean-field predictions. They are improved by the local field-corrector to take into account effects of many-body correlations. This type of analyses have been recently performed in Ref.~\cite{Tanat2015} to investigate instability of a homogeneous dipolar bilayer against the formation of density waves. The diagonalization of the density response function in the mass-symmetric bilayer written in the PRA-form leads to 
\begin{align}
 \chi_{\pm}(q,\omega)=\frac{\Pi(q,\omega)}{1-\Pi(q,\omega) W_{\pm}(q,\omega)},
 \label{chi_rpa}
\end{align}
where $W_{\pm}=W_{11}\pm W_{12}$ is the effective potential (EP) for the symmetric/antisymmetric mode, and 
\begin{align}
\Pi(q,\omega)=\frac{2 n \epsilon_q}{(\omega+i \delta)-\epsilon_q^2}
\end{align}
is the polarization function of a non-interacting system written in the so called mean-spherical approximation. In general case the frequency dependence of the EP in~(\ref{chi_rpa}) can be used to satisfy different frequency power moments.~\cite{Ark2014} Substitution of $W(q,\omega)$ with its static value in the low- or high-frequency limits [$W(q,0)$, $W(q,\infty)$] is related to the STLS~\cite{stls} and QLCA~\cite{qlca} approximations. 

In the PRA case the EP is taken as Fourier transform of the bare intralayer (interlayer) interaction potential, $V_{\al\al(\al\be)}$. The interlayer potential in the momentum space takes the form
\begin{align}
 V_{\al\be}(q)=-2 \pi D\, q\, e^{-qd}.
 \label{v12_q}
\end{align}
To remove divergence of the intralayer potential $V_{\al\al}(q)$ in the Fourier space [for a 2D dipolar system], the latter can be smoothed over the layer thickness in the $r$-space.~\cite{block2012} Alternatively, $V_{\al\al}(q)$ can be substituted by an effective potential $\tilde V$ which satisfies the FDT
\begin{align}
 S_{\al\al}(q)=-\frac{1}{\pi \rho} \int\limits_{-\infty}^{\infty} \db \omega \Im \frac{\Pi(q,\omega)/(1-e^{-\beta \omega})}{1-\Pi(q,\omega) \tilde{V}_{\al\al}(q,\omega)}.
 \label{Saa}
\end{align}
At zero and low temperature [when the dispersion relation satisfies $\hbar \omega(q) \gg k_B T$] by neglecting the frequency dependence of the EP, i.e. $\tilde V(q,\omega)\approx \tilde V(q,0)$, the integration in~(\ref{Saa}) can be performed analytically with the result
\begin{align}
 \tilde V_{\al\al}(q)=\frac{\epsilon_q}{2 \rho} \left[\frac{1}{S^2_{\al\al}(q)}-1 \right].
 \label{ueff}
\end{align}
Obviously, the similar result holds also for $\tilde W_{\pm}(q,0)$ by substitution $S_{\al\al}$ in~(\ref{Saa}) with the symmetrized static structure factor $S_{\pm}(q)$. Note, that the effective interactions, partially, take into account the exchange (Bose statistics) and many-body correlation effects via $S_{\pm}(q)$. In the following we will compare $\chi_{\pm}$ from~(\ref{chi_rpa}) for the two cases, with the direct interlayer correlations and  with the exchange effects 
\begin{align}
  &W^{d}_{\pm}(q)=\tilde V_{11}(q)\pm V_{12}(q),\label{w_d}\\
  & W^{\text{ex}}_{\pm}(q)=\frac{\epsilon_q}{2\rho} \left[\frac{1}{S^2_{\pm}(q)}-1 \right].\label{w_ex}
\end{align}
Note, that due to the relations,
\begin{align}
 &\lim \limits_{q\rightarrow \infty} S_{12}(q)=0,\quad \lim \limits_{q\rightarrow \infty} S_{\pm}(q)=\lim \limits_{q\rightarrow \infty} S_{11}(q)=1,\\
 &\lim \limits_{q\rightarrow \infty} W^{d}_{\pm}(q)= 0, \, \lim \limits_{q\rightarrow \infty} W^{\text{ex}}_{\pm}(q)=0,
\end{align}
both cases reproduce the static response function in the free-particle limit, i.e.
\begin{align}
 & \Re \chi_{\pm}(q,\omega=0)=-\frac{1}{\epsilon_q/2\rho +W_{\pm}(q)},\label{re_rpa}\\
 &\lim \limits_{q\rightarrow \infty} \Re \chi_{\pm}(q,\omega=0)=-\frac{4 m \rho}{\hbar^2 q^2}.  
\end{align}
Note, that the above approach [the ansatz~(\ref{re_rpa}) with the EP~(\ref{w_ex})] can not simultaneously reproduce the correct long-wavelength limit and the compressibility sum-rule. This can be shown by using the result in Figs.~\ref{fig:m1d01a} and~\ref{fig:rhosweightd01}b,d. In the superfluid phase the spectrum is acoustic in the long wavelength limit and, hence, in (\ref{re_rpa}),(\ref{w_ex}) we can substitute, $S_{\pm}(0)=k_BT/m c^2_{\pm}$ and $\lim \limits_{q\rightarrow 0} \Avr{w^1}_{\pm}/\Avr{w^0}_{\pm}=\lim \limits_{q\rightarrow 0} \epsilon_q/S_{\pm}(q) \approx c_{\pm} q$, with the result
\begin{align}
 \lim \limits_{q\rightarrow 0} \frac{\Re \chi_{\pm}(q,\omega=0)}{\rho}=- \lim \limits_{q\rightarrow 0}\frac{2}{q m c_{\pm}^3/k_BT- \epsilon_q}.
 \label{divergence}
\end{align}
This expression predicts a divergence of the static response function in the long wavelength limit, being in a contradiction with the exact relation
\begin{align}
\frac{\Re \chi_{\pm}(0,\omega=0)}{2\rho}=-\frac{S_{\pm}(0)}{k_BT}=\frac{1}{m c_{\pm}^2},
\end{align}
which predicts a finite value at $q=0$. This disagreement can be removed by a new effective potential in the form
\begin{align}
 W_{\pm}^{\chi}(q)=-\frac{1}{\Re \chi_{\pm}(q,0)}-\frac{\epsilon_q}{2\rho}.
 \label{w_chi}
\end{align}
This choice allows to exactly satisfy the compressibility sum-rule, however, at the expense of violating the fluctuation-dissipation theorem~(\ref{Saa}). 

We can conclude that all considered approximations are a trade of to satisfy a restricted number of frequency power moments for the spectral density.

\begin{figure}
 \begin{center} 
\hspace{-0.0cm}\includegraphics[width=0.51\textwidth]{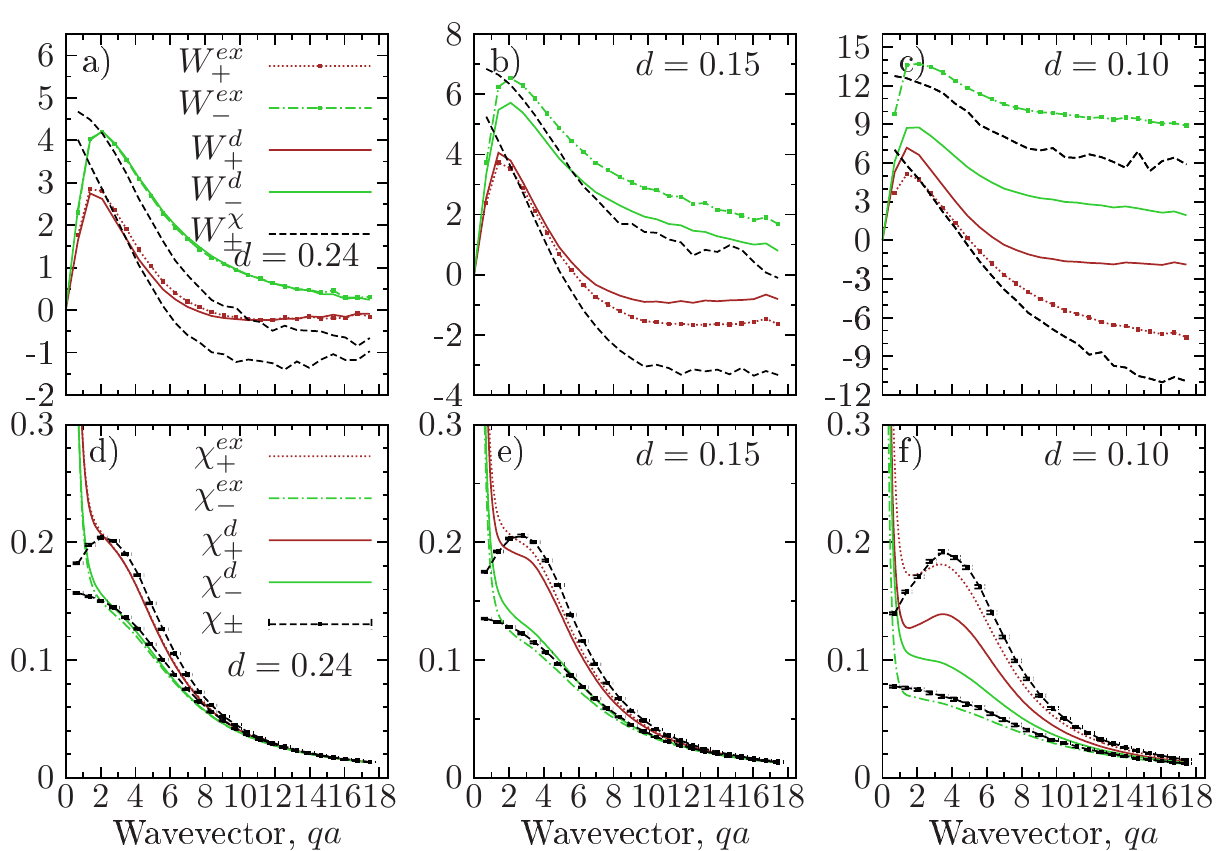}
%{../../../quasi2d/New/d55ed40/data/2d/m1new/d01/effupm}
 \end{center}
\vspace{-0.5cm} 
\caption{a)-c) Comparison of the effective potentials in the PRA-type ansatz for the symmetric (lower set of curves) and antisymmetric (upper set of curves) mode for $D=0.1$ and the layer spacing $d=0.24,0.15,0.10$. d)-f) The reference value of the static density response function, $\chi_{\pm}(q)=\abs{\Re \chi_{\pm}(q,0)}/2\rho$, versus the PRA-type approximations, $\chi^{\text{ex,d}}_{\pm}(q)$. Note, that in all cases $\chi_{+}(q)> \chi_{-}(q)$.}
 \label{fig:effupmD01}
 \end{figure}

The comparison of the effective potentials [Eqs.~(\ref{w_d}),(\ref{w_ex}) and~(\ref{w_chi})] for several interlayer spacing is shown in Fig.~\ref{fig:effupmD01}. At large spacing ($d=0.24$), the EP with the direct and exchange correlations, $W^{\text{d,ex}}$, nearly coincide, see Fig.\ref{fig:effupmD01}a. This proves that the interlayer correlations can be well described by a bare potential~(\ref{v12_q}). Simultaneously, we are able to reproduce the reference result, $\Re \chi_{\pm}(q,0)$, evaluated via~(\ref{gg2}), in a wide range of wavenumbers, see Fig.\ref{fig:effupmD01}d. As expected, at small $q$, the ansatz $\chi^{\text{d,ex}}_{\pm}(q)$ shows a divergence due to violation of the compressibility sum-rule. This can be avoided by the use of $W^{\chi}_{\pm}$ in this $q$-range. Note, that a noticeable disagreement with $\Re \chi_{\pm}(q,0)$ appears at small $q$ when $W_{\pm}^{\text{ex}}$ drops below $W_{\pm}^{\chi}$. 

Similar observations hold also for the layer spacing, $d=0.15$ and $d=0.10$, characterized by a stronger interlayer coupling. Here, the largest deviations to the exact result, $\chi_{\pm}$, are observed for the bare potential~(\ref{w_d}). For $d \leq 0.1$ this type of approximation can not be used for prediction of structural changes, as the density wave formation. The analyses of the phase transition from a homogeneous to an inhomogeneous density phase, similar to Ref.~\cite{Tanat2015}, should be restricted to larger layer spacing. The use of $W^{\text{ex}}$ in combination with $W^{\chi}$ at small $q$ can produce more accurate results. Indeed, for $qa \gtrsim 2$ we observe a nice agreement between $\chi_{\pm}$ and $\chi^{\text{ex}}_{\pm}$. For $qa \lesssim 2 $ the potential $W^{\chi}_{\pm}$ helps to avoid the divergence as observed in $\chi^{\text{d,ex}}_{\pm}$.

\begin{figure}
 \begin{center} 
\hspace{-0.0cm}\includegraphics[width=0.51\textwidth]{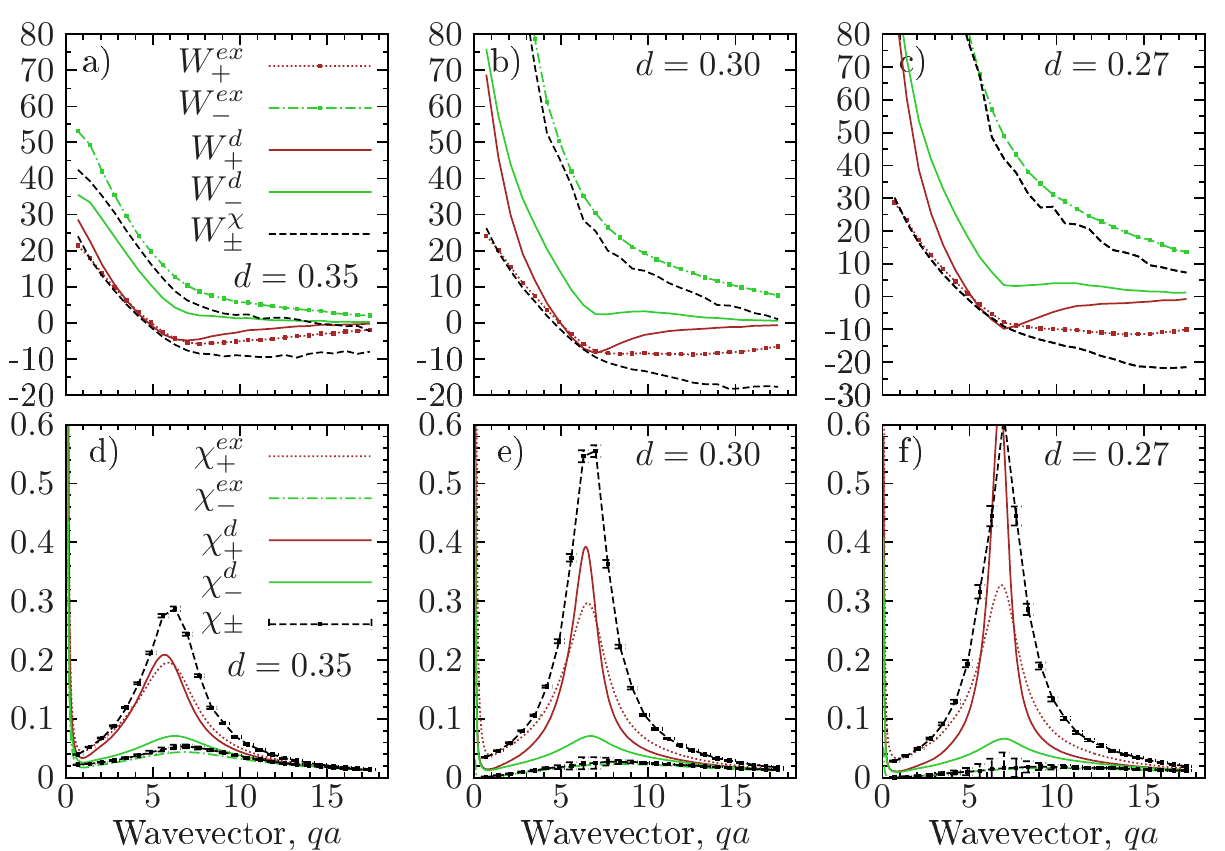}
%{../../../quasi2d/New/d55ed40/data/2d/m1new/d1/effupm}
 \end{center}
\vspace{-0.5cm} 
\caption{Same as in Fig.~\ref{fig:effupmD01} for $D=1.0$ and the layer spacing $d=0.35,0.30,0.27$.}
 \label{fig:effupmD1}
 \end{figure}

\begin{figure}
 \begin{center} 
\hspace{-0.0cm}\includegraphics[width=0.51\textwidth]{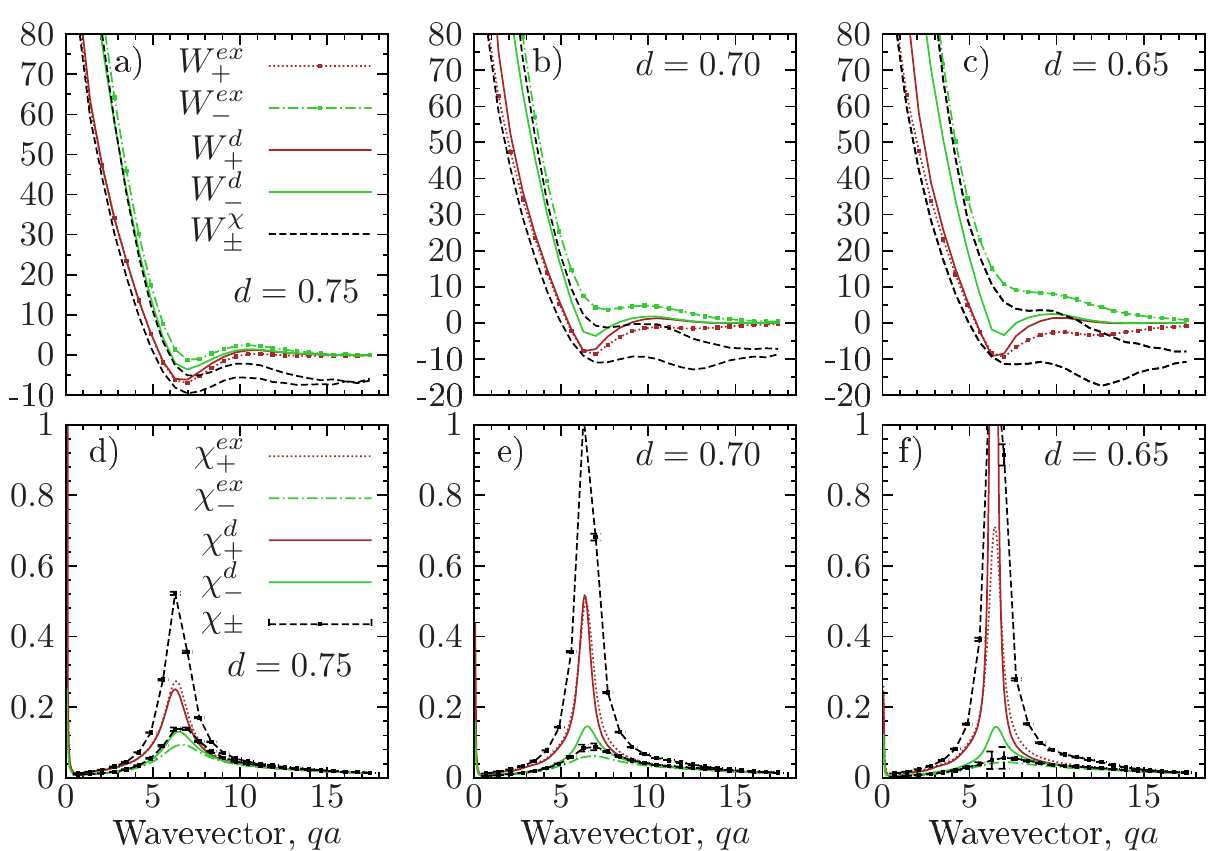}
%{../../../quasi2d/New/d55ed40/data/2d/m1new/d55/effupm}
 \end{center}
\vspace{-0.5cm} 
\caption{Same as in Fig.~\ref{fig:effupmD01} for $D=5.5$ and the layer spacing $d=0.75,0.70,0.65$. }
 \label{fig:effupmD55}
 \end{figure}

In Figs.~\ref{fig:effupmD1},\ref{fig:effupmD55} we repeat similar analyses for the dipole coupling $D=1$ and $D=5.5$. For large layer  spacing, $d=0.35$ ($D=1$) and $d=0.70, 0.75$ ($D=5.5$) the deviations between all three EP [Eqs.~(\ref{w_d}),(\ref{w_ex}) and~(\ref{w_chi})] are minimal. At smaller $d$, the predictions based on the exchange potential remain quite accurate for $qa \lesssim 7$, whereas with the pure interlayer potential~(\ref{v12_q}) demonstrate steadily increasing deviations. 

The static response function $\chi^{\text{ex}}_{\pm}$ reproduces $\chi_{\pm}$ at $qa \lesssim 2$, and shows its divergent character only for small $q$. The divergence region  shrinks significantly with the coupling strength $D$. For $D=5.5$ it can be hardly resolved and is restricted to $qa \ll 1$, see Fig.~\ref{fig:effupmD55}d,e,f. The onset of this unphysical behavior is observed when $W_{\pm}^{\text{ex}} < W_{\pm}^{\chi}$.

In general, for $D=1(5.5)$ the approximation based on the EP.~(\ref{w_d})(\ref{w_ex}) fails to quantitatively reproduce the form and the peak height of the static response function $\chi_{\pm}(q)$ for $qa > 2$. There is a significant underestimation of the halfwidth of the main peak. 

Next, we discuss the dispersion of collective modes, predicted by the RPA, by analyzing  singularities of the density response function~(\ref{re_rpa}). We end up with the result
\begin{align}
 \omega_{\pm}^2(q)=\epsilon_q^2+2 \rho \epsilon_q W_{\pm}(q).
 \label{disp_ueff}
\end{align}
The substitution of the exchange potential~(\ref{w_ex}) provides the ansatz which coincides with the Feynman mode, $\omega^{\text{ex}}_{\pm}(q)=\epsilon_q/S_{\pm}(q)=\omega^{f}_{\pm}(q)$. With the EP~(\ref{w_chi}) we get in a new estimate, $\tilde\omega^{\chi}_{\pm}=\sqrt{2\rho \epsilon_q/\Re \chi_{\pm}(q,0)}$. 

Both results can be expressed as the ratio of the frequency power moments, which similar to~(\ref{border}), form a sequence of upper bounds for a true dispersion relation
\begin{align}
&\omega_\pm(q)\leq \tilde \omega^{\chi}_{\pm}(q) \leq \omega^{f}_{\pm}(q),\\
&\omega_\pm(q) \leq \sqrt{\frac{\Avr{\omega^1}_\pm}{\Avr{\omega^{-1}}_\pm}} \leq \frac{\Avr{\omega^1}_\pm}{\Avr{\omega^0}_\pm}.
\end{align}
Note, that the similar inequality holds also for the effective potentials, $W^{\chi}_{\pm}(q) \leq W^{\text{ex}}_{\pm}(q)$, as is demonstrated in Figs.~\ref{fig:effupmD01},\ref{fig:effupmD1},\ref{fig:effupmD55}.

In Figs.~\ref{fig:m1d1a},\ref{fig:m1d55a},\ref{fig:m1d01a} we provided a comparison of $\omega^{f}_{\pm}(q)$ with the more accurate two-mode ansatz, $\omega^L_{\pm}$. The new solution introduced here as $\tilde \omega^{\chi}_{\pm}(q)$ is an improvement to $\omega^{f}_{\pm}(q)$, but is only an upper bound for $\omega^{\chi}_{\pm}(q)$ [Eq.~(\ref{w_chi0}]. This follows from the
general relation~\cite{lipp} 
\begin{align}
 \Avr{\omega^k}^2 \leq \Avr{\omega^{k-1}}\Avr{\omega^{k+1}}
\end{align}
and results in the sequence of upper bounds 
 \begin{align}
  \frac{\Avr{\omega^k}}{\Avr{\omega^{k-1}}} \leq \sqrt{\frac{\Avr{\omega^{k+1}}}{\Avr{\omega^{k-1}}}} \leq \frac{\Avr{\omega^{k+1}}}{\Avr{\omega^{k}}}.
 \end{align}
The use of the direct potential in~(\ref{disp_ueff}) does not satisfy any sum-rules, and, hence, the obtained dispersion relation does not represent either a lower or an upper bound. At strong interlayer coupling, $d=0.27$($D=1$) and  $d=0.65$($D=5.5$), corresponding results significantly overestimate the peak height of the static response function. The effective potential predicts a much deeper roton minimum (around $qa \sim 7$), being in disagreement with  more accurate and physically grounded two-mode solution $\omega^L_{\pm}(q)$.

We can summarize, that the PRA-type ansatz does not provide any new information on the dispersion relations. In its more accurate form, involving $W^{\text{ex}}_\pm$ or/and $\tilde W^{\chi}_\pm$, it reproduces the results which can be obtained by the method of moments (Sec.\ref{mom1}).

\section{Conclusion}\label{concl}

Recent progress with ultra-cold polar molecular gases motivates the analyses in the spatial geometries, where a role of the anisotropic dipole-dipole interaction is more prominent. One
example is the vertically polarized quantum gases in a quasi-2D bilayer. Such a system undergoes dramatic changes in the collective and single particle properties as the layer separation is varied.

In the present work we performed a detailed study of this problem. The diagonalization of the density response function allows to analyze the excitation spectrum in terms of the symmetric/antisymmetric mode $S_\pm(q,\omega)$ [particles in two layers oscillate in phase/out-of-phase]. The systematic analyses of these modes and their dependence on the dipolar coupling strength $D$ (controlled via dipole moment, particle mass and inlayer density), the layer spacing $d$, the superfluid fraction $\gamma_s$ and temperature are presented. The dynamic structure factor, $S_{\pm}(q,\omega)$, is reconstructed from the imaginary-time density response function via the stochastic optimization method.~\cite{fil2012} During the reconstruction the $\langle \omega^1 \rangle$ and $\langle \omega^3\rangle$ power moments are satisfied exactly, and $\langle \omega^0 \rangle$ and $\langle \omega^{-1} \rangle$ within the statistical error bars. 

For the three cases, classified here as strong, moderate and weak coupling, we discussed  characteristic features of the excitation spectrum. The predicted dispersion relations of collective modes can be used as a practical tool in experiments to readout a thermodynamic state of a dipolar gas. The stronger is the inlayer dipolar coupling $D$, the larger is the critical layer spacing $d$ for the onset of interlayer dimerization, accompanied by a fast reduction of the inlayer superfluidity. This leads to formation of the strongly bound dimers characterized by an enhanced coupling $D^\star > D$. 

In this regime the spectrum of the symmetric mode, $S_+(q,\omega)$, is strongly influenced by the interlayer dimerization and reduction of the inlayer superfluidity. With the formation of strongly bound dimer states, it demonstrates a strong rotonization (a deep roton minimum).
For $D=5.5$ we even observe the onset of dimer crystallization in a Wigner lattice. This can be identified from the behavior of the intra(inter)-layer pair correlation functions by the 
appearance of the pronounced (quasi)long-range order below some characteristic $d$-value.

The spectrum of the antisymmetric mode, $S_-(q,\omega)$, shows a clear dependence on the inlayer superfluidity. In a pure superfluid/normal phase we recover an acoustic/optical(gapped) mode, correspondingly. In contrast, in a partially superfluid phase, both are present simultaneously, and the dispersion splits into two branches corresponding to a normal and a superfluid component. We demonstrated that the spectral weight of the acoustic mode scales linearly with $\gamma_s$. This testifies that its origin is related with the fluctuations of the superfluid density. When $\gamma_s$ is reduced, the weight the acoustic mode decreases and transfers to the optical branch interpreted as a response of a normal component. The latter dominates the spectrum $S_-(q,\omega)$ when dimer states are formed.

The acoustic mode completely vanishes from $S_-(q,\omega)$, once  we repeat our analyses for distinguishable particles (``boltzmannons''). Here only the optical branch is recovered.  The gap value, $\omega_-(q \rightarrow 0)$, increases by lowering the layer spacing, and can be quite accurately predicted from the sum-rules analyses. 

In addition to the reconstruction of a full dynamic structure factor, we developed a more simplified treatment based on a generalized (canonical) solution of the momentum-problem. We have introduced the two-mode ansatz for the density response function, $\Im \chi_{\pm}(q,\omega)$, which satisfies four frequency power moments of the spectral density $S_{\pm}(q,\omega)$. The obtained dispersion relations, $\omega^L_{\pm}(q)$, quite accurately reproduce the positions of the low-frequency resonances in  $S_{\pm}(q,\omega)$ for most of the considered layer spacing $d$ and dipolar coupling $D$. The predictions become inaccurate in two cases. First, in a partially superfluid phase, when the spectrum of the antisymmetric mode splits into one acoustic and one optical branch, and when their contribution (spectral weight) to the static structure factor $S_{\pm}(q)$ becomes comparable. And second, for the symmetric mode at the wavenumbers beyond the roton. Here the spectral density shows two characteristic maxima. The first maximum is positioned slightly below the recoil energy $\epsilon_q$ and corresponds to the multiexcitation continuum. The second peak, and the corresponding low-frequency branch beyond the roton wavenumber, has been observed also in a single layer system. It can be explained by the quasiparticle decay processes into different combinations of quasiparticles with lower energy, e.g. two rotons for $D=1(5.5)$ or phonons for $D=0.1$. More detailed discussion can be found in Ref.~\cite{fil2012}

Both methods, the stochastic reconstruction and the method of moments, clearly demonstrate the presence of the gapped optical mode in the non-superfluid/partially superfluid phase of dipolar bosons. Here it is worth to mention the long-standing theoretical issue.~\cite{qlca_gap} The effects of interlayer and intralayer correlations, essential for strongly coupled bilayers, have been addressed in STLS~\cite{stls} and QLCA.~\cite{qlca} The two methods arrive at different predictions regarding the out-of-phase mode: the QLCA predicts a nonzero energy gap as $q\rightarrow 0$, while the STLS does not. Our results for $S_-(q,\omega)$ in Figs.~\ref{fig:m1d1a},~\ref{fig:m1d55a} validate that the presence of the gap can be strongly influenced, besides the many-body correlations, also by quantum statistics. The inlayer superfluidity in Bose systems can completely mask the interlayer correlations, which are responsible for formation of the gap, and lead to the acoustic dispersion in the long wavelength limit. Note, that the upper bound for the dispersion relation based on the third power moment, $\omega^3_-(q)=\sqrt{\Avr{\omega^3}_-/\Avr{\omega}_-}$, completely overlooks this physical effect and becomes  unreliable in the weakly coupled regime ($D=0.1$). Here, see Fig.~\ref{fig:m1d55a}, the gap value, $\omega^3_-(0)$, and the whole dispersion relation are shifted to the high-frequency domain being in a strong contradiction with $S_{-}(q,\omega)$. This result is expected due to the intrinsic restriction of QLCA,~\cite{qlca} originally derived for moderate and strongly interacting liquids. Indeed, for the strong coupling $D=5.5$ (see Fig.~\ref{fig:m1d55a}) the agreement with $S_{-}(q,\omega)$, at least in the phonon-maxon range and for the optical gap $\omega_-(0)$, is significantly improved.

In conclusion, we demonstrated how the interlayer dimerization in dipolar bilayers can be uniquely identified by the static, energy and dynamic characteristics. In particular, the formation of dimer states leads to: i) the increase/decrease of the thermodynamic sound speed/inlayer compressibility; ii) the shift of the acoustic dispersion to larger frequencies; iii)  the reduction of the roton gap in the symmetric mode for $D\geq 1$; iv) the increase of the optical gap of the antisymmetric mode in the non-superfluid/partially superfluid dimer phase; v) in the acoustic and roton part of the in-phase density excitation spectrum,  the excitation lifetime is increased, compared to a single layer, due to the mass ($m^\star=2m$) and the dipolar coupling ($D^\star > D$) effects. All these features should be present and observable in the partial inlayer dynamic structure factor, $S_{11}(q,\omega)=\frac{1}{2}[S_{+}(q,\omega)+S_{-}(q,\omega)]$, which is accessible with the available experimental techniques.~\cite{baranov,dysp,erbium} 

\section{Acknowledgement}
This work is supported by the Deutsche Forschungsgemeinschaft via project FI 1252/2.

\section{Appendix}
 
\subsection{Reconstruction of spectral density: method of stochastic optimization}\label{SO} 
In its core, the method of stochastic optimization~\cite{Mish,fil2012} (SO) solves, by a stochastic sampling, the minimization problem of the least deviation
\begin{align}
 &D_n[\tilde{G}_n]=\int_0^{\beta} \abs{1-\tilde{G}_n(q,\tau)/G(q,\tau)} \, d\tau,\label{dn}\\
&D_{\text{min}} \approx \sum_{\tau_i} \Delta \tau \abs{\delta G(q,\tau_i)} G^{-1}(q,\tau_i), 
\end{align}
where $\delta G(q,\tau_i)$ is the statistical error of the correlation function $G(q,\tau_i)$ numerically evaluated (e.g. by QMC) at a set of imaginary time points $\tau_i$ ($\Delta \tau=\tau_{i+1}-\tau_i$). The trial function $\tilde{G}_n$ is generated from the Laplace transform of a trial spectral density $\tilde{S}_n(q,\omega)$
\begin{align}
\tilde{G}_n(q,\tau)=\int_{-\infty}^{\infty} e^{-\tau \omega} \tilde{S}_n(q,\omega) \, d\omega. \label{gngen}
\end{align}
Similar relation holds between the dynamic structure factor and the density-density correlation function, see  Eq.~(\ref{g2}).

As a final result of the reconstruction we take a linear combination of all trial solutions (ensemble average)
\begin{align}
 S_{\text{SO}}(q,\omega)=\Avr{\tilde{S}_n(q,\omega)}
\end{align}
which satisfy the acceptance criteria
\begin{align}
D_n[\tilde{G}_n] \leq D_{\text{min}}.  
\label{accept}
\end{align}
A specific stochastic Monte Carlo sampling algorithm~\cite{Mish} is used to probe a wide class of functions parameterized into some basis set and select those which satisfy~(\ref{accept}).

The quality of the reconstructed spectra $\Avr{\tilde{S}_n(q,\omega)}$ can be judged based on the estimated deviation from the QMC data
\begin{align}
 &\delta^r G(q)=\int\limits_0^\beta \, \db \tau  \left| 1-\Avr{\tilde{G}_n(q,\tau)}/G(q,\tau)\right|,\label{gl_dG}\\
 & \Avr{\tilde{G}_n(q,\tau)}=\int_{-\infty}^{\infty} d\omega \, e^{-\tau \omega} \Avr{\tilde{S}_n(q,\omega)}.
 \label{devG}
\end{align}

In the numerical implementation it is convenient to make parametrization of trial solutions in the basis of rectangular functions
\begin{align}
&S(q,\omega)=\sum_{m=1}^{K^q} P_m^q(\omega),\\
&P_m^q(\omega)=
\begin{cases}
h_m^q,&\omega \in \left[c_m^q-\frac{w_m^q}{2},c_m^q+\frac{w_m^q}{2}\right] \\
0,&\text{otherwise},
\end{cases} 
\end{align}
here $K^q$ denotes a number of rectangles used for a wavenumber $q$. This number can significantly vary depending on a form of spectral density, e.g. either it consists of a single sharp resonance, a broad multiexcitation continuum or a combination of both. The  algorithm decides which number $K^q$ is more appropriate, and performs a stochastic sampling of the fit-parameters $\{c_m^q,w_m^q,h_m^q\}$.

With the choice of rectangular functions some important properties of the reconstructed  spectral density can be written down analytically. These include the imaginary time density-density correlation function and a set of frequency power moments
\begin{align}
G(q,\tau)=&
\begin{cases}
S(q), \quad \tau=0  \\
2\tau^{-1}\sum\limits_m h_m^q e^{-c_m^q \tau} \sinh (w_m^q \tau/2) ,&\tau\neq 0
\end{cases},\\
\Avr{\omega^0}=&\sum_{m=1}^{K^q} h_m^q \left[w_m^q+\frac{2}{\beta} e^{-\beta c_m^q}\sinh \frac{w_m^q \beta}{2} \right],\label{m0}\\
\avr{\omega^1}=& \sum\limits_{i=1}^{N} h_m^q w_m^q c_m^q- \frac{2 h_m^q}{\beta} e^{-\beta c_m^q} \left[\frac{1}{\beta}+c_m^q \right] \sinh \frac{w_m^q \beta}{2} \nonumber \\
 &+ \frac{h_m^q w_m^q}{\beta}e^{-\beta c_m^q} \cosh \frac{w_m^q \beta}{2}, \label{m2} \\
 \avr{\omega^{-1}}=& \sum\limits_{i=1}^{N} h_m^q \ln \frac{c_m^q+\frac{w_m^q}{2}}{c_m^q-\frac{w_m^q}{2}} \nonumber \\
 &-h_m^q \left[\text{Ei}[-\beta (c_m^q+\frac{w_m^q}{2})]-\text{Ei}[-\beta (c_m^q-\frac{w_m^q}{2})] \right],\label{m-1}
\end{align}
with $\text{Ei}(x)=\int_{-\infty}^{x} \db t\, e^{t}/t$.

The expression for the third-frequency-moment takes the form
 \begin{align}
\Avr{\omega^3}&=\int\limits_0^{\infty}\db \omega\, \omega^3(1-e^{-\beta \omega})\sum_{m=1}^{K^q} P_m^q(\omega)\nonumber\\
  &=\sum_{m=1}^{K^q} \left(I^q_{m}+J^q_{m}\right),\label{m3}
 \end{align}
where the last two terms are expressed as
 \begin{align}
&I^q_m= h_m^q \left(\omega_+^4 -\omega_{-}^4\right)/4,\\
&J^q_m=h_m^q \left[F(\omega_+)-F(\omega_-)\right]/\beta^4,\\
&F(\omega)=e^{-\beta \omega} \left[ 6 +6 \beta \omega +3 (\beta \omega)^2 +(\beta \omega)^3 \right].
 \end{align}
Here we have introduced the end-points, $\omega_{\pm}=c_m^q\pm\omega_m^q/2$. All power moments get their wavenumber-dependence via the $q$-dependent fit parameters $\{c_m^q,w_m^q,h_m^q\}$.

An important improvement of the method is to include information available from the frequency power moments, see Sec.~\ref{one-comp}. During the Monte Carlo sampling, we in addition minimize the deviations of the first and third power moments~(\ref{m2}),(\ref{m3}) from their reference values known from~(\ref{w1}),(\ref{w3}). This is done by  inclusion of two additional deviation measures $\delta^r  \Avr{\omega^k}(q)$ ($k=1,3$) in the acceptance criteria~(\ref{accept}). This procedure makes the reconstruction results more stable with respect to the statistical noise $\delta G(q,\tau_i)$ and significantly reduces a class of possible solutions. Below we demonstrate accuracy of the reconstruction procedure on two examples.

\subsection{Examples of the reconstruction}\label{improv} 
 
 \begin{figure}
 \begin{center} 
 \hspace{-0.50cm}\includegraphics[width=0.495\textwidth]{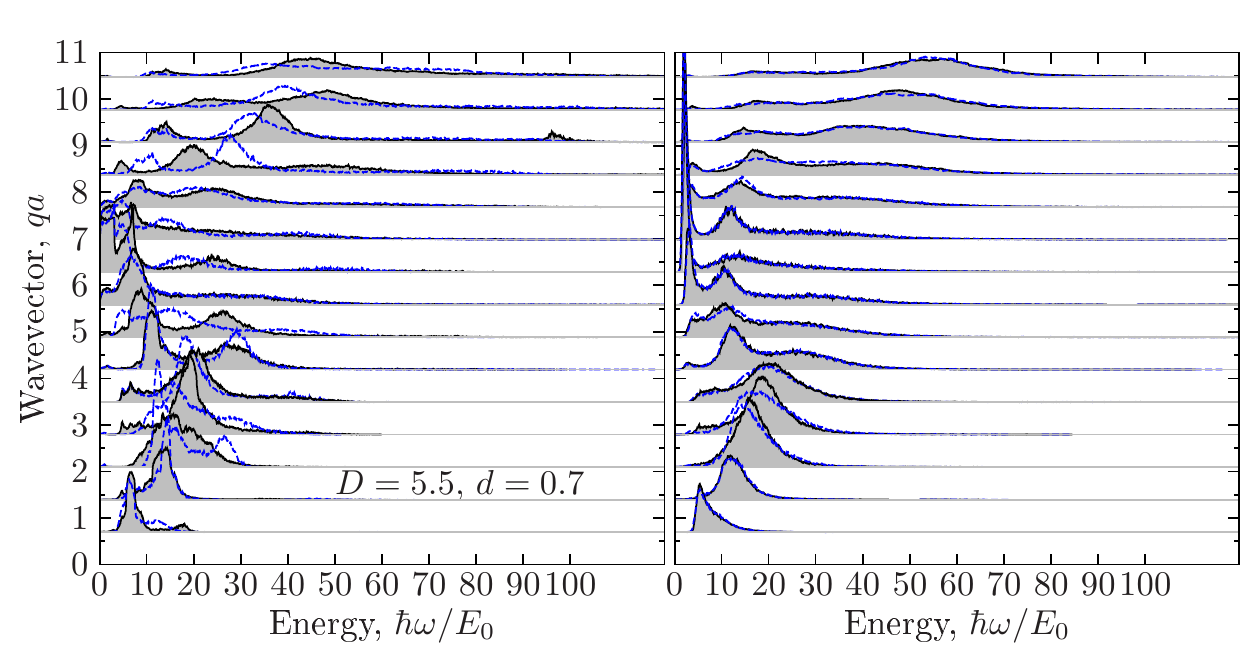}
 %{../../../quasi2d/New/d55ed40/data/2d/m1new/d55/d55l07b1/sto/w1w3figs/specd55l07}
 \end{center}
 \vspace{-0.7cm}
 \caption{Comparison of the dynamic structure factor $S_{\al\al}(q_n,\omega)$ ($\al=1,2$) in the mass-symmetric bilayer for a set of wavenumbers $q_n=2\pi n/L$ ($L=9$). Left(right) panel is without(with) minimization of $\delta^r  \Avr{\omega^k}(q)$ ($k=1,3$). The reconstruction on the left panel demonstrates much larger deviations between  $S_{11}(q,\omega)$ and $S_{22}(q,\omega)$. Here the statistical errors  in the density correlation function allows for a broader class of different trial solutions which satisfy~(\ref{accept}).}
 \label{fig:spec-now3}
 \end{figure}
 
For the mass-symmetric bilayer, the stability of the reconstruction with respect to the statistical noise $\delta G$ can be checked by comparison of the dynamic structure factor $S_{\al\al}(q,\omega)$ of two layers ($\al=1,2$). They should coincide due to the symmetry reason. In general, the reconstruction results can differ as the density correlation function $G_{\al\al}(q,\tau)$ is evaluated independently for each layer. A similarity/discrepancy of the reconstructed spectral densities $S_{11}$ and $S_{22}$ will characterize quality of the reconstruction and the effect of statistical noise. In the following we present results for a dipolar bilayer specified by the parameters $\{\beta=1, D=5.5\}$ at the layer spacing $d=0.7$.

\begin{figure}
 \begin{center} 
 \hspace{-0.40cm}\includegraphics[width=0.48\textwidth]{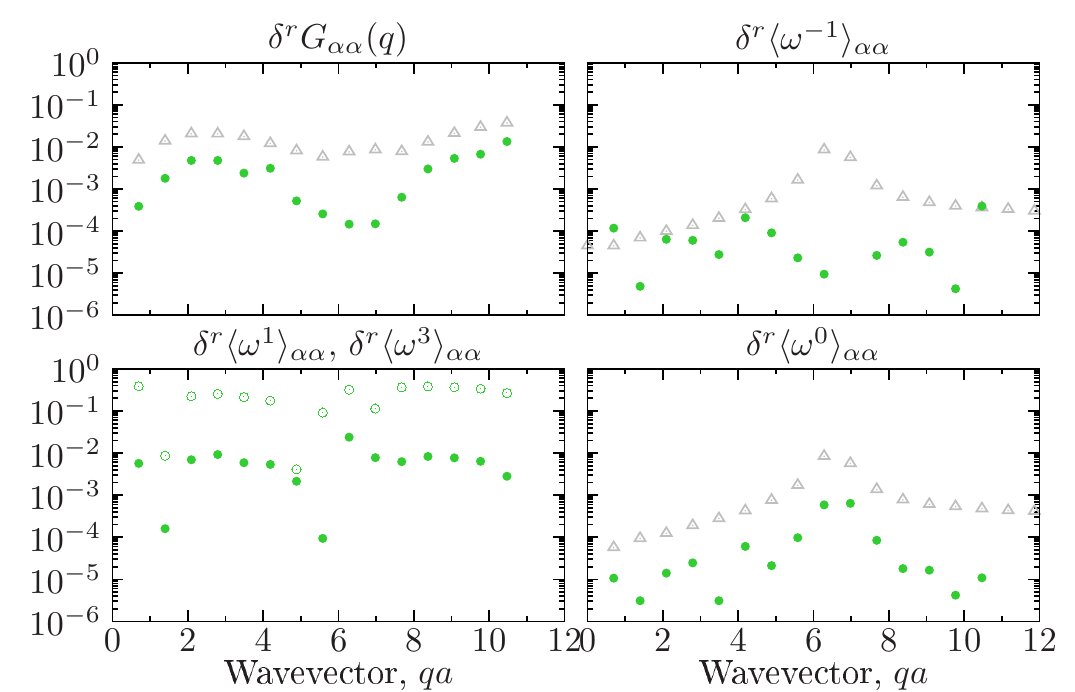}
 %{../../../quasi2d/New/d55ed40/data/2d/m1new/d55/d55l07b1/sto/now3/errgtau}
 \end{center}
 \vspace{-0.6cm}
 \caption{Absolute relative deviations of the spectral power moments~(\ref{m0})-(\ref{m3}) from the reference values~(\ref{f-sums-equation}). The upper bound is given the relative errors of~(\ref{f-sums-equation}) shown by open gray triangles. The largest deviations are observed in the third power moment (shown by open circles), up to $\delta^r \Avr{\omega^3} \sim 0.38$ ($38\%$).}
 \label{fig:errfsums-now3}
 \end{figure}
 
The results are shown in Fig.~\ref{fig:spec-now3}. In the first case, the reconstruction is performed with no constrains on the power moments $\Avr{\omega}$ and $\Avr{\omega^3}$. The recovered spectrum $S_{\al\al}(q,\omega)$ ($\al=1,2$) corresponds to the ensemble average which produces the best fit to $G_{\al\al}(q,\tau)$ within the statistical error bars. From Fig.~\ref{fig:spec-now3}(left panel) we find that the spectral densities are quite different. In particular, for the wavenumber $q_3$ two possible solutions are found during the reconstruction. One with a broad peak and the second with two narrow peaks. Both spectral densities fit within the statistical error bars the corresponding correlation functions, $G_{\al\al}(q_n,\tau)\pm \delta G_{\al\al}(q_n,\tau)$. The similar situation is observed for the wavenumbers $q_4$ and $q_7$. For $q_{12}\sim 8.5$ there are two peaks in both spectral densities with a broad continuum at high frequencies, but the peak position is different.
 
To characterize the obtained results we can evaluate the spectral power moments and compare with the reference values~(\ref{f-sums-equation}).
The first moment, i.e. the $f$-sum rule, is known explicitly, $\Avr{\omega^1}=q^2/2m$. Other moments, $\Avr{\omega^k} (k=-1,0,3)$, are estimated numerically and contain  statistical relative errors $\delta^r$ shown in Fig.~\ref{fig:errfsums-now3} by the open grey triangles. The $\Avr{\omega^3}$-sum rule can be quite accurately estimated via~(\ref{w3aa}) and has the relative error below $10^{-6}$.
The relative errors present an upper bound for possible deviations with the spectral moments evaluated via~(\ref{m0})-(\ref{m3}). These deviations for different wavenumbers are presented in Fig.~\ref{fig:errfsums-now3}. The deviations in $\Avr{\omega^k}$ ($k=-1,0$) are well below the upper bound. The reconstructed spectra, however, fail to satisfy the first and third power moment. For $\Avr{\omega^3}$ the relative error can reach up to $38\%$. Note, that $\Avr{\omega^3}$-sum rule is determined by the high-frequency behavior of the spectral density specified by the high-frequency resonances (if any) and the multiexcitation continuum. Hence, large errors in $\delta^r\Avr{\omega^3}$ signal that these features are not reproduced correctly. This information, however, is very important for the bilayers where the high-frequency behavior is dominated by the out-of-phase density excitations. In contrast, the $\Avr{\omega^{0(-1)}}$-sum rules are dominated by the low-frequency phonon-maxon-roton branch due to the in-phase excitations. Hence, the accuracy of the reconstruction, based on the criteria~(\ref{gl_dG}), is guaranteed only for the low-frequency domain. All high-frequency features in $S(q,\omega)$ are exponentially damped, see Eq.~(\ref{gngen}), and, correspondingly, make only a small contribution to the imaginary-time density correlation function used for the fit. This contribution is disturbed significantly due to the presence of the statistical noise (same order of magnitude), and can not be resolved correctly during the reconstruction.

\begin{figure}
 \begin{center} 
 \hspace{-0.40cm}\includegraphics[width=0.48\textwidth]{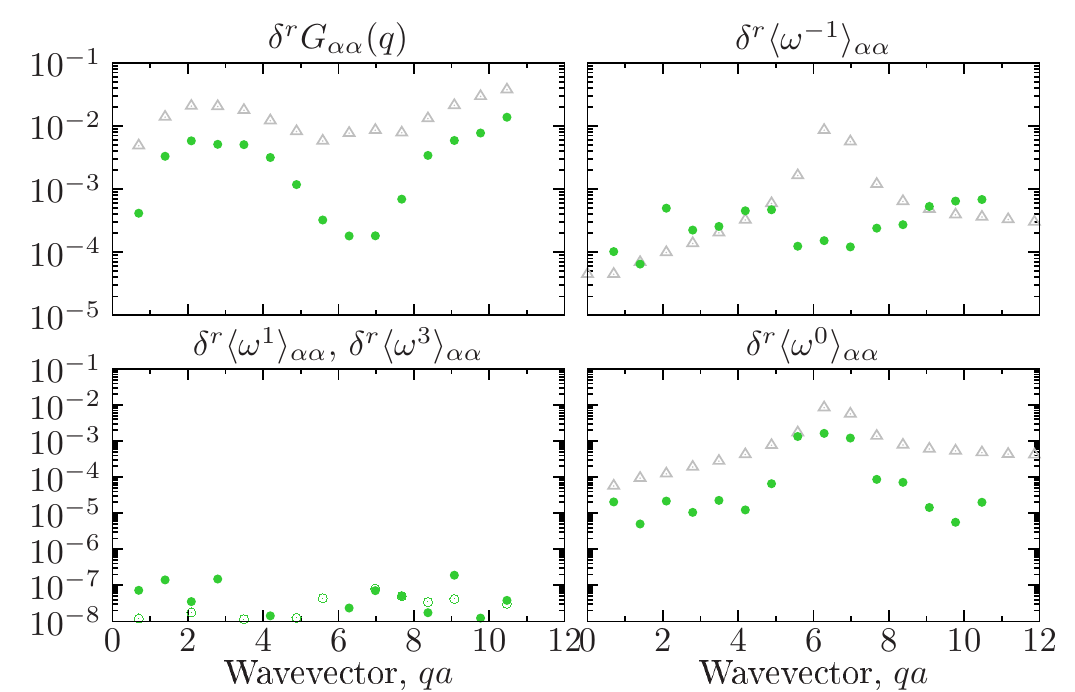}
 %{../../../quasi2d/New/d55ed40/data/2d/m1new/d55/d55l07b1/sto/w3/errgtau}
 \end{center}
 \vspace{-0.6cm}
 \caption{The same as in Fig.~\ref{fig:errfsums-now3}  with additional minimization of $\delta^r  \Avr{\omega^k}(q)$ ($k=1,3$).}
 \label{fig:errfsums-w1w3}
 \end{figure}

In the second example, we perform the reconstruction when in the stochastic algorithm two additional deviation measures, $\delta^r  \Avr{\omega^k}(q)$ ($k=1,3$), are minimized. With the annealing steps used in the stochastic optimization, these moments are satisfied with the relative error $\delta^r \lesssim 10^{-7}$, see Fig.~\ref{fig:errfsums-w1w3}. The improved spectral densities $S_{11}$ and $S_{22}$ are shown in Fig.~\ref{fig:spec-now3}(right panel) and demonstrate much better convergence for the same level of statistical noise $\delta G$ as in the first example. In particular, in the roton region the spectral shape demonstrates more systematic behavior and shaper resonances. This behavior is reproduced for both layers. The relative deviations from all available frequency power moments are within the allowed statistical errors, see Fig.~\ref{fig:errfsums-w1w3}.

To conclude, the addition of the first and third power moments in the reconstruction algorithm can significantly improves the convergence in both low- and high-frequency domains. This approach is always advantages over the minimization of the single deviation measure determined solely by the imaginary-time correlation function~(\ref{accept}).

\end{document}